\documentclass[journal]{IEEEtran}

\usepackage[pdftex]{graphicx} 

\usepackage{amsmath, amssymb, amsthm}
\usepackage{booktabs}
\usepackage{rotating}
\usepackage{multirow}
\usepackage{url}
\usepackage{color}
\usepackage{enumerate}
\usepackage{orcidlink}

\usepackage{cite}

\usepackage{algorithmic}
\usepackage{algorithm}

\newcommand{\algorithmicinitialize}{\textbf{Initialize:}}
\newcommand{\INITIALIZE}{\item[\algorithmicinitialize]}

\usepackage[caption=false,font=normalsize,labelfont=sf,textfont=sf]{subfig}

\usepackage{xcolor}
\newcommand{\argmin}{\mathop{\rm argmin}\limits}

\def\x{{\mathbf x}}

\def\X{{\mathbf X}}

\DeclareMathOperator{\prox}{prox}

\DeclareMathOperator{\sgn}{sgn}

\def\Xcal{{\boldsymbol{\mathcal X}}}
\def\Bcal{{\boldsymbol{\mathcal B}}}
\def\Acal{{\boldsymbol{\mathcal A}}}
\def\Scal{{\boldsymbol{\mathcal S}}}
\def\Ncal{{\boldsymbol{\mathcal N}}}
\def\Lcal{{\boldsymbol{\mathcal L}}}
\def\Ycal{{\boldsymbol{\mathcal Y}}}

\def\Ocal{{\boldsymbol{\mathcal O}}}

\def\Vcal{{\boldsymbol{\mathcal V}}}

\def\Tcal{{\boldsymbol{\mathcal T}}}

\def\Ordnung{{\mathcal O}}

\def\real{{\mathbb R}}

\def\regu{{R}}
\def\proj{{P}}
\def\reshapeX{{\mathrm{mat}}}
\def\tempY#1{\widetilde{#1}}

\def\op{{\mathrm{op}}}

\def\Lfrak{{\mathfrak{L}}}
\def\Dfrak{{\mathfrak{D}}}
\def\Ifrak{{\mathfrak{I}}}

\def\Afrak{{\mathfrak{A}}}

\global\long\def\ValBest#1{\textbf{#1}}
\global\long\def\ValSecond#1{\underline{#1}}

\global\long\def\ValThird#1{\textcolor{black}{#1}}
\global\long\def\ValFourth#1{\textcolor{black}{#1}}
\global\long\def\ValFifth#1{\textcolor{black}{#1}}

\begin{document}
\bstctlcite{IEEEexample:BSTcontrol}

\title{Joint Background--Anomaly--Noise Decomposition \\ for Robust Hyperspectral Anomaly Detection \\ via Constrained Convex Optimization}
\author{Koyo~Sato
,~\IEEEmembership{Graduate Student Member,~IEEE,} 
and 
Shunsuke~Ono
,~\IEEEmembership{Senior Member,~IEEE}
\thanks{K. Sato is with the Department of Computer Science, Institute of Science Tokyo, Yokohama, 226-8503, Japan (e-mail:
sato.k.bc04@m.isct.ac.jp).}
\thanks{S. Ono is with the Department of Computer Science, Institute of Science Tokyo, Yokohama, 226-8503, Japan (e-mail: ono@comp.isct.ac.jp).}
\thanks{This work was supported in part by JST FOREST under Grant JPMJFR232M and JST AdCORP under Grant JPMJKB2307, and in part by JSPS KAKENHI under Grant 22H03610, 22H00512, 23H01415, 23K17461, 24K03119, 24K22291, 25H01296, and 25K03136, and in part by JST SPRING, Japan under Grant JPMJSP2180.
(Corresponding author: Koyo Sato.)
}
\thanks{
Data is available online at https://github.com/MDI-ScienceTokyo/Joint-Background-Anomaly-Noise-Decomposition-for-Robust-Hyperspectral-Anomaly-Detection.
}}

\markboth{Journal of \LaTeX\ Class Files,~Vol.~13, No.~9, September~2014}%
{Shell \MakeLowercase{\textit{et al.}}: Bare Demo of IEEEtran.cls for Journals}

\maketitle

\begin{abstract}
We propose a novel hyperspectral (HS) anomaly detection method that is robust to various types of noise. 
Most existing HS anomaly detection methods are designed without explicit consideration of noise or are based on the assumption of Gaussian noise.
However, in real-world situations, observed HS images are often degraded by various types of noise, such as sparse noise and stripe noise, due to sensor failure or calibration errors, significantly affecting the detection performance. 
To address this problem, this article establishes a robust HS anomaly detection method with a mechanism that can properly remove mixed noise while separating background and anomaly parts. 
Specifically, we newly formulate a constrained convex optimization problem to decompose background and anomaly parts, and three types of noise from a given HS image. 
Then, we develop an efficient algorithm based on a preconditioned variant of a primal-dual splitting method to solve this problem.
Experimental results using seven real HS datasets demonstrate that the proposed method achieves detection accuracy comparable to state-of-the-art methods on original images and exhibits significantly higher robustness in scenarios where various types of mixed noise are added.
\end{abstract}


%
\IEEEpeerreviewmaketitle

\section{Introduction}
\label{sec:introduction}
\IEEEPARstart{H}{yperspectral} (HS) images are three-dimensional data comprising two spatial dimensions and one spectral dimension, containing hundreds of contiguous spectral bands covering both visible and near-infrared wavelengths.
Such rich spectral information enables detailed material discrimination that cannot be achieved with conventional RGB or multispectral images. 
This advantage has led to extensive research on HS image analysis techniques, including classification, unmixing, and anomaly detection~\cite{Borengasser2007HSIApplications, Grahn2007Techniques, Thenkabail2016VegetationOverview, Lu2020AgricultureOverview}.

HS anomaly detection is a fundamental task that aims to identify background and anomaly parts within a given HS image.
The background part consists of background pixels that are widely distributed across the image and share similar spectral signatures, whereas the anomaly part contains anomalies whose spectral signatures differ significantly from those of the surrounding background pixels.
In general, anomalies appear as a small set of spatially localized pixels, and what constitutes an anomaly varies with the application scenario~\cite{ADsurvey_2022}.
For instance, in maritime scenes, seawater serves as the background, while ships or people drifting on the sea are regarded as anomalies.
In agricultural monitoring, crops across the field represent the background, whereas diseased regions or immature fruits are treated as anomalies.
Given these application-dependent characteristics, HS anomaly detection plays an essential role in identifying regions of interest without requiring prior knowledge, and has been widely applied to various real-world scenarios, such as search and rescue operations, environmental monitoring, geological exploration, and military defense~\cite{AD_ex_2002,AD_overview_2010,AD_overview_2014,AD_overview_2017,ADsurvey_2022,ADreview_deep_2022}.

Existing HS anomaly detection methods are roughly classified into three groups: statistics-based, decomposition-based, and deep learning-based methods.
Statistics-based methods~\cite{GRX_1990, LRX_2013, zhao2014robust, zhao2015beyond,zhao2016robust, 2SGLRT_2022} are designed based on statistical assumptions about a background part, and detect anomalies as deviations from it.
Decomposition-based methods~\cite{
LRASR_2015, CRD_2015, zhao2017hyperspectral, GTVLRR_2020, PCA-TLRSR_2023, HADGSM_2024, MTVLRR_2024, TDNNTA_2025, liu2025exploiting, ADLR_2018, GoDecAD_2016, LSDMMog_2021, DECNN_2021, DSR_2023, HyADD_2024, MERAETV_2024, TGODECAD_2025, AHMID_2023
} estimate background and anomaly parts by solving an optimization problem that exploits their structural differences.
Deep learning-based methods~\cite{
sparseAE_2015, sparseAE_2019, GAED_2022, RGAE_2022, zhao2024novel,
BS3LNet_2023, yang2024multi,
FS2CCTrans_2025, TransGCF_2026
} employ neural networks to reconstruct a background part and extract anomalies from reconstruction residuals.
Various architectures have been proposed, including autoencoder-based methods~\cite{sparseAE_2015, sparseAE_2019, GAED_2022, RGAE_2022, zhao2024novel},
blind-spot networks~\cite{BS3LNet_2023, yang2024multi}, and recent transformer-based and graph-convolutional-network-based architectures~\cite{FS2CCTrans_2025, TransGCF_2026}.
Beyond these three groups, a learning-free frequency-domain method has also been proposed to detect anomalies using unpredictive residuals in the wavelet and DCT domains~\cite{zhou2022learning}.
Most of these methods are based on the assumption that HS images consist solely of background and anomaly parts, and achieve satisfactory detection performance under such idealized assumptions.

In practical applications, however, HS images are inevitably contaminated with various types of noise due to sensor failure, scanning mechanisms, calibration errors, and other factors. 
Representative types of noise include thermal noise and quantization noise, which are typically modeled as Gaussian noise; impulse noise and missing pixels, which are categorized as sparse noise due to their isolated and random nature; and stripe noise, which manifests as regular linear patterns across the image~\cite{noise_2018}.
These types of noise distort the spectral signatures of HS images and pose a fundamental challenge to accurately separating background and anomaly parts~\cite{EAS_2022}.
In particular, although sparse and stripe noise lack spectral continuity, they often form isolated or periodic patterns similar to anomalies, making their discrimination difficult and increasing the risk of false alarms.
Therefore, it is essential to appropriately handle such mixed noise to ensure robust and reliable anomaly detection.

To mitigate the adverse effect of such mixed noise, several studies have explored two main approaches.
The first approach follows a sequential framework that performs denoising prior to anomaly detection.
As a representative example, the abundance and dictionary-based low-rank decomposition (ADLR)~\cite{ADLR_2018} has been proposed. 
This method extracts abundance vectors while suppressing noise via spectral unmixing, and subsequently utilizes them for anomaly detection.
However, such two-step strategies may suppress subtle spectral signatures essential for anomaly detection during the denoising process, leading to degraded detection performance.

As another approach, several studies have proposed joint frameworks for simultaneous noise suppression and anomaly detection~\cite{GoDecAD_2016, LSDMMog_2021, DECNN_2021, DSR_2023, HyADD_2024, MERAETV_2024, TGODECAD_2025}.
Such integration improves the robustness of anomaly detection under noisy conditions by avoiding the information loss inherent in sequential processing.
However, most of them are designed based on the assumption that the noise superimposed on HS images can be modeled either as Gaussian noise or as a single noise component, without explicitly distinguishing between different noise types. 
Given the distinct characteristics of the noise types discussed above, explicitly modeling each as an independent component is expected to improve detection robustness.

To address mixed noise, the antinoise hierarchical mutual-incoherence-induced discriminative learning (AHMID)~\cite{AHMID_2023} has been proposed.
This method jointly estimates background and anomaly parts along with Gaussian and stripe noise from a given HS image by solving an optimization problem.
Although such an explicit modeling of mixed noise improves robustness, AHMID has several limitations.
First, since the background part is modeled as the product of a dictionary and its coefficient matrix, the detection performance is sensitive to the quality of the preconstructed dictionary.
Second, the algorithmic behavior tends to be unstable because both the dictionary and the coefficient matrix are updated as optimization variables, which requires alternating optimization.
Furthermore, the inclusion of multiple regularization terms for all components in its objective function leads to an interdependence among hyperparameters, which makes parameter tuning laborious.

These limitations motivate a natural question:
\textit{Can we develop a stable and robust HS anomaly detection method that requires no preprocessing and simplifies parameter tuning?}
To address this question, in this article, we propose a novel HS anomaly detection method that can accurately extract an anomaly part from a given HS image corrupted by various types of noise.
Specifically, we newly formulate a constrained convex
optimization problem to decompose background and anomaly
parts, and Gaussian, sparse, and stripe noise from a given HS image. 
Then, we design an efficient solver based on a preconditioned variant of a primal-dual splitting method (P-PDS)~\cite{DP-PDS_2011} with the operator-norm-based design method of variable-wise diagonal preconditioning (OVDP)~\cite{naganuma2023_DPDS}.
The main contributions of this article are as follows:
\begin{itemize}

\item \textit{(Robustness to mixed noise):} 
Most joint methods are based on the assumption that the noise superimposed on HS images is Gaussian or can be represented by a single component.
On the other hand, in the proposed method, Gaussian, sparse, 
and stripe noise are explicitly modeled as three independent 
components.
This allows each noise type to be addressed according to 
its own characteristics.
In addition, unlike AHMID, the background part is characterized without a preconstructed dictionary, eliminating both the need for preprocessing and the dependency on dictionary quality.
These advantages allow the proposed method to maintain high detection accuracy even under realistic and degraded observation conditions.

\item \textit{(Reduction of interdependent hyperparameters):}
The objective function of AHMID includes regularization terms for all components.
On the other hand, in the proposed formulation, instead of adding terms characterizing the various types of noise to the objective function, they are imposed as hard constraints.
This transforms complex interdependent hyperparameters into independent parameters that can be easily set.
The advantages of such constrained formulations have been addressed in the literature of signal recovery, e.g., in~\cite{afonso2010augmented,chierchia2015epigraphical,ono2015signal,ono2017primal,ono2017l_}.

\item \textit{(Stable algorithm design with automatic stepsize selection):} 
Unlike AHMID, which requires alternating optimization, the proposed algorithm is developed based on P-PDS with OVDP~\cite{naganuma2023_DPDS}.
This method can automatically determine the appropriate stepsizes, ensuring stable convergence while simplifying practical implementation.

\item \textit{(Computational efficiency):}
The background part is characterized by total variation (TV) regularization. 
While the optimization process involving the widely used nuclear norm requires a high-cost singular value decomposition at each iteration, the TV regularization can be computed via simple soft-thresholding operations with variable splitting. 
Consequently, the proposed method achieves high computational efficiency.
\end{itemize}

The remainder of this article is organized as follows. 
In Sec.~\ref{sec:preliminaries}, we introduce several mathematical tools required for the proposed method.
Sec.~\ref{sec:proposed_method} presents the problem formulation and optimization algorithm of the proposed method.
In Sec.~\ref{sec:experiments}, we demonstrate the superiority of the proposed method over existing methods including state-of-the-art ones through comprehensive experiments.
Finally, Sec.~\ref{sec:conclusion} concludes this article.

The preliminary version of this work, without mathematical details, the generalization of a background part modeling, more extensive experiments, or deeper discussion, has appeared in conference proceedings~\cite{sato2023robust}.

\section{Preliminaries}
\label{sec:preliminaries}
In this section, we introduce minimal mathematical tools required for the proposed method.
Readers interested in more details are referred to~\cite{bauschke2011convex,beck2017first}.
The notations and definitions used in this article are given in Table~\ref{tab:notation}.

\begin{table}[t]
  \centering
  \caption{Notations and Definitions.}
  \label{tab:notation}
  \scalebox{0.925}{
  \begin{tabular}{cc}
    \toprule
    Notations & Definitions \\
    \cmidrule(lr){1-1} \cmidrule(lr){2-2}
    
    $\real$ & set of real numbers \vspace{1mm} \\
    
    $x$ & scalar, $x \in \real$ \vspace{1mm} \\ 
    
    $\x$ & vector, $\x \in \real^{d_1}$ \vspace{1mm} \\ 
    
    $x_{i}$ & $i$-th element of a vector $\x$ \vspace{1mm} \\ 

    $\| \x \|_2$ & $\ell_2$-norm of a vector $\x$, $\| \x \|_2 := \sqrt{\sum_i x_i^2}$ \vspace{1mm} \\
    
    $\X$ & matrix, $\X \in \real^{d_1 \times d_2}$ \vspace{1mm} \\ 
    
    $\Xcal$ & tensor, $\Xcal \in \real^{d_1 \times d_2 \times d_3}$ \vspace{1mm} \\ 
    
    $x_{i,j,k}, [\Xcal]_{i,j,k}$ & $(i,j,k)$-th element of a tensor $\Xcal$ \vspace{1mm} \\ 
    
    $[\Xcal]_{i,j,:}$ & $(i,j)$-th tube of a tensor $\Xcal$,  $[\Xcal]_{i,j,:} \in \real^{d_{3}}$ \vspace{1mm} \\ 
    
    $\Ocal$ & zero tensor \vspace{1mm} \\ 
    
    \multirow{2}{*}{$\| \Xcal \|_1$} & $\ell_1$-norm of a tensor $\Xcal$, \\ 
    & $\| \Xcal \|_1 := \sum_{i,j,k} |x_{i,j,k}|$ \vspace{1mm} \\ 
    
    \multirow{2}{*}{$\| \Xcal \|_F$} & Frobenius norm of a tensor $\Xcal$, \\ 
    & $\| \Xcal \|_F := \sqrt{\sum_{i,j,k} x_{i,j,k}^2}$ \vspace{1mm} \\ 
    
    \multirow{2}{*}{$\| \Xcal \|_{2,1}$} & $\ell_{2,1}$-norm of a tensor $\Xcal$, \\
    & $\| \Xcal \|_{2,1} := \sum_{i,j} \sqrt{\sum_k x_{i,j,k}^2}$ \vspace{1mm} \\ 
    
    \multirow{3}{*}{$\Dfrak_v$} & vertical difference operator, \\
    & $[\Dfrak_v (\Xcal)]_{i,j,k} :=
    \begin{cases}
      x_{(i+1),j,k} - x_{i,j,k},  & (1 \leq i < d_1) \\
      0,  & (i = d_1)
    \end{cases}$ \vspace{1mm} \\ 
    
    \multirow{3}{*}{$\Dfrak_h$} & horizontal difference operator, \\
    & $[\Dfrak_h (\Xcal)]_{i,j,k} :=
    \begin{cases}
      x_{i,(j+1),k} - x_{i,j,k}, & (1 \leq j < d_2) \\
      0, & (j = d_2)
    \end{cases}$ \vspace{1mm} \\ 
    
    \multirow{3}{*}{$\Dfrak_b$} & spectral difference operator, \\
    & $[\Dfrak_b (\Xcal)]_{i,j,k} :=
    \begin{cases}
      x_{i,j,(k+1)} - x_{i,j,k}, & (1 \leq k < d_3) \\
      0, & (k = d_3)
    \end{cases}$ \vspace{1mm} \\ 
    
    $\Lfrak^\ast$ & adjoint operator of a linear operator $\Lfrak$ \vspace{1mm} \\ 

    \multirow{4}{*}{$\Dfrak_v^*$} & \textcolor{black}{adjoint operator of $\Dfrak_v$,} \\
    & $[\Dfrak_v^*(\Xcal)]_{i,j,k} :=
    \begin{cases}
      - x_{i,j,k}, & (i = 1),\\
      x_{(i-1),j,k} - x_{i,j,k}, & (1 < i < d_1),\\
      x_{(i-1),j,k}, & (i = d_1)
    \end{cases}$ \vspace{1mm} \\

    $\Lfrak_1 \circ \Lfrak_2$ & composition of linear operators $\Lfrak_1$ and $\Lfrak_2$ \vspace{1mm} \\ 
    
    \multirow{2}{*}{$\| \Lfrak \|_{\op}$} &  operator norm of a linear operator, \\
    & $\| \Lfrak \|_{\op} := \sup_{\Xcal \neq \Ocal} \frac{\| \Lfrak (\Xcal) \|_F}{\| \Xcal \|_F}$ \vspace{1mm} \\ 
    
    \multirow{2}{*}{$\mathcal{B}_{F,\varepsilon}^{\Ycal}$} & Frobenius norm ball with center $\Ycal$ and radius $\varepsilon$, \\ 
    & $\mathcal{B}_{F,\varepsilon}^{\Ycal} := \{ \Xcal \in \real^{d_1 \times d_2 \times d_3} | \| \Xcal - \Ycal \|_F \leq \varepsilon \}$ \vspace{1mm} \\ 
    
    \multirow{2}{*}{$\mathcal{B}_{1,\alpha}$} & $\ell_1$-norm ball with center $\Ocal$ and radius $\alpha$, \\
    & $\mathcal{B}_{1,\alpha} := \{ \Xcal \in \real^{d_1 \times d_2 \times d_3} | \| \Xcal \|_1 \leq \alpha \}$ \vspace{0mm} \\
    
    \bottomrule
  \end{tabular}
  }
\end{table}

\subsection{Proximal Tools}
\label{subsec:proximal_tools}

A function $f : \mathbb{R}^D \to (-\infty, +\infty]$ is called proper if its domain is nonempty,
$f(\mathcal{X}) > -\infty$ for all $\mathcal{X} \in \mathbb{R}^{D ( = d_1 \times d_2 \times d_3)}$,
and there exists at least one $\mathcal{X} \in \mathbb{R}^D$ such that $f(\mathcal{X}) < +\infty$.
It is said to be lower-semicontinuous if, for any $\alpha \in \mathbb{R}$, the sublevel set 
$\{ \mathcal{X} \in \mathbb{R}^D : f(\mathcal{X}) \leq \alpha \}$ is closed.
Moreover, $f$ is convex if, for any $\mathcal{X}, \mathcal{Y} \in \mathbb{R}^D$ 
and $\lambda \in [0, 1]$, the following inequality holds:
$f(\lambda \mathcal{X} + (1 - \lambda) \mathcal{Y}) \leq \lambda f(\mathcal{X}) + (1 - \lambda) f(\mathcal{Y})$.
A function that is proper, lower-semicontinuous, and convex is called a proper lower-semicontinuous convex function.

Let $\Gamma_0(\real^D)$ be the set of all proper lower-semicontinuous convex functions on $\real^D$.
For any $\gamma > 0$, the proximity operator of a function $f \in \Gamma_0(\real^D)$ is defined by
\begin{equation}
  \label{eq:prox1}
  \mathrm{prox}_{\gamma f} (\Xcal) := \argmin_{\Ycal \in \real^D} f(\Ycal) + \frac{1}{2 \gamma} \| \Xcal - \Ycal\|_F^2.
\end{equation}

Let $C$ be a nonempty closed convex set\footnote{
A set $C \subset \real^D$ is said to be convex if $\lambda \Xcal + (1 - \lambda)\Ycal \in C$ for any $\Xcal,\Ycal \in C$ and $\lambda \in [0, 1]$.
}.
Then, the indicator function $\iota_C \in \Gamma_0(\real^D)$ of $C$ is defined by
\begin{equation}
  \label{eq:indicator}
  \iota_{C} ( \Xcal ) := 
  \begin{cases}
    0, \: & \mathrm{if} \: \Xcal \in C; \\
    \infty, \: & \mathrm{otherwise}.
  \end{cases}
\end{equation}
The proximity operator of an indicator function $\iota_{C}$ equals the metric projection onto $C$, i.e.,
\begin{align}
  \label{eq:prox4}
  \mathrm{prox}_{\gamma \iota_{C}} (\Xcal) &= \argmin_{\Ycal \in C} \iota_{C}(\Ycal) + \frac{1}{2\gamma}\| \Xcal - \Ycal \|_F^2 \nonumber \\ 
  &= \argmin_{\Ycal \in C} \| \Xcal - \Ycal \|_F =: P_C(\Xcal).
\end{align}

\subsection{Preconditioned Variant of Primal-Dual Splitting Method (P-PDS)}
\label{subsec:P-PDS}
A Primal-Dual Splitting method (PDS)~\cite{PDS_2011} is an efficient algorithm for solving convex optimization problems of the form: 
\begin{align}
    \label{eq:PDS_model}
    &\min_{\substack{\Xcal_1, \ldots, \Xcal_N, \\ \Ycal_1, \ldots , \Ycal_M}} \sum_{i=1}^{N} f_i(\Xcal_i) + \sum_{j=1}^{M} g_j(\Ycal_j), \nonumber \\
    &\mathrm{s.t.} \,\,\,\, \Ycal_1 = \sum_{i=1}^{N} \Lfrak_{1,i}(\Xcal_{i}), \: \ldots ,\: \Ycal_M = \sum_{i=1}^{N} \Lfrak_{M,i}(\Xcal_{i}),
\end{align}
where $f_i \in \Gamma_0(\real^{D_i}) \: (i=1, \ldots, N)$ and $g_j \in \Gamma_0(\real^{D_j}) \: (j=1, \ldots, M)$ are proximable\footnote{
If the proximity operator of a function $f \in \Gamma_0(\real^D)$ is efficiently computable, we call $f$ \textit{proximable}.
} proper lower-semicontinuous convex functions,
and $\Lfrak_{j,i} \: (i=1, \ldots, N, \: j=1, \ldots, M)$ are linear operators.
The stepsizes of the standard PDS must be set manually within a range that satisfies the convergence conditions. 
On the other hand, Preconditioned variants of PDS (P-PDS)~\cite{DP-PDS_2011,naganuma2023_DPDS} can automatically determine the appropriate stepsizes based on the problem structure and converge faster in general than the standard PDS. 
Among them, we adopt P-PDS with Operator-norm-based design method of Variable-wise Diagonal Preconditioning (OVDP)~\cite{naganuma2023_DPDS}.
This method solves Prob.~(\ref{eq:PDS_model}) by the following iterative procedures:
\begin{align}
    \label{eq:PDS_alg}
    \left \lfloor
    \begin{array}{l}
        \Xcal_{1}^{(n+1)} \leftarrow \prox_{\gamma_{x_1} f_1} (\Xcal_1^{(n)} - \gamma_{x_1}(\sum_{j=1}^{M} \Lfrak_{j,1}^{\ast}(\Ycal_j^{(n)})); \\ 
        \vdots \\ 
        \Xcal_{N}^{(n+1)} \leftarrow \prox_{\gamma_{x_N} f_N} (\Xcal_N^{(n)} - \gamma_{x_N}(\sum_{j=1}^{M} \Lfrak_{j,N}^{\ast}(\Ycal_j^{(n)})); \vspace{0.5mm} \\ 
        \tempY{\Ycal}_{1} \leftarrow \Ycal_1^{(n)} + \gamma_{y_1}(\sum_{i=1}^{N} \Lfrak_{1,i} (2\Xcal_i^{(n+1)} - \Xcal_i^{(n)})); \vspace{0.5mm} \\ 
        \Ycal_{1}^{(n+1)} \leftarrow \tempY{\Ycal}_{1} - \gamma_{y_1}\prox_{\frac{1}{\gamma_{y_1}} g_1} ( \frac{1}{\gamma_{y_1}} \tempY{\Ycal}_{1} ); \\ 
        \vdots \\ 
        \tempY{\Ycal}_{M} \leftarrow \Ycal_M^{(n)} + \gamma_{y_M}(\sum_{i=1}^{N} \Lfrak_{M,i} (2\Xcal_i^{(n+1)} - \Xcal_i^{(n)})); \vspace{0.5mm} \\ 
        \Ycal_{M}^{(n+1)} \leftarrow \tempY{\Ycal}_{M} - \gamma_{y_M}\prox_{\frac{1}{\gamma_{y_M}} g_M} ( \frac{1}{\gamma_{y_M}} \tempY{\Ycal}_{M} ); \\ 
        n \leftarrow n+1;
    \end{array}
    \right.
\end{align}
where $\gamma_{x_i} \: (i=1, \ldots, N)$ and $\gamma_{y_j} \: (j=1, \ldots, M)$ are the stepsizes, which are automatically determined as follows:
\begin{equation}
    \label{eq:PDS_gamma}
    \gamma_{x_i} = \frac{1}{\sum_{j=1}^{M} \mu_{j,i}^2}, \, \gamma_{y_j} = \frac{1}{N},
\end{equation}
where $\mu_{j,i} \: (i=1, \ldots, N, \: j=1, \ldots, M)$ are the upper bounds of $\| \Lfrak_{j,i} \|_{\op}$ (see Table~\ref{tab:notation} for the definition of the operator norm of a linear operator).

\section{Proposed Method}
\label{sec:proposed_method}

\begin{figure*}[!t]
  \centering
  \includegraphics[keepaspectratio, scale = 0.56]{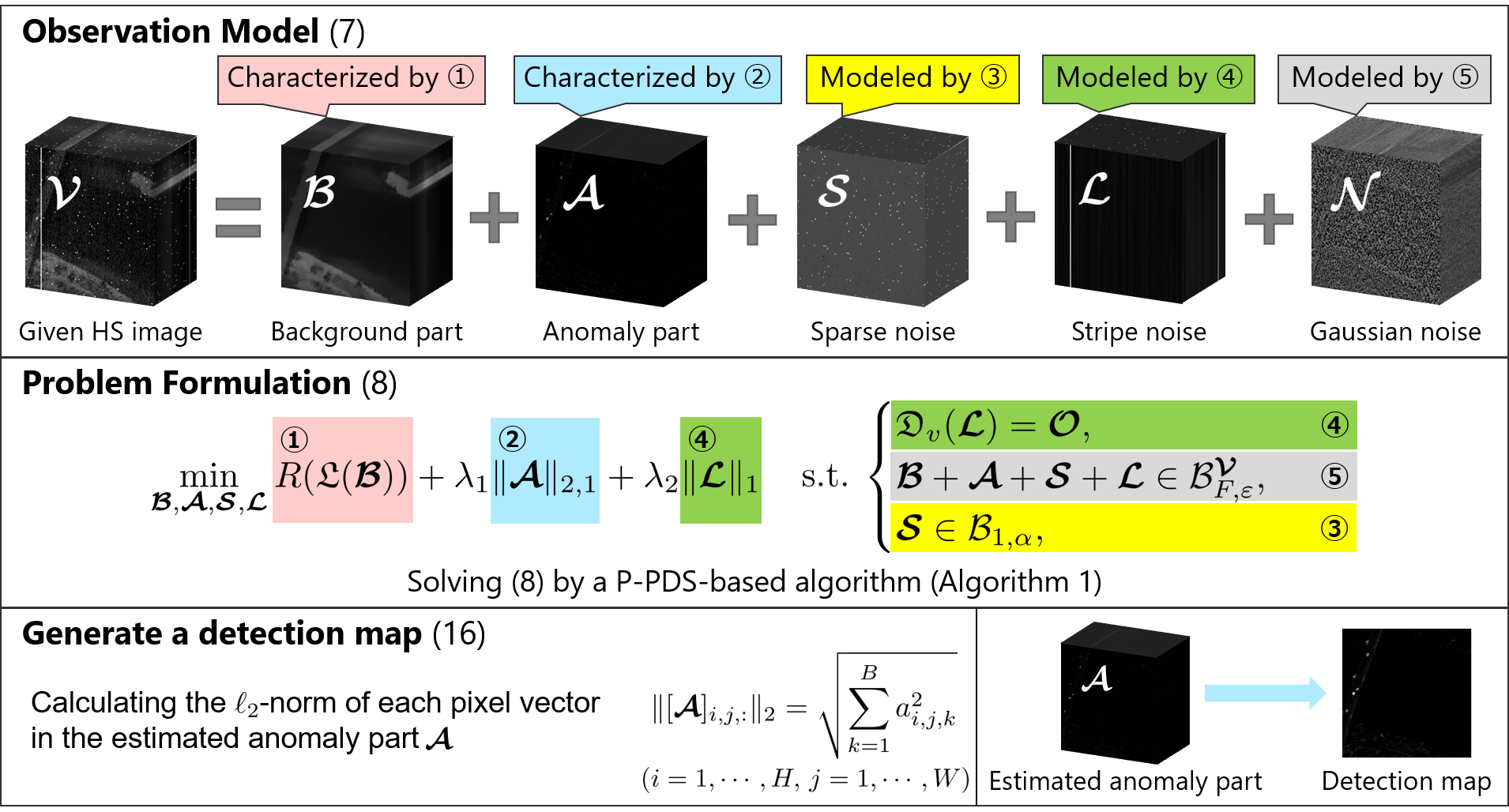}
  \caption{Overview of the proposed method.}
  \label{fig:flowchart}
\end{figure*}

An overview of the proposed method is illustrated in Fig.~\ref{fig:flowchart}. 
In this section, to explicitly handle mixed noise superimposed on an HS image, we first introduce an observation model that includes three types of noise.
Then, based on the model, we formulate the HS anomaly detection problem as a constrained convex optimization problem and derive a P-PDS-based algorithm to efficiently solve it.
Finally, we give several specific designs of functions that characterize a background part, and the computational complexity of the proposed method with each of them.

\subsection{Problem Formulation}
\label{subsec:problem_formulation}
We consider the following observation model:
\begin{equation} 
\label{eq:proposed-model} 
\Vcal = \Bcal + \Acal + \Scal + \Lcal + \Ncal ,
\end{equation}
where $\Vcal \in \mathbb{R}^{H \times W \times B}$ is a given HS image ($H$ and $W$ are the height and width of the HS image, and $B$ is the number of the spectral bands), 
$\Bcal \in \mathbb{R}^{H \times W \times B}$ is a background part,
$\Acal \in \mathbb{R}^{H \times W \times B}$ is an anomaly part,
$\Scal \in \mathbb{R}^{H \times W \times B}$ is sparse noise,
$\Lcal \in \mathbb{R}^{H \times W \times B}$ is stripe noise,
and $\Ncal \in \mathbb{R}^{H \times W \times B}$ is Gaussian noise.

Based on this model, we formulate the HS anomaly detection problem with mixed noise removal as the following constrained convex optimization problem:
\begin{align}
    \label{eq:proposed-optimization-problem}
    &\min_{\Bcal, \Acal, \Scal, \Lcal}  \regu(\Lfrak (\Bcal)) + \lambda_1 \| \Acal \|_{2,1} + \lambda_2 \|\Lcal\|_1, \nonumber \\
    & \mathrm{s.t.} \:
    \begin{cases} 
        \Dfrak_v(\Lcal) = \Ocal, \\
        \Bcal + \Acal + \Scal + \Lcal \in \mathcal{B}_{F,\varepsilon}^{\Vcal}, \\
        \Scal \in \mathcal{B}_{1,\alpha}, \\
    \end{cases}
\end{align}
where $\lambda_1 \geq 0$ and $\lambda_2 \geq 0$ are hyperparameters. 
Here, $R$ is a non-differentiable convex function whose proximity operator can be computed efficiently, 
and $\mathfrak{L}$ is a linear operator.
For the definitions of $\mathcal{B}_{F,\varepsilon}^{\Vcal}$ and $\mathcal{B}_{1,\alpha}$, see Table~\ref{tab:notation}.
The role of each term and constraint in Prob.~\eqref{eq:proposed-optimization-problem} is summarized as follows:
\begin{itemize}
    \item The first term is a general form of a suitably-chosen function to characterize the spatial continuity and/or the spectral correlation of the background part. 
    We introduce some specific examples of the function design in Sec.~\ref{subsec:regularization_term}.

    \item The second term models the spatial sparsity of the anomaly part.
    This is based on the fact that anomalies are small objects with a low probability of existence in the spatial domain.
    
    \item The third term and the first constraint characterize stripe noise that is superimposed with constant intensity in one direction.
    In this article, without loss of generality, we assume that this noise is generated only in the vertical direction. 
    Specifically, the third term adjusts the sparsity of stripe noise, while the first constraint, called the flatness constraint, models the constant intensity.
    The advantages of such characterizations for stripe noise are described in~\cite{naganuma_Denoising_2022}. 
    
    \item The second constraint imposes a data-fidelity condition on the given HS image.
    Since Gaussian noise is spatially and spectrally distributed with bounded energy, the Frobenius norm ball, whose radius $\varepsilon$ is adjusted according to the Gaussian noise intensity, captures this property.

    \item The third constraint models sparse noise, where the upper bound $\alpha$ of the $\ell_1$-norm is adjusted based on the probability of its occurrence.
    While anomalies are spatially sparse but spectrally continuous, sparse noise occurs irregularly without spectral continuity.
    By characterizing these distinct types of sparsity, the two components can be separated.

\end{itemize}
As described above, each of the three noise components is characterized by a tailored mechanism reflecting its structural property.
Since they differ fundamentally in structure, no single constraint can capture all of them simultaneously. 
Therefore, modeling each as an independent component is essential for accurate decomposition.

\subsection{Optimization Algorithm}
\label{subsec:optimization}
We develop an efficient solver for Prob.~\eqref{eq:proposed-optimization-problem} based on P-PDS with OVDP~\cite{naganuma2023_DPDS}.
First, using the indicator functions $\iota_{\mathcal{B}_{1,\alpha}}$, $\iota_{\mathcal{B}_{F,\varepsilon}^{\Vcal}}$, and $\iota_{\{ \Ocal \}}$ (see Eq.~\eqref{eq:indicator} for the definition),
we rewrite Prob.~\eqref{eq:proposed-optimization-problem} as the following equivalent problem:
\begin{align}
    \label{eq:proposed_indicator}
        \min_{\substack{\Bcal, \Acal, \Scal, \Lcal, \\ \Ycal_1, \Ycal_2, \Ycal_3 }} 
        & \regu(\Ycal_{1})
        + \lambda_1 \| \Acal \|_{2,1} 
        + \lambda_2 \|\Lcal\|_1 \nonumber \\
        &
        + \iota_{\{ \Ocal \}} (\Ycal_2)
        + \iota_{\mathcal{B}_{F,\varepsilon}^{\Vcal}} (\Ycal_3) 
        + \iota_{\mathcal{B}_{1,\alpha}} (\Scal),
        \nonumber\\
        \mathrm{s.t.} \quad
        & \begin{cases} 
            \Ycal_1 = \Lfrak (\Bcal), \\
            \Ycal_2 = \Dfrak_v(\Lcal), \\
            \Ycal_3 = \Bcal + \Acal + \Scal + \Lcal,
        \end{cases} 
\end{align}
where $\Ycal_1$, $\Ycal_2$ and $\Ycal_3$ are auxiliary variables.
Specifically, $\Ycal_1$ is associated with the background characterization term, $\Ycal_2$ corresponds to the flatness constraint $\Dfrak_v(\Lcal)=\Ocal$, and $\Ycal_3$ is related to the data-fidelity constraint 
$\Bcal + \Acal + \Scal + \Lcal \in \mathcal{B}_{F,\varepsilon}^{\Vcal}$.

Then, by defining,
\begin{align}
    \label{eq:proposed_func}
    & f_1(\Bcal) := 0, \: f_2(\Acal) := \lambda_1 \| \Acal \|_{2,1}, \nonumber \\
    & f_3(\Scal) := \iota_{\mathcal{B}_{1,\alpha}} (\Scal), \: f_4(\Lcal) := \lambda_2 \|\Lcal\|_1, \nonumber \\ 
    & g_1(\Ycal_1) := \regu(\Ycal_1), 
    \: g_2(\Ycal_2) := \iota_{\{ \Ocal \}} (\Ycal_2), \nonumber \\
    & g_3(\Ycal_3) := \iota_{\mathcal{B}_{F,\varepsilon}^{\Vcal}} (\Ycal_3), 
\end{align}
Prob.~\eqref{eq:proposed_indicator} can be seen as Prob.~(\ref{eq:PDS_model}), so that we can apply P-PDS with OVDP to Prob.~\eqref{eq:proposed_indicator}.
We show the detailed algorithm in Algorithm~\ref{alg:Proposed}.
Following Eq.~\eqref{eq:PDS_gamma}, the stepsizes $\gamma_B$, $\gamma_A$, $\gamma_S$, $\gamma_L$,
$\gamma_{Y_1}$, $\gamma_{Y_2}$, $\gamma_{Y_3}$ are automatically determined as follows:
\begin{align}
    \label{eq:proposed_gamma}
    & \gamma_B = \frac{1}{1+ \| \Lfrak \|_{\op}^2}, \: \gamma_A = 1, \: \gamma_S = 1, \: \gamma_L = \frac{1}{5}, \nonumber \\
    & \gamma_{Y_1} = \gamma_{Y_2} = \gamma_{Y_3} = \frac{1}{4}.
\end{align}

In what follows, we explain how to compute each step of Algorithm~\ref{alg:Proposed}.
The computations of the proximity operators of the $\ell_{2,1}$-norm in Step 4 and the $\ell_1$-norm in Step 6 are given as follows:
\begin{align}
  \label{eq:prox-l12}
  [\mathrm{prox}_{\gamma \| \cdot \|_{2,1}} (\Xcal)]_{i,j,k} &= \max \Bigl\{1 - \frac{\gamma}{\| [\Xcal]_{i, j, :} \|_{2}} , 0 \Bigr\}[\Xcal]_{i,j,k},
    \\
  \label{eq:prox-l1}
  [\mathrm{prox}_{\gamma \| \cdot \|_{1}} (\Xcal)]_{i,j,k} &= \sgn([\Xcal]_{i,j,k}) \max\{| [\Xcal]_{i,j,k} | - \gamma, 0\}.
\end{align}
In addition, the proximity operators of $\iota_{\{ \Ocal \}}$ in Step 10 and $\iota_{\mathcal{B}_{F,\varepsilon}^{\Ycal}}$ in Step 12 are the metric projections onto $\Ocal$ and $\mathcal{B}_{F,\varepsilon}^{\Ycal}$, respectively. 
Their computations are given by
\begin{align}
  \label{eq:prox-l0}
  \mathrm{prox}_{\gamma \iota_{\{\Ocal\}}} (\Xcal) &= P_{\{\Ocal\}} (\Xcal) = \Ocal,
  \\
  \label{eq:prox-l2ball}
  \mathrm{prox}_{\gamma \iota_{\mathcal{B}_{F,\varepsilon}^{\Ycal}}} (\Xcal) &= P_{\mathcal{B}_{F,\varepsilon}^{\Ycal}} (\Xcal) \nonumber \\ &= 
  \begin{cases}
      \Xcal, \: & \mathrm{if} \: \Xcal \in \mathcal{B}_{F,\varepsilon}^{\Ycal}; \\
      \Ycal + \frac{\varepsilon (\Xcal-\Ycal)}{\| \Xcal - \Ycal \|_F}, 
      \: & \mathrm{otherwise}.
  \end{cases}
\end{align}
For the computation of the proximity operator of $\iota_{\mathcal{B}_{1,\alpha}}$ in Step 5, we use a fast $\ell_1$-ball projection algorithm~\cite{Proj_L1ball_2016}.

After estimating the anomaly part $\Acal$, we generate a 2D detection map by calculating the $\ell_2$-norm of each pixel vector as follows: 
\begin{equation} 
\label{eq:generate} 
\| [\Acal]_{i,j,:} \|_2 = \sqrt{\sum_{k=1}^B a_{i,j,k}^2} , \; (i = 1, \cdots,H , \, j = 1, \cdots, W).
\end{equation}

\begin{algorithm}[!t]
  \caption{Proposed algorithm for solving Prob.~\eqref{eq:proposed-optimization-problem}}
  \label{alg:Proposed}
  \begin{algorithmic}[1]
    \REQUIRE{$\Vcal$, $\lambda_1$, $\lambda_2$, $\varepsilon$, $\alpha$}
    \INITIALIZE{
      $\Bcal^{(0)} = \Ocal$, $\Acal^{(0)} = \Ocal$, $\Scal^{(0)} = \Ocal$, $\Lcal^{(0)} = \Ocal$,
      $\Ycal_1^{(0)} = \Ocal$, $\Ycal_2^{(0)} = \Ocal$, $\Ycal_3 = \Ocal$, 
      $\gamma_B$, $\gamma_A$, $\gamma_S$, $\gamma_L$,
      $\gamma_{Y_1}$, $\gamma_{Y_2}$, $\gamma_{Y_3}$
      }
    \STATE $n = 0;$
    \WHILE{stopping conditions are not met,} \vspace{0.5mm}
    
    \STATE $\Bcal^{(n+1)} \leftarrow \Bcal^{(n)} - \gamma_B (\Lfrak^* (\Ycal_{1}^{(n)}) +  \Ycal_3^{(n)}); \vspace{0.5mm} $
    
    \STATE $\Acal^{(n+1)} \leftarrow \prox_{\gamma_A \lambda_1 \| \cdot \|_{2,1}} (\Acal^{(n)} - \gamma_A\Ycal_3^{(n)}); \vspace{0.5mm} $
    
    \STATE $\Scal^{(n+1)} \leftarrow \proj_{\mathcal{B}_{1,\alpha}} (\Scal^{(n)} - \gamma_S\Ycal_3^{(n)}); \vspace{0.5mm} $
    
    \STATE $\Lcal^{(n+1)} \leftarrow \prox_{\gamma_L \lambda_2 \| \cdot \|_1} (\Lcal^{(n)} - \gamma_L (\Dfrak_v^* (\Ycal_2^{(n)}) + \Ycal_3^{(n)}); \vspace{0.5mm} $
    
    \STATE $\tempY{\Ycal}_{1} \leftarrow \Ycal_{1}^{(n)} + \gamma_{Y_{1}} \Lfrak (2\Bcal^{(n+1)} - \Bcal^{(n)}); \vspace{0.5mm} $
    
    \STATE $\Ycal_{1}^{(n+1)} \leftarrow \tempY{\Ycal}_{1} - \gamma_{Y_{1}}\prox_{\frac{1}{\gamma_{Y_{1}}} \regu} (\frac{\tempY{\Ycal}_{1}}{\gamma_{Y_{1}}}); \vspace{0.5mm} $

    \STATE $\tempY{\Ycal}_2 \leftarrow \Ycal_2^{(n)} + \gamma_{Y_2} \Dfrak_v (2\Lcal^{(n+1)} - \Lcal^{(n)}); \vspace{0.5mm} $
    
    \STATE $\Ycal_2^{(n+1)} \leftarrow \tempY{\Ycal}_2 - \gamma_{Y_2} P_{\{ \Ocal \}} (\frac{\tempY{\Ycal}_2}{\gamma_{Y_2}}); \vspace{0.5mm} $
    
    \STATE $\tempY{\Ycal}_3 \leftarrow \Ycal_3^{(n)} + \gamma_{Y_3} ( 2( \Bcal^{(n+1)} + \Acal^{(n+1)} + \Scal^{(n+1)} + \Lcal^{(n+1)} ) - ( \Bcal^{(n)} + \Acal^{(n)} + \Scal^{(n)} + \Lcal^{(n)}) ); \vspace{0.5mm}$
    
    \STATE $\Ycal_3^{(n+1)} \leftarrow \tempY{\Ycal}_3 - \gamma_{Y_3} P_{\mathcal{B}_{F,\varepsilon}^{\Vcal}} (\frac{\tempY{\Ycal}_3}{\gamma_{Y_3}}); \vspace{0.5mm} $
    
    \STATE $n \leftarrow n+1;$
    
    \ENDWHILE
    
    \ENSURE{ $\Bcal^{(n)}$, $\Acal^{(n)}$, $\Scal^{(n)}$, $\Lcal^{(n)}$}
  \end{algorithmic}
\end{algorithm}

\subsection{Specific Designs of Background Characterization Function}
\label{subsec:regularization_term}
We give some examples of $\regu(\Lfrak(\Bcal))$ that characterizes the background part in Prob.~\eqref{eq:proposed-optimization-problem}.

\subsubsection{Hyperspectral Total Variation (HTV)~\cite{HTV_2012}}
HTV models the spatial piecewise smoothness of the background part by promoting the group sparsity of vertical and horizontal neighborhood differences across all bands.
We define a spatial difference operator as
\begin{align}
    \label{eq:spatial_difference_op}
    [\Dfrak(\Xcal)]_{i,j,k} := 
    \begin{cases}
        [\Dfrak_v (\Xcal)]_{i,j,k}, & (1 \leq k \leq d_3), \\
        [\Dfrak_h (\Xcal)]_{i,j,k-d_3}, & (d_3 < k \leq 2 d_3),
    \end{cases}
\end{align}
where $\Dfrak_v$ and $\Dfrak_h$ denote the vertical and horizontal difference operators, respectively (see Table~\ref{tab:notation} for more details).
The definition of HTV is given by
\begin{equation}
    \label{eq:HTV2}
    \| \Xcal \|_\mathrm{HTV} := \| \Dfrak (\Xcal) \|_{2,1}.
\end{equation}
Then, we can see that HTV is a special case of $\regu(\Lfrak(\Bcal))$ by letting $\regu = \| \cdot \|_{2,1}$ and $\Lfrak = \Dfrak$.
Note that the computation of the proximity operator of the $\ell_{2,1}$-norm is shown in Eq.~\eqref{eq:prox-l12}.

\subsubsection{Spatio-Spectral Total Variation (SSTV)~\cite{SSTV_2016}}
SSTV can promote the spatial and spectral piecewise smoothness of the background part.
SSTV is defined, by using the vertical and horizontal differences of spectral differences, as
\begin{equation}
    \label{eq:SSTV}
    \| \Xcal \|_\mathrm{SSTV} := \| \Dfrak( \Dfrak_b (\Xcal) ) \|_{1},
\end{equation}
where $\Dfrak_b$ denotes the spectral difference operator (see Table~\ref{tab:notation} for the definition).
Then, we can see that SSTV is a special case of $\regu(\Lfrak(\Bcal))$ by letting $\regu = \| \cdot \|_{1}$ and $\Lfrak = \Dfrak \circ \Dfrak_b$.
Note that the computation of the proximity operator of the $\ell_{1}$-norm is shown in Eq.~\eqref{eq:prox-l1}.

\subsubsection{Hybrid Spatio-Spectral Total Variation (HSSTV)~\cite{HSSTV_2020}}
HSSTV is a hybrid of HTV and SSTV.
We define a spatial-spectral difference operator as
\begin{align}
    \label{eq:spatial_spectral_difference_op}
    [\Afrak_\omega(\Xcal)]_{i,j,k} := 
    \begin{cases}
        [\Dfrak(\Dfrak_b(\Xcal))]_{i,j,k}, & (1 \leq k \leq 2d_3), \\
        \omega [\Dfrak (\Xcal)]_{i,j,k-2d_3}, & (2 d_3 < k \leq 4 d_3), \\
    \end{cases}
\end{align}
where $\omega > 0$ is a hyperparameter.
Then, HSSTV is defined by
\begin{align}
    \label{eq:HSSTV}
    \| \Xcal \|_\mathrm{HSSTV} := \| \Afrak_\omega(\Xcal) \|_{1}.
\end{align}
Here, we can see that HSSTV is a special case of $\regu(\Lfrak(\Bcal))$ by letting $\regu = \| \cdot \|_{1}$ and $\Lfrak = \Afrak_\omega$.
The computation of the proximity operator of the $\ell_{1}$-norm is shown in Eq.~\eqref{eq:prox-l1}.

\subsubsection{Nuclear Norm}
Due to the high spectral correlation between background pixels, the background part exhibits low-rank characteristics. 
To model this, we can also use a low-rank approximation using the nuclear norm.
For $\X \in \mathbb{R}^{M \times N} (M \leq N)$, the nuclear norm of $\X$ is given by
\begin{align}
    \label{eq:nuclear_norm}
    \| \X \|_* := \sum_{i=1}^M \sigma_i(\X),
\end{align}
where $\sigma_i(\cdot)$ is the $i$-th largest singular value of $\X$.
Here, we define $\reshapeX{(\Xcal)} : \real^{H \times W \times B} \rightarrow \real^{B \times HW}$ as an operator that reshapes a three-dimensional HS image cube into a matrix. 
Then, we can see that the nuclear norm is a special case of $\regu(\Lfrak(\Bcal))$ by letting $\regu = \| \cdot \|_{*}$ and $\Lfrak = \reshapeX$.

Let the singular value decomposition of $\X \in \mathbb{R}^{M \times N} (M \leq N)$ be $\X = \mathbf{U}\mathbf{\Sigma}\mathbf{V}^\top $.
The computation of the proximity operator of the nuclear norm is given as follows:
\begin{equation}
  \label{eq:prox-nuclear}
  \mathrm{prox}_{\gamma \| \cdot \|_{*}} (\X) = \mathbf{U} \mathbf{\widetilde{\Sigma}}_\gamma \mathbf{V}^\top ,
\end{equation}
where $\widetilde{\mathbf{\Sigma}}_\gamma$ is a diagonal matrix whose diagonal elements are given by $[\mathrm{diag}(\widetilde{\mathbf{\Sigma}}_\gamma)]_{i} = \max{ \{\sigma_i(\X)-\gamma, 0 \} }$ for $i = 1, \ldots , M$.

\subsubsection{Stepsize Choices for Each Background Characterization}
\label{subsubsec:stepsize_choices}
Finally, we derive the choices of the stepsize $\gamma_B$ in Eq.~\eqref{eq:proposed_gamma} for each background characterization.
The operator norm of the identity operator is 1.
In addition, from~\cite{Chambolle_2004}, each difference operator satisfies $\| \Dfrak_v \|_\op \leq 2$, $\| \Dfrak_h \|_\op \leq 2$, $\| \Dfrak_b \|_\op \leq 2$, and $\| \Dfrak \|_\op \leq 2\sqrt{2}$, respectively.
Furthermore, from the submultiplicity of the operator norm, we have 
$\| \Dfrak \circ \Dfrak_b \|_\op \leq \| \Dfrak \|_\op \| \Dfrak_b \|_\op \leq 4\sqrt{2}$. 
Substituting these values and upper bounds into Eq.~\eqref{eq:PDS_gamma}, we can determine the stepsize $\gamma_B$ as shown in Table~\ref{tab:stepsize}.

In what follows, we provide a detailed derivation for the case of HSSTV as a representative example.
For any $\Xcal$, from the definition of $\Afrak_{\omega}$ in Eq.~\eqref{eq:spatial_spectral_difference_op}, we have
\begin{align}
\| \Afrak_{\omega}(\Xcal) \|_F^2
=
\| \Dfrak(\Dfrak_b(\Xcal)) \|_F^2
+
\omega^2 \| \Dfrak(\Xcal) \|_F^2.
\label{eq:HSSTV_norm}
\end{align}
Therefore, the operator norm of $\Afrak_{\omega}$ is bounded as
\begin{align}
\| \Afrak_{\omega} \|_{\op}^2
\leq
\| \Dfrak \circ \Dfrak_b \|_{\op}^2
+
\omega^2 \| \Dfrak \|_{\op}^2 
\leq
32 + 8\omega^2.
\end{align}
According to Eq.~\eqref{eq:PDS_gamma} and Eq.~\eqref{eq:proposed_indicator}, the stepsize $\gamma_B$ is determined by the squared operator-norm bounds of all linear operators associated with the variable $\Bcal$.
Specifically, $\Bcal$ is involved in both the constraint $ \Ycal_1 = \Afrak_{\omega}(\Bcal)$ and the constraint $\Ycal_3 = \Bcal + \Acal + \Scal + \Lcal$.
Therefore, the corresponding contributions are $\| \Afrak_{\omega} \|_{\op}^2 \leq 32 + 8\omega^2$ and $\| \Ifrak \|_{\op}^2 = 1$, where $\Ifrak$ denotes the identity operator.
By substituting these bounds into Eq.~\eqref{eq:PDS_gamma}, we obtain
\begin{align}
\gamma_B = \frac{1}{(32 + 8\omega^2) + 1}
= \frac{1}{33 + 8\omega^2}.
\end{align}

\begin{table}[!t]
  \centering
  \caption{Specific Function $\regu$, Linear Operator $\Lfrak$, and Stepsize $\gamma_B$ with Respect to Each Background Characterization.}
  \label{tab:stepsize}
  \scalebox{1}{
  \begin{tabular}{cccc}
    \toprule
    Regularizations & $\regu$ & $\Lfrak$ & $\gamma_B$ in Eq.~\eqref{eq:proposed_gamma} \\ 
    \cmidrule(lr){1-1} \cmidrule(lr){2-2} \cmidrule(lr){3-3} \cmidrule(lr){4-4} 
    HTV~\cite{HTV_2012}     & $\| \cdot \|_{2,1}$ & $\Dfrak$                & $\frac{1}{9}$            \vspace{1.5mm} \\ 
    SSTV~\cite{SSTV_2016}   & $\| \cdot \|_{1}$   & $\Dfrak \circ \Dfrak_b$ & $\frac{1}{33}$           \vspace{1.5mm} \\ 
    HSSTV~\cite{HSSTV_2020} & $\| \cdot \|_{1}$   & $\Afrak_\omega$         & $\frac{1}{33+8\omega^2}$ \vspace{1.5mm} \\ 
    Nuclear Norm            & $\| \cdot \|_{*}$   & $ \reshapeX$            & $\frac{1}{2}$            \vspace{0mm}   \\
    \bottomrule
  \end{tabular}
  }
\end{table}

\subsection{Computational Complexity}
\label{subsec:computational_complexity}
Table~\ref{tab:computational_cost} shows the computational complexity of the operation for a tensor $\Xcal \in \real^{H \times W \times B}$ and a matrix $\X \in \real^{B \times HW}$ used in the proposed method.
Let $N = HWB$, the computational complexity for each step of Algorithm~\ref{alg:Proposed} is as follows:
\begin{itemize}
    \item Steps 3, 4, 6, 7, 9, 11 and 12: $\Ordnung (N)$.

    \item Step 5: $\Ordnung (N \log N)$.

    \item Step 10: $\Ordnung (1)$.

    \item Step 8: $\Ordnung (N)$ when HTV, SSTV, or HSSTV is used to characterize the background part; and $\Ordnung (BN)$ when the nuclear norm is used.
\end{itemize}
From the above, the overall computational complexity for each iteration of the proposed method using HTV, SSTV, or HSSTV is $\Ordnung (N \log N)$, and since $\log N << B$ in general, the proposed method using the nuclear norm is $\Ordnung (BN)$.

\begin{table}[!t]
  \centering
  \caption{Computational Complexity of Each Operation.}
  \label{tab:computational_cost}
  \scalebox{1}{
  \begin{tabular}{cc}
    \toprule
    Operations & $\Ordnung$-notations \\ 
    \cmidrule(lr){1-1} \cmidrule(lr){2-2} 
    $\Dfrak(\Xcal)$ & $\Ordnung(N)$ \vspace{1mm} \\
    
    $\Dfrak( \Dfrak_b (\Xcal) )$ & $\Ordnung(N)$ \vspace{1mm} \\

    $\Afrak_\omega(\Xcal)$ & $\Ordnung(N)$ \vspace{1mm} \\

    $\mathrm{prox}_{\gamma \| \cdot \|_{2,1}} (\Xcal)$ in \eqref{eq:prox-l12} & $\Ordnung(N)$ \vspace{1mm} \\

    $\mathrm{prox}_{\gamma \| \cdot \|_{1}} (\Xcal)$ in \eqref{eq:prox-l1} & $\Ordnung(N)$ \vspace{1mm} \\

    $P_{\mathcal{B}_{F,\varepsilon}^{\Ycal}} (\Xcal)$ in \eqref{eq:prox-l2ball} & $\Ordnung(N)$ \vspace{1mm} \\

    $P_{\{\Ocal\}} (\Xcal)$ in \eqref{eq:prox-l0} & $\Ordnung(1)$ \vspace{1mm} \\

    $\proj_{\mathcal{B}_{1,\alpha}}  (\Xcal)$ in~\cite{Proj_L1ball_2016} & $\Ordnung(N \log N)$ \vspace{1mm} \\
    
    $\mathrm{prox}_{\gamma \| \cdot \|_{*}} (\X)$ in~\eqref{eq:prox-nuclear} & $\Ordnung(BN)$ \vspace{0mm} \\
    
    \bottomrule
  \end{tabular}
  }
\end{table}

\section{Experiments}
\label{sec:experiments}
In this section, we demonstrate the effectiveness of the proposed method through comprehensive experiments using seven HS anomaly detection datasets.
Specifically, we verify the following two aspects:
\begin{itemize}
    \item The proposed method achieves competitive detection performance on the original datasets.

    \item The proposed method is much more robust against various types of noise than existing methods.
\end{itemize}

All experiments were conducted using MATLAB R2021a on a 64-bit Windows 11 PC with an Intel Core i9-10900K, 32GB of RAM, and an NVIDIA GeForce RTX 3090.

\begin{figure*}[!t]
    \centering
    \begin{minipage}{0.02\hsize}
    	\centerline{\rotatebox{90}{\small{Pseudocolor}}}
    \end{minipage}
    \begin{minipage}{0.1325\hsize}
        \centering
        \includegraphics[keepaspectratio, scale = 0.45]{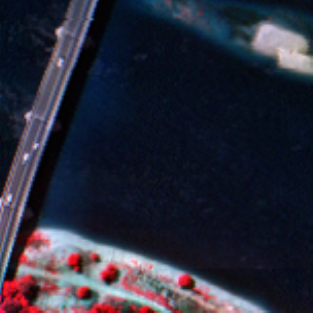}
    \end{minipage}
    \begin{minipage}{0.1325\hsize}
        \centering
        \includegraphics[keepaspectratio, scale = 0.675]{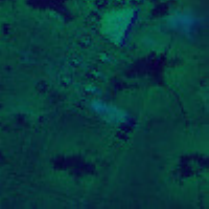}
    \end{minipage}
    \begin{minipage}{0.1325\hsize}
        \centering
        \includegraphics[keepaspectratio, scale = 0.675]{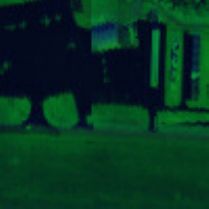}
    \end{minipage}
    \begin{minipage}{0.1325\hsize}
        \centering
        \includegraphics[keepaspectratio, scale = 0.675]{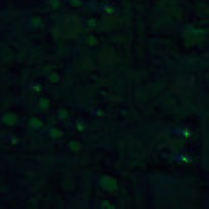}
    \end{minipage}
    \begin{minipage}{0.1325\hsize}
        \centering
        \includegraphics[keepaspectratio, scale = 0.675]{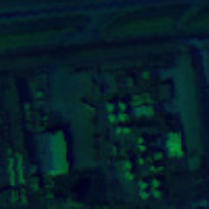}
    \end{minipage}
    \begin{minipage}{0.1325\hsize}
        \centering
        \includegraphics[keepaspectratio, scale = 0.675]{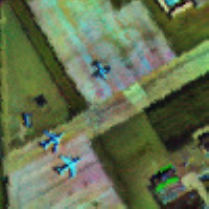}
    \end{minipage}
    \begin{minipage}{0.1325\hsize}
        \centering
        \includegraphics[keepaspectratio, scale = 0.45]{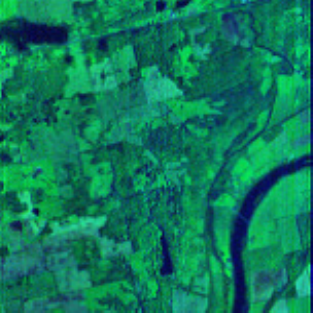}
    \end{minipage}
    
    \vspace{1.5mm}
    
    \begin{minipage}{0.02\hsize}
      \centerline{\rotatebox{90}{\small{Ground Truth}}}
    \end{minipage}
    \begin{minipage}{0.1325\hsize}
        \centering
        \includegraphics[keepaspectratio, scale = 0.45]{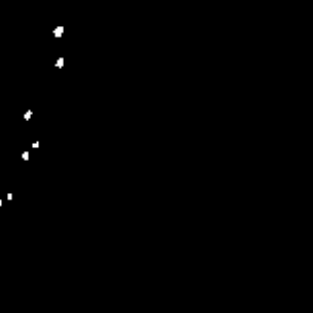}
    \end{minipage}
    \begin{minipage}{0.1325\hsize}
        \centering
        \includegraphics[keepaspectratio, scale = 0.675]{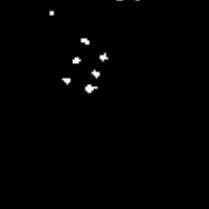}
    \end{minipage}
    \begin{minipage}{0.1325\hsize}
          \centering
          \includegraphics[keepaspectratio, scale = 0.675]{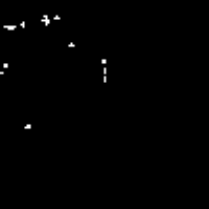}
    \end{minipage}
    \begin{minipage}{0.1325\hsize}
          \centering
          \includegraphics[keepaspectratio, scale = 0.675]{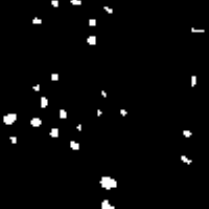}
    \end{minipage}
    \begin{minipage}{0.1325\hsize}
          \centering
          \includegraphics[keepaspectratio, scale = 0.675]{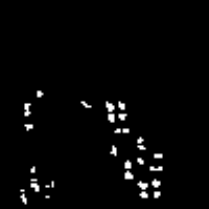}
    \end{minipage}
    \begin{minipage}{0.1325\hsize}
          \centering
          \includegraphics[keepaspectratio, scale = 0.675]{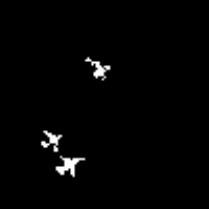}
    \end{minipage}
    \begin{minipage}{0.1325\hsize}
          \centering
          \includegraphics[keepaspectratio, scale = 0.45]{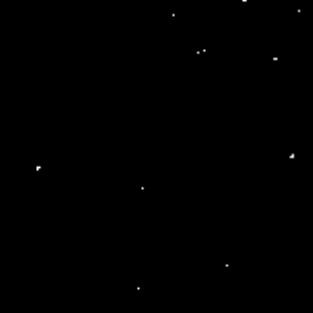}
    \end{minipage}
    
    \vspace{1mm}
    
    \begin{minipage}{0.02\hsize}
        ~
    \end{minipage}
    \begin{minipage}{0.1325\hsize}
        \centering
        \footnotesize{Pavia Centre}
    \end{minipage}
    \begin{minipage}{0.1325\hsize}
        \centering
         \footnotesize{Texas Coast}
    \end{minipage}
    \begin{minipage}{0.1325\hsize}
         \centering
        \footnotesize{Gainesville}
    \end{minipage}
    \begin{minipage}{0.1325\hsize}
         \centering
        \footnotesize{Los Angeles I}
    \end{minipage}
    \begin{minipage}{0.1325\hsize}
        \centering
        \footnotesize{Los Angeles I\hspace{-1.2pt}I}
    \end{minipage}
    \begin{minipage}{0.1325\hsize}
        \centering
        \footnotesize{San Diego}
    \end{minipage}
    \begin{minipage}{0.1325\hsize}
        \centering
        \footnotesize{Hyperion}
    \end{minipage}
    
    \caption{Pseudocolor images and ground truths for each dataset.}
    \label{fig:datasets}
\end{figure*}
\begin{table}[!t]
  \centering
  \caption{Details of the HS Images.}
  \label{tab:data}
  \scalebox{0.87}{
  \begin{tabular}{ccccc}
    \toprule
    Data & Sensor & Time & Resolution & Size \\ 
    \cmidrule(lr){1-1} \cmidrule(lr){2-2} \cmidrule(lr){3-3} \cmidrule(lr){4-4} \cmidrule(lr){5-5}
    Pavia Centre                  & ROSIS-03 & ------        & 1.3 m  & $150 \times 150 \times 102$ \vspace{1mm} \\ 
    Texas Coast                   & AVIRIS   & Aug. 29, 2010 & 17.2 m & $100 \times 100 \times 204$ \vspace{1mm} \\ 
    Gainesville                   & AVIRIS   & Sep. 4, 2010  & 3.5 m  & $100 \times 100 \times 191$ \vspace{1mm} \\
    Los Angeles I                 & AVIRIS   & Nov. 9, 2011  & 7.1 m  & $100 \times 100 \times 205$ \vspace{1mm} \\ 
    Los Angeles I\hspace{-1.2pt}I & AVIRIS   & Nov. 9, 2011  & 7.1 m  & $100 \times 100 \times 205$ \vspace{1mm} \\ 
    San Diego                     & AVIRIS   & ------        & 3.5 m  & $100 \times 100 \times 189$ \vspace{1mm} \\
    Hyperion                      & Hyperion   & 2008          & 30 m & $150 \times 150 \times 155$ \vspace{0mm} \\
    
    \bottomrule
  \end{tabular}
  }
\end{table}

\subsection{Experimental Setup}

\begin{table}[!t]
  \centering
  \caption{Noise Settings. Here, $\sigma$ Denotes the Standard Deviation of Gaussian Noise, $S_p$ And $S_l$ Represent the Ratios of Salt-And-Pepper And Stripe Noise, Respectively. Note That the Intensity of Stripe Noise Is Uniformly Random in the Range [-0.3, 0.3], And "---" Indicates That No Noise Was Added.}
  \label{tab:noise_case}
  \scalebox{1}{
  \begin{tabular}{cccc}
    \toprule
    Cases & Gaussian & salt-and-pepper & stripe \\ 
    \cmidrule(lr){1-1} \cmidrule(lr){2-2} \cmidrule(lr){3-3} \cmidrule(lr){4-4}

    Case 1 & --- & --- & --- \vspace{1mm} \\

    Case 2 & $\sigma = 0.03$ & --- & --- \vspace{1mm} \\

    Case 3 & --- & $S_p = 0.03$ & $S_l = 0.03$ \vspace{1mm} \\

    Case 4 & $\sigma = 0.01$ & $S_p = 0.01$ & $S_l = 0.01$ \vspace{1mm} \\

    Case 5 & $\sigma = 0.05$ & $S_p = 0.05$ & $S_l = 0.05$ \vspace{0mm} \\
    
    \bottomrule
  \end{tabular}
  }
\end{table}
\begin{table}[!t]
  \centering
  \caption{Hyperparameter Settings for Each Method.}
  \label{tab:para}
  \scalebox{0.88}{
  \begin{tabular}{cc}
    \toprule
    Methods & Parameters \\ 
    \cmidrule(lr){1-1} \cmidrule(lr){2-2} 
    \multirow{2}{*}{\shortstack{2S-GLRT \\ \cite{2SGLRT_2022}}} & $\omega_{in} \in \{3, 5, 7, 9, 11, 13, 15\}$, \\
    & $\omega_{out} \in \{5, 7, 9, 11, 13, 15, 17, 19, 21, 23, 25\}$ \vspace{1.5mm}\\ 

    \multirow{2}{*}{\shortstack{GAED \\ \cite{GAED_2022}}} &  $c \in 7$, \:  $l_a \in [0.1, 0.4]$, \: $\beta \in [1, 10]$, \: $r_{\mathrm{iter}} \in [300, 500]$, \\
    &  Dimension of the Middle-Hidden Layer $\in 25$ \vspace{1.5mm}\\

    \multirow{3}{*}{\shortstack{RGAE \\ \cite{RGAE_2022}}} & $\mathrm{n_{hid}} \in \{20, 40, 60, 80, 100, 120, 140, 160\}$,\\
    & $S \in \{50, 100, 150, 300, 500\}$,\\
    & $\lambda \in \{10^{-4}, 10^{-3}, 10^{-2}, 10^{-1}\}$ \vspace{1.5mm}\\

    \multirow{3}{*}{\shortstack{ADLR \\ \cite{ADLR_2018}}} & $\mathrm{c} \in \{5, 10, 15, 20, 25\}$,\\
    & $\mathrm{bw} \in \{0.2, 0.3, 0.4, 0.5\}$,\\
    & $\lambda \in \{0.01, 0.02, 0.03, 0.05, 0.1\}$ \vspace{1.5mm}\\
    
    \multirow{4}{*}{\shortstack{GTVLRR \\ \cite{GTVLRR_2020}}} & $M = 6$,\; $P = 20$,\\
    & $\lambda \in \{0.005, 0.05, 0.1, 0.3, 0.5, 0.7, 1\},$ \\
    & $\beta \in \{0.005, 0.05, 0.1, 0.2, 0.4, 0.7, 1\},$ \\
    & $\gamma \in \{0.005, 0.01, 0.02, 0.05, 0.1, 0.2, 0.5\}$ \vspace{1.5mm}\\ 

    \multirow{4}{*}{\shortstack{LSDM-MoG \\ \cite{LSDMMog_2021}}} & $t_0 = 10^{-3}$,\; $\mu_0 = 0$,\\
    & $\alpha_{01}, \ldots, \alpha_{0K}, \beta_0, a_0, b_0, c_0, d_0 = 10^{-6},$ \\
    & $l_0 \in \{10, 20, 30, 40, 50, 60, 70, 80, 90, 100\},$ \\
    & $K \in \{1, 2, 3, 4, 5, 6, 7, 8, 9, 10\}$ \vspace{1.5mm}\\

    \multirow{2}{*}{\shortstack{PCA-TLRSR \\ \cite{PCA-TLRSR_2023}}} & $\mathrm{dim} \in \{5, 10, 15\}$,\\
    & $\lambda, \lambda^{\prime} \in \{0.001, 0.005, 0.01, 0.05, 0.1, 0.2, 0.3, 0.4, 0.5\}$ \vspace{1.5mm}\\

    \multirow{2}{*}{\shortstack{AHMID \\ \cite{AHMID_2023}}} & $\mathrm{layer} \in 5$,\: $\alpha \in 1$,\\
    & $\lambda \in [0.01, 0.02, 0.03, 0.04, 0.05, 0.06, 0.07, 0.08, 0.09, 0.1]$ \vspace{1.5mm}\\

    MTVLRR \cite{MTVLRR_2024} & $\lambda \in [0.1, 0.2, 0.3, 0.4, 0.5, 0.6, 0.7, 0.8, 0.9]$ \vspace{1.5mm}\\

    Ours & $\lambda_1 \in \{0.01, 0.05, 0.1, 0.25, 0.5, 0.75, 1, 1.5, 2\}$, \\ 
    (HTV)& $\lambda_2 \in \{0, 0.001, 0.01, 0.025, 0.05, 0.075, 0.1, 0.25, 0.5, 1\}$ \vspace{1.5mm}\\ 
    
    Ours & $\lambda_1 \in \{0.1, 0.25, 0.5, 0.75, 1, 1.25, 1.5, 5, 10\}$, \\ 
    (SSTV) & $\lambda_2 \in \{0, 0.001, 0.01, 0.025, 0.05, 0.075, 0.1, 0.25, 0.5, 1\}$ \vspace{1.5mm}\\ 
    
    \multirow{3}{*}{\shortstack{Ours \\ (HSSTV)}} & $\omega = 0.05$,\\
     & $\lambda_1 \in \{0.05, 0.1, 0.25, 0.5, 0.75, 1, 1.25, 1.5, 2\}$, \\ 
     & $\lambda_2 \in \{0, 0.001, 0.005, 0.0075, 0.01, 0.025, 0.05, 0.1, 0.5, 1\}$ \vspace{1.5mm}\\ 
    
    Ours & $\lambda_1 \in \{0.005, 0.01, 0.025, 0.05, 0.075, 0.1, 0.25, 0.5, 1\}$, \\ 
    (Nuclear) & $\lambda_2 \in \{0, 0.001, 0.01, 0.025, 0.05, 0.075, 0.1, 0.25, 0.5, 1\}$ \vspace{0mm} \\ 
    
    \bottomrule
  \end{tabular}
  }
\end{table}

We used seven HS anomaly detection datasets from~\cite{HSdata_2017} and ~\cite{MTVLRR_2024}.
Fig.~\ref{fig:datasets} shows the pseudocolor images and ground truths of these datasets.
The details of each dataset are shown in Table~\ref{tab:data}.
The pixel values in each HS image were normalized to the range $[0,1]$.

We compared the proposed method with ten existing HS anomaly detection methods from classical to state-of-the-art ones.
Specifically, we included the statistics-based methods: 
global Reed-Xiaoli detector (GRX)~\cite{GRX_1990} and 
two-step generalized likelihood ratio test (2S-GLRT)~\cite{2SGLRT_2022}; 
the deep learning-based methods:
guided autoencoder detection (GAED)~\cite{GAED_2022} and
robust graph autoencoder (RGAE)~\cite{RGAE_2022};
and the representation-based methods:
abundance- and dictionary-based low-rank decomposition (ADLR)~\cite{ADLR_2018}, 
graph and total variation regularized low-rank representation (GTVLRR)~\cite{GTVLRR_2020},  
low-rank and sparse decomposition with mixture of Gaussian (LSDM-MoG)~\cite{LSDMMog_2021}, 
principal component analysis-based tensor low-rank and sparse representation (PCA-TLRSR)~\cite{PCA-TLRSR_2023}, 
antinoise hierarchical mutual-incoherence-induced discriminative learning (AHMID)~\cite{AHMID_2023}, and
merging total variation into low-rank representation (MTVLRR)~\cite{MTVLRR_2024}.

Most existing HS anomaly detection methods are designed without explicit consideration of noise or are based on the assumption of Gaussian noise.
However, in real-world scenarios, HS images can be contaminated not only with thermal noise and quantization noise (typically modeled as Gaussian noise), but also with sparse noise caused by sensor defects or data transmission errors, and stripe noise arising from line-scanning procedures or calibration issues~\cite{noise_2018}.

Therefore, to evaluate the detection performance of the existing and proposed methods in various scenarios, we designed five noise contamination cases, as shown in Table~\ref{tab:noise_case}.
Case 1 serves as a baseline scenario without additional noise.
In Cases 2 and 3, we added Gaussian noise and non-Gaussian noise, respectively, to evaluate their individual effects on detection performance.
Finally, we introduced mixed-noise scenarios combining both noise types in Cases 4 and 5 to examine the performance of each method under more complex conditions.

\subsection{Evaluation Metrics}
To evaluate the detection performance, we used three types of areas under the receiver operating characteristic (ROC) curve (AUC) metrics~\cite{3DROC_2021}: $\mathrm{AUC}_{(P_D, P_F)}$, the area under $\mathrm{ROC}_{(P_D, P_F)}$; $\mathrm{AUC}_{(P_D, \tau)}$, the area under $\mathrm{ROC}_{(P_D, \tau)}$; and $\mathrm{AUC}_{(P_F, \tau)}$, the area under $\mathrm{ROC}_{(P_F, \tau)}$. 
Here, $P_D$, $P_F$, and $\tau$ denote the probability of detection, probability of false alarm, and threshold value, respectively.
The closer the values of $\mathrm{AUC}_{(P_D, P_F)}$ and $\mathrm{AUC}_{(P_D, \tau)}$ are to 1, the better the detection performance and anomaly detectability, respectively. 
In contrast, the closer the value of $\mathrm{AUC}_{(P_F, \tau)}$ is to 0, the better the background suppressibility.

\subsection{Parameter Setting}

The hyperparameters for each method were set to the values that maximized the $\mathrm{AUC}_{(P_D, P_F)}$ value within the ranges shown in Table~\ref{tab:para}.

For the proposed method, the stopping condition for Algorithm~\ref{alg:Proposed} was defined as
\begin{equation}
  \label{eq:stop_condition}
  \frac{\| \Tcal^{(n+1)} - \Tcal^{(n)} \|_F}{\| \Tcal^{(n)} \|_F} \leq r_{\mathrm{tol}},
\end{equation}
where $\Tcal^{(n)} = \Bcal^{(n)} + \Acal^{(n)} + \Scal^{(n)} + \Lcal^{(n)}$.
The maximum number of iterations was set to 10,000.

\begin{figure}[t]
\centering
\begin{minipage}[t]{0.48\linewidth}
    \centering
    \includegraphics[width=\linewidth]{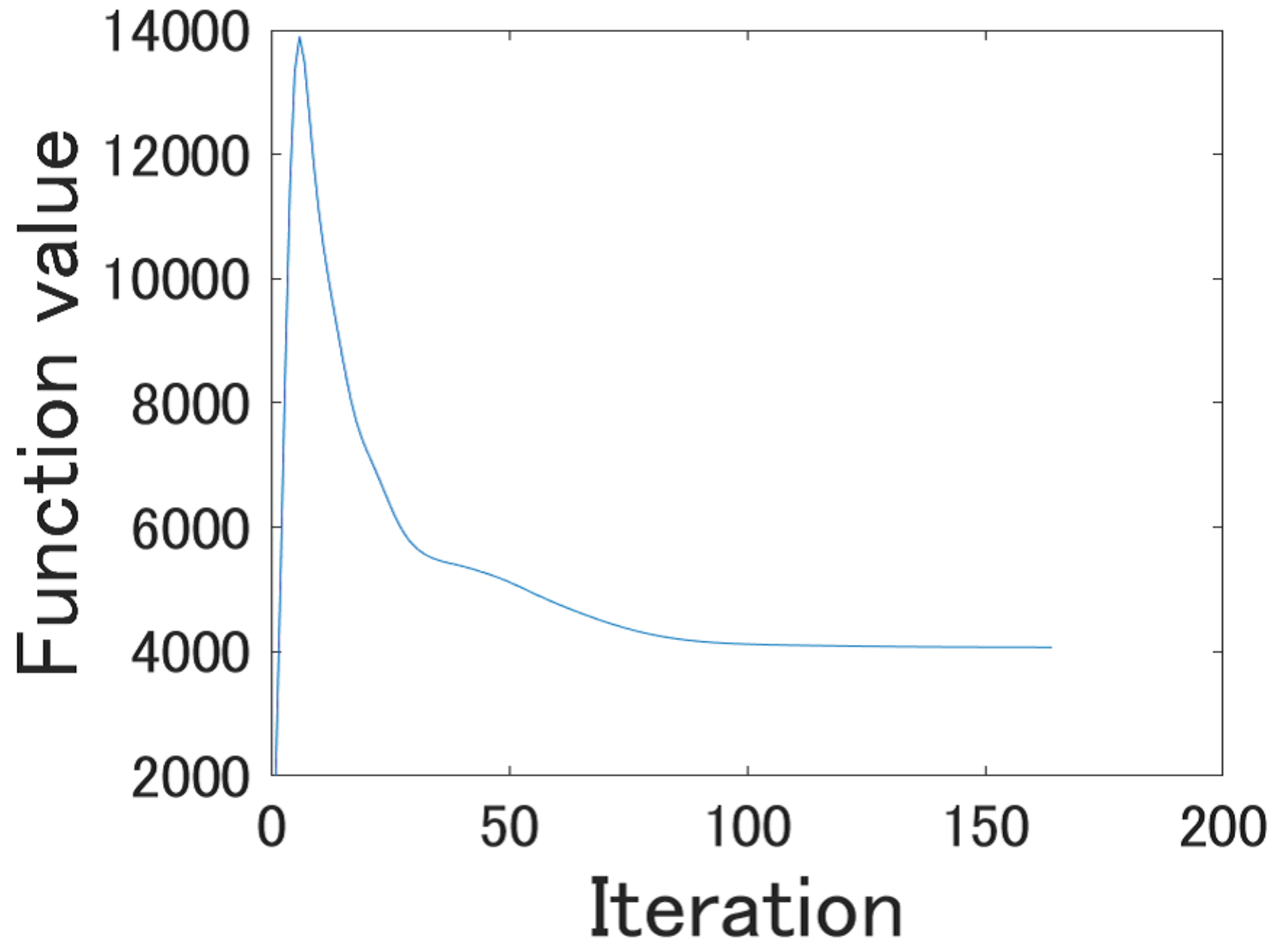}
\end{minipage}
\hfill
\begin{minipage}[t]{0.48\linewidth}
    \centering
    \includegraphics[width=\linewidth]{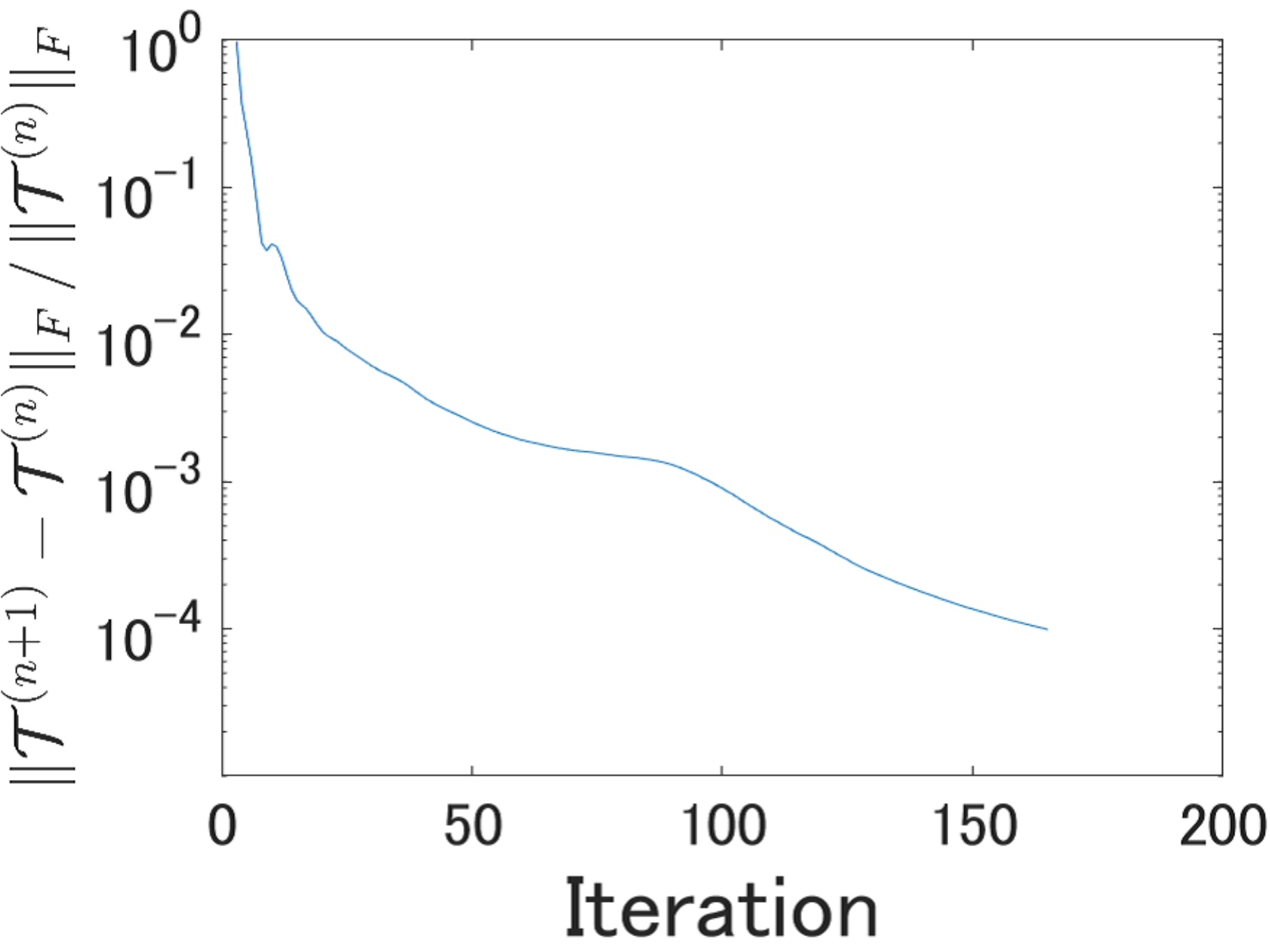}
\end{minipage}

\caption{Convergence behavior of the proposed method using HTV for Pavia Centre in Case 5.}
\label{fig:convergence}
\end{figure}

Fig.~\ref{fig:convergence} shows the evolution of the objective function value and the relative change of $\Tcal^{(n)}$. 
The objective function value is already stabilized when the relative change reaches the order of $10^{-4}$, indicating sufficient convergence. 
A similar tendency was observed across other datasets and noise conditions. 
Based on these observations, $r_{\mathrm{tol}} = 10^{-4}$ was adopted in this article.

Regarding the remaining parameters, $\varepsilon$ and $\alpha$ in 
Prob.~\eqref{eq:proposed-optimization-problem} were determined as
\begin{align}
    \label{eq:epsilon_alpha}
    \varepsilon = \eta\sigma  \sqrt{HWB(1-S_p)},  \quad \alpha = \frac{1}{2}\eta S_p HWB ,
\end{align}
with $\eta=0.9$.

\subsection{Experimental Results}

\subsubsection{Case 1 (Original datasets)}

\begin{table*}[!t]
    \centering
    \caption{$\mathrm{AUC}_{(P_D, P_F)}$, $\mathrm{AUC}_{(P_D, \tau)}$, And $\mathrm{AUC}_{(P_F, \tau)}$ Values of All The Detectors For Each Dataset In Case 1. \\ (The Best And Second-Best Values Are Highlighted in Bold And Underlined, Respectively.)}
    \label{tab:AUCs_Case1}
    \scalebox{0.75}{
    \begin{tabular}{cccccccccccccccc}
        \toprule
        Datasets & Metrics & \multicolumn{14}{c}{Methods} \\ 
        \cmidrule(lr){1-1} \cmidrule(lr){2-2} \cmidrule(lr){3-16}
        & & GRX & 2S-GLRT & GAED & RGAE & ADLR & GTVLRR & LSDM-MoG & PCA-TLRSR & AHMID & MTVLRR & Ours & Ours & Ours & Ours \\  
        & & \cite{GRX_1990} & \cite{2SGLRT_2022} & \cite{GAED_2022} & \cite{RGAE_2022} & \cite{ADLR_2018} &\cite{GTVLRR_2020} & \cite{LSDMMog_2021} & \cite{PCA-TLRSR_2023} & \cite{AHMID_2023} & \cite{MTVLRR_2024} & (HTV) & (SSTV) & (HSSTV) & (Nuclear) \\
     
        \midrule    
        \multirow{3}{*}{\shortstack{Pavia \\ Centre}} 
         & $\mathrm{AUC}_{(P_D, P_F)}$  & 0.9538 & \ValSecond{0.9868} & 0.9436 & 0.9166 & 0.9035 & \ValFifth{0.9829} & 0.9603 & 0.9720 & 0.9189 & \ValFourth{0.9836} & \ValBest{0.9907} & 0.9497 & \ValThird{0.9843} & 0.9498 \\ 
         & $\mathrm{AUC}_{(P_D, \tau)}$  & 0.1343 & 0.1607 & 0.0893 & 0.1452 & \ValSecond{0.3754} & 0.2262 & \ValBest{0.3849} & \ValThird{0.3476} & 0.0208 & 0.2174 & \ValFourth{0.3293} & 0.1537 & \ValFifth{0.2718} & 0.0941 \\ 
         & $\mathrm{AUC}_{(P_F, \tau)}$  & 0.0233 & 0.0186 & \ValThird{0.0065} & 0.0236 & 0.1487 & 0.0246 & 0.0627 & 0.0881 & \ValSecond{0.0022} & \ValFifth{0.0162} & 0.0181 & 0.0183 & \ValFourth{0.0161} & \ValBest{0.0012} \\ 
        \cmidrule(lr){1-16}

        \multirow{3}{*}{\shortstack{Texas \\ Coast}} 
         & $\mathrm{AUC}_{(P_D, P_F)}$  & \ValFifth{0.9907} & \ValSecond{0.9970} & 0.9811 & 0.9827 & 0.9801 & 0.9881 & \ValThird{0.9958} & \ValFourth{0.9926} & 0.9795 & 0.9597 & \ValBest{0.9978} & 0.9833 & 0.9888 & 0.9895 \\ 
         & $\mathrm{AUC}_{(P_D, \tau)}$  & 0.3143 & 0.1784 & 0.3694 & 0.3760 & \ValBest{0.9681} & \ValSecond{0.6571} & \ValThird{0.6302} & \ValFifth{0.5622} & 0.4639 & 0.4901 & 0.5484 & \ValFourth{0.6275} & 0.3349 & 0.3101 \\ 
         & $\mathrm{AUC}_{(P_F, \tau)}$  & 0.0556 & \ValBest{0.0055} & \ValFourth{0.0169} & \ValThird{0.0168} & 0.4732 & 0.1138 & 0.1233 & 0.1183 & 0.1215 & 0.0830 & 0.0336 & 0.2159 & \ValFifth{0.0232} & \ValSecond{0.0073} \\ 
        \cmidrule(lr){1-16}
         
        \multirow{3}{*}{\shortstack{Gainesville}}
         & $\mathrm{AUC}_{(P_D, P_F)}$  & 0.9513 & 0.9486 & 0.9610 & 0.8219 & 0.9743 & \ValThird{0.9926} & \ValFourth{0.9838} & \ValSecond{0.9928} & 0.9690 & 0.9742 & \ValBest{0.9950} & \ValFifth{0.9826} & 0.9816 & 0.9733 \\ 
         & $\mathrm{AUC}_{(P_D, \tau)}$  & 0.0963 & 0.0387 & 0.1196 & 0.1024 & \ValFourth{0.4188} & \ValBest{0.5042} & 0.4130 & \ValFifth{0.4144} & 0.2058 & 0.2122 & \ValSecond{0.4721} & 0.4042 & \ValThird{0.4280} & 0.0998 \\ 
         & $\mathrm{AUC}_{(P_F, \tau)}$  & 0.0351 & \ValBest{0.0036} & \ValThird{0.0167} & 0.0381 & 0.0294 & 0.0655 & 0.1208 & 0.0365 & 0.0446 & 0.0338 & \ValFourth{0.0218} & \ValFifth{0.0291} & 0.0377 & \ValSecond{0.0099} \\ 
        \cmidrule(lr){1-16}
        
        \multirow{3}{*}{\shortstack{Los \\ Angeles I}}
         & $\mathrm{AUC}_{(P_D, P_F)}$  & 0.9887 & 0.9265 & 0.9938 & 0.9948 & \ValSecond{0.9965} & 0.9923 & 0.9950 & 0.9874 & \ValBest{0.9966} & 0.9848 & \ValSecond{0.9965} & \ValSecond{0.9965} & \ValSecond{0.9965} & 0.9962 \\ 
         & $\mathrm{AUC}_{(P_D, \tau)}$  & 0.0891 & 0.0466 & 0.0379 & 0.0392 & \ValBest{0.7502} & 0.1235 & 0.1410 & 0.1109 & 0.0638 & 0.1570 & \ValFourth{0.1885} & \ValSecond{0.1913} & \ValThird{0.1912} & \ValFifth{0.1726} \\ 
         & $\mathrm{AUC}_{(P_F, \tau)}$  & 0.0114 & \ValSecond{0.0034} & \ValBest{0.0006} & \ValBest{0.0006} & 0.0672 & 0.0143 & \ValFourth{0.0039} & 0.0105 & \ValFifth{0.0046} & 0.0133 & 0.0363 & 0.0405 & 0.0404 & 0.0306 \\ 
        \cmidrule(lr){1-16}
        
        \multirow{3}{*}{\shortstack{Los \\ Angeles I\hspace{-1.2pt}I}}
         & $\mathrm{AUC}_{(P_D, P_F)}$  & \ValFifth{0.9692} & 0.9406 & 0.9391 & 0.9572 & 0.9051 & 0.9290 & 0.9613 & \ValThird{0.9834} & 0.9677 & 0.9045 & \ValBest{0.9890} & \ValSecond{0.9843} & \ValFourth{0.9829} & 0.9664 \\ 
         & $\mathrm{AUC}_{(P_D, \tau)}$  & 0.1461 & 0.0680 & 0.2409 & 0.2536 & \ValBest{0.8487} & 0.3058 & \ValSecond{0.4358} & \ValFourth{0.3774} & 0.0869 & 0.2900 & \ValThird{0.4202} & \ValFifth{0.3697} & 0.2796 & 0.1994 \\ 
         & $\mathrm{AUC}_{(P_F, \tau)}$  & 0.0437 & \ValSecond{0.0064} & 0.0190 & \ValFourth{0.0180} & 0.4947 & 0.0989 & 0.0875 & 0.0908 & \ValBest{0.0026} & 0.1052 & 0.0331 & 0.0317 & \ValFifth{0.0180} & \ValThird{0.0109} \\ 
        \cmidrule(lr){1-16}
        
        \multirow{3}{*}{\shortstack{San \\ Diego}}
         & $\mathrm{AUC}_{(P_D, P_F)}$  & 0.9403 & 0.9074 & \ValThird{0.9902} & \ValSecond{0.9921} & \ValFourth{0.9892} & \ValBest{0.9927} & 0.9709 & 0.9827 & 0.9263 & \ValFifth{0.9884} & 0.9866 & 0.9603 & 0.9678 & 0.9738 \\ 
         & $\mathrm{AUC}_{(P_D, \tau)}$  & 0.1778 & 0.0002 & 0.1882 & 0.1823 & \ValBest{0.7943} & \ValThird{0.4235} & \ValSecond{0.5505} & 0.3189 & 0.2425 & \ValFourth{0.4159} & \ValFifth{0.3861} & 0.2851 & 0.2352 & 0.1633 \\ 
         & $\mathrm{AUC}_{(P_F, \tau)}$  & 0.0589 & \ValBest{0.0007} & \ValThird{0.0080} & \ValSecond{0.0076} & 0.3182 & 0.0604 & 0.2011 & 0.0445 & 0.0383 & 0.0247 & 0.0334 & 0.0486 & \ValFifth{0.0145} & \ValFourth{0.0101} \\ 
        \cmidrule(lr){1-16} 

        \multirow{3}{*}{\shortstack{Hyperion}}
         & $\mathrm{AUC}_{(P_D, P_F)}$  & 0.9978 & 0.9934 & 0.9772 & 0.9416 & 0.9925 & \ValThird{0.9990} & \ValBest{0.9999} & \ValSecond{0.9997} & 0.9843 & \ValFifth{0.9983} & 0.9976 & 0.9818 & 0.9848 & \ValFourth{0.9983} \\ 
         & $\mathrm{AUC}_{(P_D, \tau)}$  & 0.2356 & 0.0255 & 0.2193 & 0.2412 & \ValBest{0.9635} & 0.2838 & 0.3989 & 0.3716 & \ValThird{0.6092} & \ValFourth{0.4367} & 0.3867 & \ValSecond{0.6909} & \ValFifth{0.4310} & 0.1249 \\ 
         & $\mathrm{AUC}_{(P_F, \tau)}$  & \ValFifth{0.0331} & \ValSecond{0.0003} & \ValThird{0.0124} & 0.0418 & 0.4378 & \ValFourth{0.0254} & 0.1179 & 0.0359 & 0.3073 & 0.0373 & 0.0383 & 0.4285 & 0.0756 & \ValBest{0.0001} \\ 

        \bottomrule
 
    \end{tabular}
    }
    \\
\end{table*}
\begin{figure*}[!t]
    \centering

    \begin{minipage}{0.03\hsize}
        \centerline{{\rotatebox{90}{\small{\shortstack{(a) Texas Coast (Case 1)}}}}}
    \end{minipage}
    \begin{minipage}{0.95\hsize}
        \begin{minipage}{0.115\hsize}
            \includegraphics[keepaspectratio, scale = 0.555]{figs/Original/Urban1/pseudocolor-eps-converted-to.pdf}
        \end{minipage}
        \begin{minipage}{0.115\hsize}
            \includegraphics[keepaspectratio, scale = 0.555]{figs/Original/Urban1/GT-eps-converted-to.pdf}
        \end{minipage}
        \begin{minipage}{0.115\hsize}
            \includegraphics[keepaspectratio, scale = 0.555]{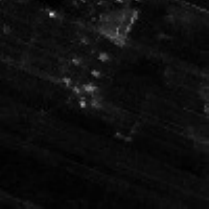}
        \end{minipage}
        \begin{minipage}{0.115\hsize}
            \includegraphics[keepaspectratio, scale = 0.555]{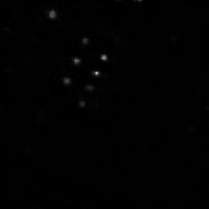}
        \end{minipage}
        \begin{minipage}{0.115\hsize}
            \includegraphics[keepaspectratio, scale = 0.555]{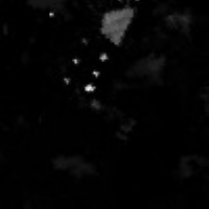}
        \end{minipage}
        \begin{minipage}{0.115\hsize}
            \includegraphics[keepaspectratio, scale = 0.555]{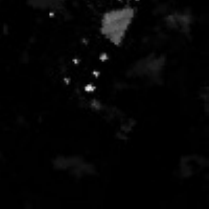}
        \end{minipage}
        \begin{minipage}{0.115\hsize}
            \includegraphics[keepaspectratio, scale = 0.555]{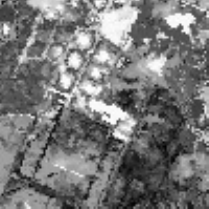}
        \end{minipage}
        \begin{minipage}{0.115\hsize}
            \includegraphics[keepaspectratio, scale = 0.555]{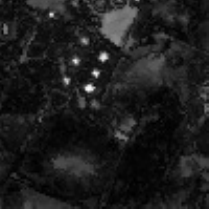}
        \end{minipage}
    
        \vspace{0.5mm}
    
        \begin{minipage}{0.115\hsize}
            \centerline{\footnotesize{Pseudocolor}}
        \end{minipage}
        \begin{minipage}{0.115\hsize}
            \centerline{\footnotesize{Ground Truth}}
        \end{minipage}
        \begin{minipage}{0.115\hsize}
            \centerline{\footnotesize{GRX}}
        \end{minipage}
        \begin{minipage}{0.115\hsize}
            \centerline{\footnotesize{2S-GLRT}}
        \end{minipage}
        \begin{minipage}{0.115\hsize}
            \centerline{\footnotesize{GAED}}
        \end{minipage}
        \begin{minipage}{0.115\hsize}
            \centerline{\footnotesize{RGAE}}
        \end{minipage}
        \begin{minipage}{0.115\hsize}
            \centerline{\footnotesize{ADLR}}
        \end{minipage}
        \begin{minipage}{0.115\hsize}
            \centerline{\footnotesize{GTVLRR}}
        \end{minipage}
    
        \vspace{1mm}
    
        \begin{minipage}{0.115\hsize}
            \includegraphics[keepaspectratio, scale = 0.555]{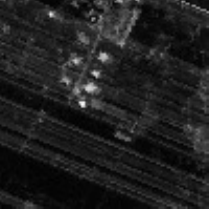}
        \end{minipage}
        \begin{minipage}{0.115\hsize}
            \includegraphics[keepaspectratio, scale = 0.555]{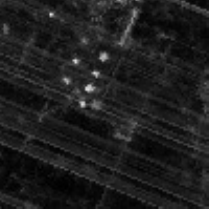}
        \end{minipage}
        \begin{minipage}{0.115\hsize}
            \includegraphics[keepaspectratio, scale = 0.555]{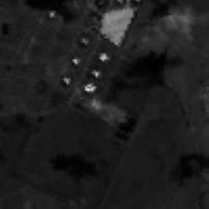}
        \end{minipage}
        \begin{minipage}{0.115\hsize}
            \includegraphics[keepaspectratio, scale = 0.555]{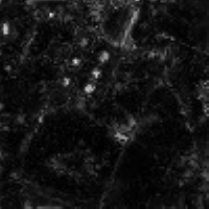}
        \end{minipage}
        \begin{minipage}{0.115\hsize}
            \includegraphics[keepaspectratio, scale = 0.555]{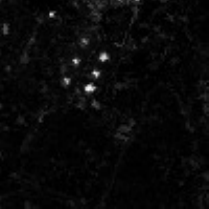}
        \end{minipage}
        \begin{minipage}{0.115\hsize}
            \includegraphics[keepaspectratio, scale = 0.555]{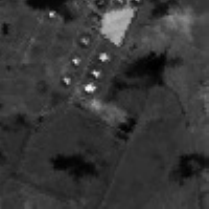}
        \end{minipage}
        \begin{minipage}{0.115\hsize}
            \includegraphics[keepaspectratio, scale = 0.555]{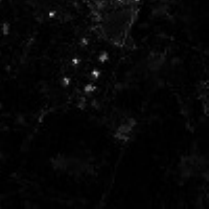}
        \end{minipage}
        \begin{minipage}{0.115\hsize}
            \includegraphics[keepaspectratio, scale = 0.555]{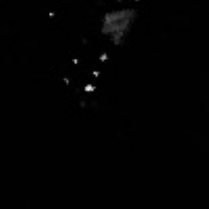}
        \end{minipage}
    
        \vspace{0.5mm}
    
        \begin{minipage}{0.115\hsize}
            \centerline{\footnotesize{LSDM-MoG}}
        \end{minipage}
        \begin{minipage}{0.115\hsize}
            \centerline{\footnotesize{PCA-TLRSR}}
        \end{minipage}
        \begin{minipage}{0.115\hsize}
            \centerline{\footnotesize{AHMID}}
        \end{minipage}
        \begin{minipage}{0.115\hsize}
            \centerline{\footnotesize{MTVLRR}}
        \end{minipage}
        \begin{minipage}{0.115\hsize}
            \centerline{\footnotesize{Ours (HTV)}}
        \end{minipage}
        \begin{minipage}{0.115\hsize}
            \centerline{\footnotesize{Ours (SSTV)}}
        \end{minipage}
        \begin{minipage}{0.115\hsize}
            \centerline{\footnotesize{Ours (HSSTV)}}
        \end{minipage}
        \begin{minipage}{0.115\hsize}
            \centerline{\footnotesize{Ours (Nuclear)}}
        \end{minipage}
    \end{minipage}

    \vspace{2mm}

    \begin{minipage}{0.03\hsize}
        \centerline{{\rotatebox{90}{\small{\shortstack{(b) Gainesville (Case 1)}}}}}
    \end{minipage}
    \begin{minipage}{0.95\hsize}
        \begin{minipage}{0.115\hsize}
            \includegraphics[keepaspectratio, scale = 0.555]{figs/Original/Urban3/pseudocolor-eps-converted-to.pdf}
        \end{minipage}
        \begin{minipage}{0.115\hsize}
            \includegraphics[keepaspectratio, scale = 0.555]{figs/Original/Urban3/GT-eps-converted-to.pdf}
        \end{minipage}
        \begin{minipage}{0.115\hsize}
            \includegraphics[keepaspectratio, scale = 0.555]{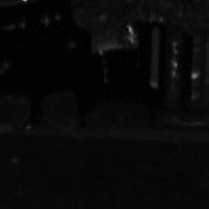}
        \end{minipage}
        \begin{minipage}{0.115\hsize}
            \includegraphics[keepaspectratio, scale = 0.555]{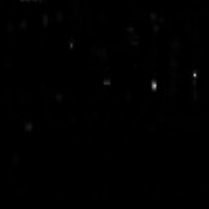}
        \end{minipage}
        \begin{minipage}{0.115\hsize}
            \includegraphics[keepaspectratio, scale = 0.555]{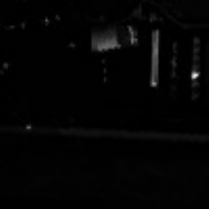}
        \end{minipage}
        \begin{minipage}{0.115\hsize}
            \includegraphics[keepaspectratio, scale = 0.555]{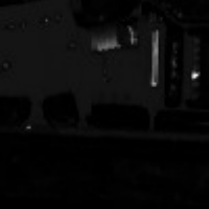}
        \end{minipage}
        \begin{minipage}{0.115\hsize}
            \includegraphics[keepaspectratio, scale = 0.555]{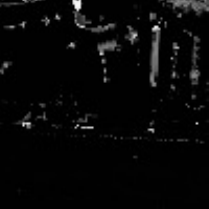}
        \end{minipage}
        \begin{minipage}{0.115\hsize}
            \includegraphics[keepaspectratio, scale = 0.555]{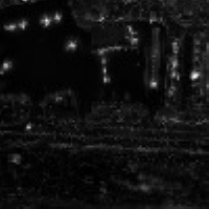}
        \end{minipage}
    
        \vspace{0.5mm}
    
        \begin{minipage}{0.115\hsize}
            \centerline{\footnotesize{Pseudocolor}}
        \end{minipage}
        \begin{minipage}{0.115\hsize}
            \centerline{\footnotesize{Ground Truth}}
        \end{minipage}
        \begin{minipage}{0.115\hsize}
            \centerline{\footnotesize{GRX}}
        \end{minipage}
        \begin{minipage}{0.115\hsize}
            \centerline{\footnotesize{2S-GLRT}}
        \end{minipage}
        \begin{minipage}{0.115\hsize}
            \centerline{\footnotesize{GAED}}
        \end{minipage}
        \begin{minipage}{0.115\hsize}
            \centerline{\footnotesize{RGAE}}
        \end{minipage}
        \begin{minipage}{0.115\hsize}
            \centerline{\footnotesize{ADLR}}
        \end{minipage}
        \begin{minipage}{0.115\hsize}
            \centerline{\footnotesize{GTVLRR}}
        \end{minipage}
    
        \vspace{1mm}
    
        \begin{minipage}{0.115\hsize}
            \includegraphics[keepaspectratio, scale = 0.555]{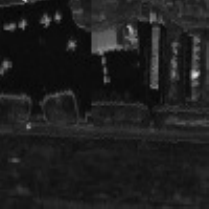}
        \end{minipage}
        \begin{minipage}{0.115\hsize}
            \includegraphics[keepaspectratio, scale = 0.555]{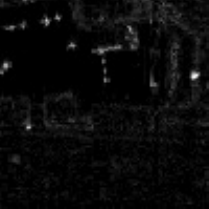}
        \end{minipage}
        \begin{minipage}{0.115\hsize}
            \includegraphics[keepaspectratio, scale = 0.555]{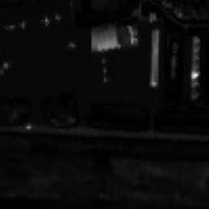}
        \end{minipage}
        \begin{minipage}{0.115\hsize}
            \includegraphics[keepaspectratio, scale = 0.555]{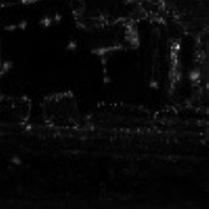}
        \end{minipage}
        \begin{minipage}{0.115\hsize}
            \includegraphics[keepaspectratio, scale = 0.555]{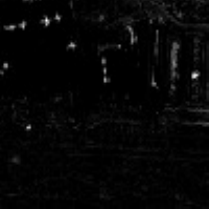}
        \end{minipage}
        \begin{minipage}{0.115\hsize}
            \includegraphics[keepaspectratio, scale = 0.555]{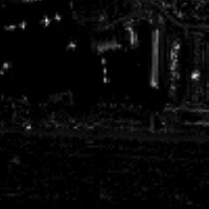}
        \end{minipage}
        \begin{minipage}{0.115\hsize}
            \includegraphics[keepaspectratio, scale = 0.555]{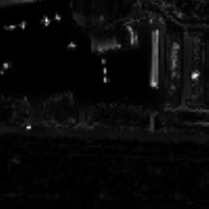}
        \end{minipage}
        \begin{minipage}{0.115\hsize}
            \includegraphics[keepaspectratio, scale = 0.555]{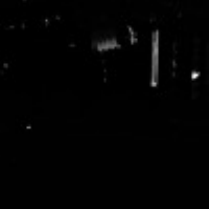}
        \end{minipage}
    
        \vspace{0.5mm}
    
        \begin{minipage}{0.115\hsize}
            \centerline{\footnotesize{LSDM-MoG}}
        \end{minipage}
        \begin{minipage}{0.115\hsize}
            \centerline{\footnotesize{PCA-TLRSR}}
        \end{minipage}
        \begin{minipage}{0.115\hsize}
            \centerline{\footnotesize{AHMID}}
        \end{minipage}
        \begin{minipage}{0.115\hsize}
            \centerline{\footnotesize{MTVLRR}}
        \end{minipage}
        \begin{minipage}{0.115\hsize}
            \centerline{\footnotesize{Ours (HTV)}}
        \end{minipage}
        \begin{minipage}{0.115\hsize}
            \centerline{\footnotesize{Ours (SSTV)}}
        \end{minipage}
        \begin{minipage}{0.115\hsize}
            \centerline{\footnotesize{Ours (HSSTV)}}
        \end{minipage}
        \begin{minipage}{0.115\hsize}
            \centerline{\footnotesize{Ours (Nuclear)}}
        \end{minipage}
    \end{minipage}

    \vspace{2mm}

    \begin{minipage}{0.03\hsize}
        \centerline{{\rotatebox{90}{\small{\shortstack{(c) Los Angeles I (Case 1)}}}}}
    \end{minipage}
    \begin{minipage}{0.95\hsize}
        \begin{minipage}{0.115\hsize}
            \includegraphics[keepaspectratio, scale = 0.555]{figs/Original/Urban4/pseudocolor-eps-converted-to.pdf}
        \end{minipage}
        \begin{minipage}{0.115\hsize}
            \includegraphics[keepaspectratio, scale = 0.555]{figs/Original/Urban4/GT-eps-converted-to.pdf}
        \end{minipage}
        \begin{minipage}{0.115\hsize}
            \includegraphics[keepaspectratio, scale = 0.555]{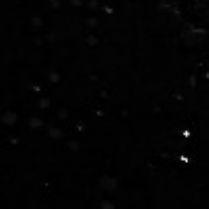}
        \end{minipage}
        \begin{minipage}{0.115\hsize}
            \includegraphics[keepaspectratio, scale = 0.555]{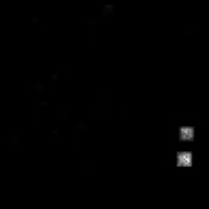}
        \end{minipage}
        \begin{minipage}{0.115\hsize}
            \includegraphics[keepaspectratio, scale = 0.555]{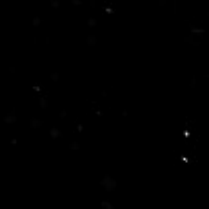}
        \end{minipage}
        \begin{minipage}{0.115\hsize}
            \includegraphics[keepaspectratio, scale = 0.555]{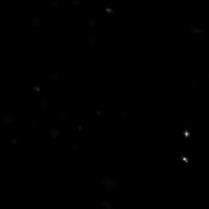}
        \end{minipage}
        \begin{minipage}{0.115\hsize}
            \includegraphics[keepaspectratio, scale = 0.555]{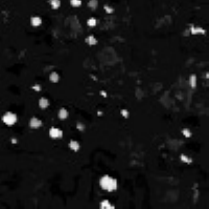}
        \end{minipage}
        \begin{minipage}{0.115\hsize}
            \includegraphics[keepaspectratio, scale = 0.555]{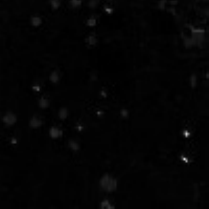}
        \end{minipage}
    
        \vspace{0.5mm}
    
        \begin{minipage}{0.115\hsize}
            \centerline{\footnotesize{Pseudocolor}}
        \end{minipage}
        \begin{minipage}{0.115\hsize}
            \centerline{\footnotesize{Ground Truth}}
        \end{minipage}
        \begin{minipage}{0.115\hsize}
            \centerline{\footnotesize{GRX}}
        \end{minipage}
        \begin{minipage}{0.115\hsize}
            \centerline{\footnotesize{2S-GLRT}}
        \end{minipage}
        \begin{minipage}{0.115\hsize}
            \centerline{\footnotesize{GAED}}
        \end{minipage}
        \begin{minipage}{0.115\hsize}
            \centerline{\footnotesize{RGAE}}
        \end{minipage}
        \begin{minipage}{0.115\hsize}
            \centerline{\footnotesize{ADLR}}
        \end{minipage}
        \begin{minipage}{0.115\hsize}
            \centerline{\footnotesize{GTVLRR}}
        \end{minipage}
    
        \vspace{1mm}
    
        \begin{minipage}{0.115\hsize}
            \includegraphics[keepaspectratio, scale = 0.555]{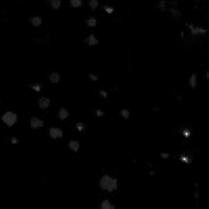}
        \end{minipage}
        \begin{minipage}{0.115\hsize}
            \includegraphics[keepaspectratio, scale = 0.555]{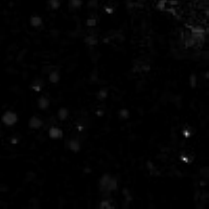}
        \end{minipage}
        \begin{minipage}{0.115\hsize}
            \includegraphics[keepaspectratio, scale = 0.555]{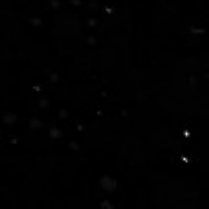}
        \end{minipage}
        \begin{minipage}{0.115\hsize}
            \includegraphics[keepaspectratio, scale = 0.555]{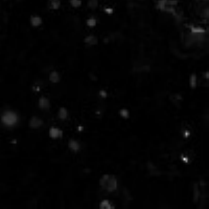}
        \end{minipage}
        \begin{minipage}{0.115\hsize}
            \includegraphics[keepaspectratio, scale = 0.555]{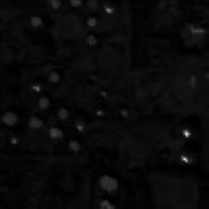}
        \end{minipage}
        \begin{minipage}{0.115\hsize}
            \includegraphics[keepaspectratio, scale = 0.555]{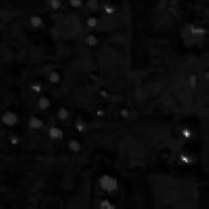}
        \end{minipage}
        \begin{minipage}{0.115\hsize}
            \includegraphics[keepaspectratio, scale = 0.555]{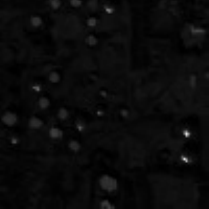}
        \end{minipage}
        \begin{minipage}{0.115\hsize}
            \includegraphics[keepaspectratio, scale = 0.555]{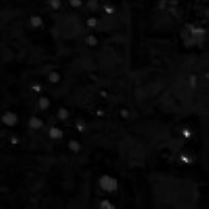}
        \end{minipage}
    
        \vspace{0.5mm}
    
        \begin{minipage}{0.115\hsize}
            \centerline{\footnotesize{LSDM-MoG}}
        \end{minipage}
        \begin{minipage}{0.115\hsize}
            \centerline{\footnotesize{PCA-TLRSR}}
        \end{minipage}
        \begin{minipage}{0.115\hsize}
            \centerline{\footnotesize{AHMID}}
        \end{minipage}
        \begin{minipage}{0.115\hsize}
            \centerline{\footnotesize{MTVLRR}}
        \end{minipage}
        \begin{minipage}{0.115\hsize}
            \centerline{\footnotesize{Ours (HTV)}}
        \end{minipage}
        \begin{minipage}{0.115\hsize}
            \centerline{\footnotesize{Ours (SSTV)}}
        \end{minipage}
        \begin{minipage}{0.115\hsize}
            \centerline{\footnotesize{Ours (HSSTV)}}
        \end{minipage}
        \begin{minipage}{0.115\hsize}
            \centerline{\footnotesize{Ours (Nuclear)}}
        \end{minipage}
    \end{minipage}

    \vspace{2mm}

    \begin{minipage}{0.03\hsize}
        \centerline{{\rotatebox{90}{\small{\shortstack{(d) Los Angeles I\hspace{-1.2pt}I (Case 1)}}}}}
    \end{minipage}
    \begin{minipage}{0.95\hsize}
        \begin{minipage}{0.115\hsize}
            \includegraphics[keepaspectratio, scale = 0.555]{figs/Original/Urban5/pseudocolor-eps-converted-to.pdf}
        \end{minipage}
        \begin{minipage}{0.115\hsize}
            \includegraphics[keepaspectratio, scale = 0.555]{figs/Original/Urban5/GT-eps-converted-to.pdf}
        \end{minipage}
        \begin{minipage}{0.115\hsize}
            \includegraphics[keepaspectratio, scale = 0.555]{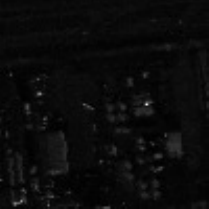}
        \end{minipage}
        \begin{minipage}{0.115\hsize}
            \includegraphics[keepaspectratio, scale = 0.555]{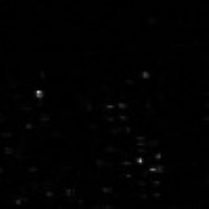}
        \end{minipage}
        \begin{minipage}{0.115\hsize}
            \includegraphics[keepaspectratio, scale = 0.555]{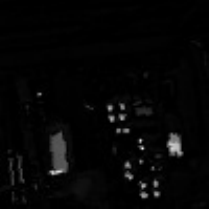}
        \end{minipage}
        \begin{minipage}{0.115\hsize}
            \includegraphics[keepaspectratio, scale = 0.555]{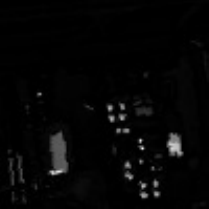}
        \end{minipage}
        \begin{minipage}{0.115\hsize}
            \includegraphics[keepaspectratio, scale = 0.555]{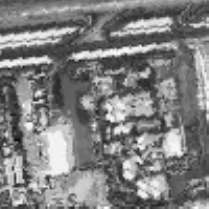}
        \end{minipage}
        \begin{minipage}{0.115\hsize}
            \includegraphics[keepaspectratio, scale = 0.555]{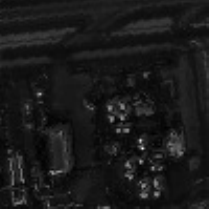}
        \end{minipage}
    
        \vspace{0.5mm}
    
        \begin{minipage}{0.115\hsize}
            \centerline{\footnotesize{Pseudocolor}}
        \end{minipage}
        \begin{minipage}{0.115\hsize}
            \centerline{\footnotesize{Ground Truth}}
        \end{minipage}
        \begin{minipage}{0.115\hsize}
            \centerline{\footnotesize{GRX}}
        \end{minipage}
        \begin{minipage}{0.115\hsize}
            \centerline{\footnotesize{2S-GLRT}}
        \end{minipage}
        \begin{minipage}{0.115\hsize}
            \centerline{\footnotesize{GAED}}
        \end{minipage}
        \begin{minipage}{0.115\hsize}
            \centerline{\footnotesize{RGAE}}
        \end{minipage}
        \begin{minipage}{0.115\hsize}
            \centerline{\footnotesize{ADLR}}
        \end{minipage}
        \begin{minipage}{0.115\hsize}
            \centerline{\footnotesize{GTVLRR}}
        \end{minipage}
    
        \vspace{1mm}
    
        \begin{minipage}{0.115\hsize}
            \includegraphics[keepaspectratio, scale = 0.555]{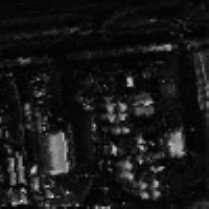}
        \end{minipage}
        \begin{minipage}{0.115\hsize}
            \includegraphics[keepaspectratio, scale = 0.555]{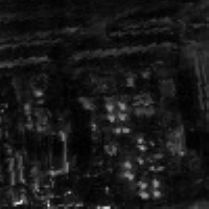}
        \end{minipage}
        \begin{minipage}{0.115\hsize}
            \includegraphics[keepaspectratio, scale = 0.555]{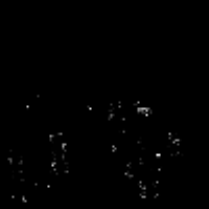}
        \end{minipage}
        \begin{minipage}{0.115\hsize}
            \includegraphics[keepaspectratio, scale = 0.555]{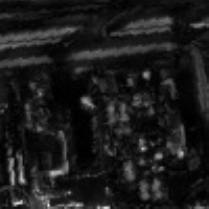}
        \end{minipage}
        \begin{minipage}{0.115\hsize}
            \includegraphics[keepaspectratio, scale = 0.555]{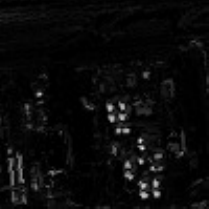}
        \end{minipage}
        \begin{minipage}{0.115\hsize}
            \includegraphics[keepaspectratio, scale = 0.555]{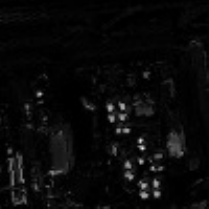}
        \end{minipage}
        \begin{minipage}{0.115\hsize}
            \includegraphics[keepaspectratio, scale = 0.555]{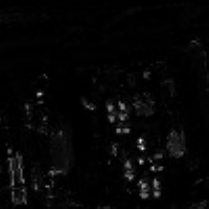}
        \end{minipage}
        \begin{minipage}{0.115\hsize}
            \includegraphics[keepaspectratio, scale = 0.555]{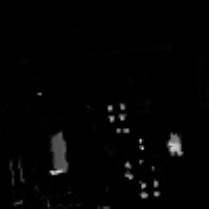}
        \end{minipage}
    
        \vspace{0.5mm}
    
        \begin{minipage}{0.115\hsize}
            \centerline{\footnotesize{LSDM-MoG}}
        \end{minipage}
        \begin{minipage}{0.115\hsize}
            \centerline{\footnotesize{PCA-TLRSR}}
        \end{minipage}
        \begin{minipage}{0.115\hsize}
            \centerline{\footnotesize{AHMID}}
        \end{minipage}
        \begin{minipage}{0.115\hsize}
            \centerline{\footnotesize{MTVLRR}}
        \end{minipage}
        \begin{minipage}{0.115\hsize}
            \centerline{\footnotesize{Ours (HTV)}}
        \end{minipage}
        \begin{minipage}{0.115\hsize}
            \centerline{\footnotesize{Ours (SSTV)}}
        \end{minipage}
        \begin{minipage}{0.115\hsize}
            \centerline{\footnotesize{Ours (HSSTV)}}
        \end{minipage}
        \begin{minipage}{0.115\hsize}
            \centerline{\footnotesize{Ours (Nuclear)}}
        \end{minipage}
    \end{minipage}

    \vspace{2mm}

    \begin{minipage}{0.03\hsize}
        \centerline{{\rotatebox{90}{\small{\shortstack{(e) San Diego (Case 1)}}}}}
    \end{minipage}
    \begin{minipage}{0.95\hsize}
        \begin{minipage}{0.115\hsize}
            \includegraphics[keepaspectratio, scale = 0.555]{figs/Original/SanDiego/pseudocolor-eps-converted-to.pdf}
        \end{minipage}
        \begin{minipage}{0.115\hsize}
            \includegraphics[keepaspectratio, scale = 0.555]{figs/Original/SanDiego/GT-eps-converted-to.pdf}
        \end{minipage}
        \begin{minipage}{0.115\hsize}
            \includegraphics[keepaspectratio, scale = 0.555]{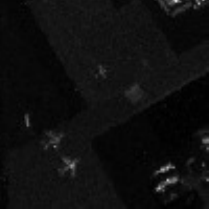}
        \end{minipage}
        \begin{minipage}{0.115\hsize}
            \includegraphics[keepaspectratio, scale = 0.555]{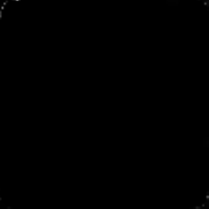}
        \end{minipage}
        \begin{minipage}{0.115\hsize}
            \includegraphics[keepaspectratio, scale = 0.555]{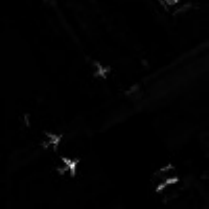}
        \end{minipage}
        \begin{minipage}{0.115\hsize}
            \includegraphics[keepaspectratio, scale = 0.555]{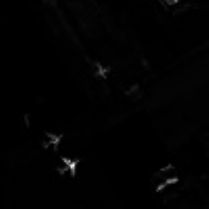}
        \end{minipage}
        \begin{minipage}{0.115\hsize}
            \includegraphics[keepaspectratio, scale = 0.555]{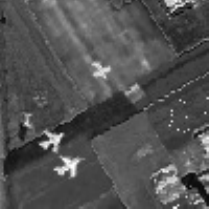}
        \end{minipage}
        \begin{minipage}{0.115\hsize}
            \includegraphics[keepaspectratio, scale = 0.555]{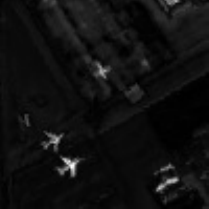}
        \end{minipage}
    
        \vspace{0.5mm}
    
        \begin{minipage}{0.115\hsize}
            \centerline{\footnotesize{Pseudocolor}}
        \end{minipage}
        \begin{minipage}{0.115\hsize}
            \centerline{\footnotesize{Ground Truth}}
        \end{minipage}
        \begin{minipage}{0.115\hsize}
            \centerline{\footnotesize{GRX}}
        \end{minipage}
        \begin{minipage}{0.115\hsize}
            \centerline{\footnotesize{2S-GLRT}}
        \end{minipage}
        \begin{minipage}{0.115\hsize}
            \centerline{\footnotesize{GAED}}
        \end{minipage}
        \begin{minipage}{0.115\hsize}
            \centerline{\footnotesize{RGAE}}
        \end{minipage}
        \begin{minipage}{0.115\hsize}
            \centerline{\footnotesize{ADLR}}
        \end{minipage}
        \begin{minipage}{0.115\hsize}
            \centerline{\footnotesize{GTVLRR}}
        \end{minipage}
    
        \vspace{1mm}
    
        \begin{minipage}{0.115\hsize}
            \includegraphics[keepaspectratio, scale = 0.555]{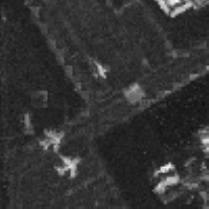}
        \end{minipage}
        \begin{minipage}{0.115\hsize}
            \includegraphics[keepaspectratio, scale = 0.555]{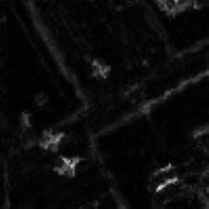}
        \end{minipage}
        \begin{minipage}{0.115\hsize}
            \includegraphics[keepaspectratio, scale = 0.555]{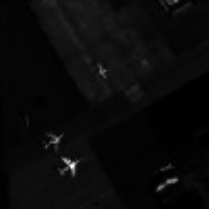}
        \end{minipage}
        \begin{minipage}{0.115\hsize}
            \includegraphics[keepaspectratio, scale = 0.555]{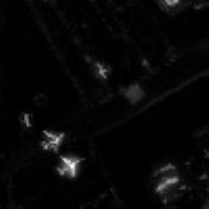}
        \end{minipage}
        \begin{minipage}{0.115\hsize}
            \includegraphics[keepaspectratio, scale = 0.555]{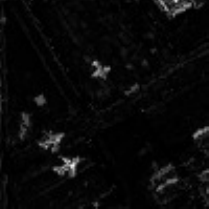}
        \end{minipage}
        \begin{minipage}{0.115\hsize}
            \includegraphics[keepaspectratio, scale = 0.555]{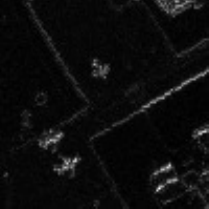}
        \end{minipage}
        \begin{minipage}{0.115\hsize}
            \includegraphics[keepaspectratio, scale = 0.555]{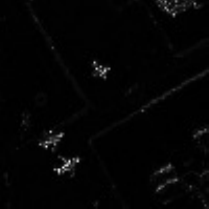}
        \end{minipage}
        \begin{minipage}{0.115\hsize}
            \includegraphics[keepaspectratio, scale = 0.555]{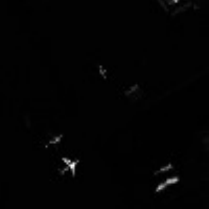}
        \end{minipage}
    
        \vspace{0.5mm}
    
        \begin{minipage}{0.115\hsize}
            \centerline{\footnotesize{LSDM-MoG}}
        \end{minipage}
        \begin{minipage}{0.115\hsize}
            \centerline{\footnotesize{PCA-TLRSR}}
        \end{minipage}
        \begin{minipage}{0.115\hsize}
            \centerline{\footnotesize{AHMID}}
        \end{minipage}
        \begin{minipage}{0.115\hsize}
            \centerline{\footnotesize{MTVLRR}}
        \end{minipage}
        \begin{minipage}{0.115\hsize}
            \centerline{\footnotesize{Ours (HTV)}}
        \end{minipage}
        \begin{minipage}{0.115\hsize}
            \centerline{\footnotesize{Ours (SSTV)}}
        \end{minipage}
        \begin{minipage}{0.115\hsize}
            \centerline{\footnotesize{Ours (HSSTV)}}
        \end{minipage}
        \begin{minipage}{0.115\hsize}
            \centerline{\footnotesize{Ours (Nuclear)}}
        \end{minipage}
    \end{minipage}

    \caption{Resulting detection maps for Texas Coast, Gainesville, Los Angeles I, Los Angeles I\hspace{-1.2pt}I, and San Diego in Case 1 generated by all the detectors.}
    \label{fig:DetectionMap_Case1}

\end{figure*}

\begin{figure*}[!t]
    \centering

    \begin{minipage}{0.03\hsize}
        \centerline{{\rotatebox{90}{\small{\shortstack{Hyperion (Case 1)}}}}}
    \end{minipage}
    \begin{minipage}{0.95\hsize}
        \begin{minipage}{0.115\hsize}
            \includegraphics[keepaspectratio, scale = 0.37]{figs/Original/Hyperion/pseudocolor-eps-converted-to.pdf}
        \end{minipage}
        \begin{minipage}{0.115\hsize}
            \includegraphics[keepaspectratio, scale = 0.37]{figs/Original/Hyperion/GT-eps-converted-to.pdf}
        \end{minipage}
        \begin{minipage}{0.115\hsize}
            \includegraphics[keepaspectratio, scale = 0.37]{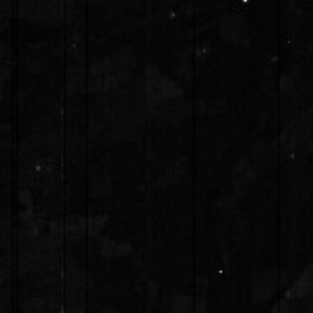}
        \end{minipage}
        \begin{minipage}{0.115\hsize}
            \includegraphics[keepaspectratio, scale = 0.37]{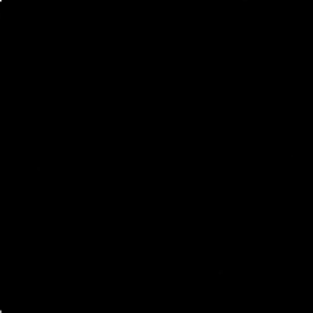}
        \end{minipage}
        \begin{minipage}{0.115\hsize}
            \includegraphics[keepaspectratio, scale = 0.37]{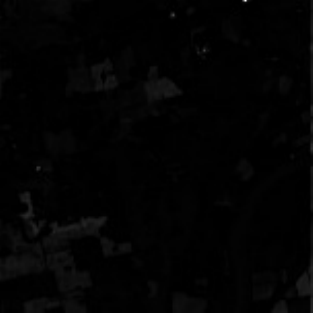}
        \end{minipage}
        \begin{minipage}{0.115\hsize}
            \includegraphics[keepaspectratio, scale = 0.37]{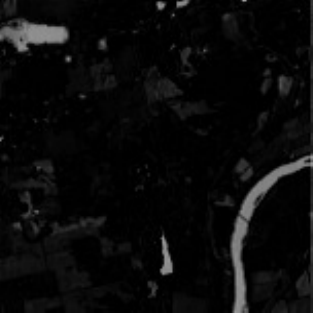}
        \end{minipage}
        \begin{minipage}{0.115\hsize}
            \includegraphics[keepaspectratio, scale = 0.37]{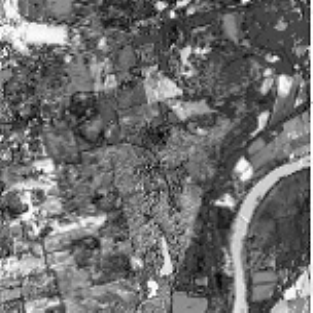}
        \end{minipage}
        \begin{minipage}{0.115\hsize}
            \includegraphics[keepaspectratio, scale = 0.37]{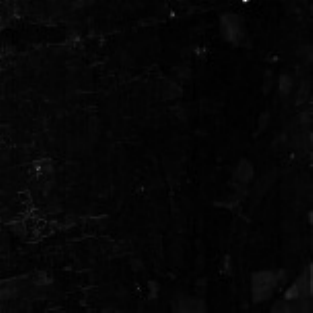}
        \end{minipage}
    
        \vspace{0.5mm}
    
        \begin{minipage}{0.115\hsize}
            \centerline{\footnotesize{Pseudocolor}}
        \end{minipage}
        \begin{minipage}{0.115\hsize}
            \centerline{\footnotesize{Ground Truth}}
        \end{minipage}
        \begin{minipage}{0.115\hsize}
            \centerline{\footnotesize{GRX}}
        \end{minipage}
        \begin{minipage}{0.115\hsize}
            \centerline{\footnotesize{2S-GLRT}}
        \end{minipage}
        \begin{minipage}{0.115\hsize}
            \centerline{\footnotesize{GAED}}
        \end{minipage}
        \begin{minipage}{0.115\hsize}
            \centerline{\footnotesize{RGAE}}
        \end{minipage}
        \begin{minipage}{0.115\hsize}
            \centerline{\footnotesize{ADLR}}
        \end{minipage}
        \begin{minipage}{0.115\hsize}
            \centerline{\footnotesize{GTVLRR}}
        \end{minipage}
    
        \vspace{1mm}
    
        \begin{minipage}{0.115\hsize}
            \includegraphics[keepaspectratio, scale = 0.37]{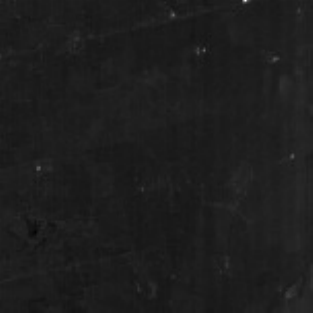}
        \end{minipage}
        \begin{minipage}{0.115\hsize}
            \includegraphics[keepaspectratio, scale = 0.37]{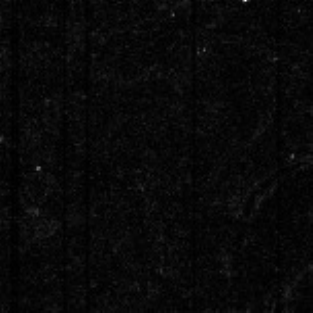}
        \end{minipage}
        \begin{minipage}{0.115\hsize}
            \includegraphics[keepaspectratio, scale = 0.37]{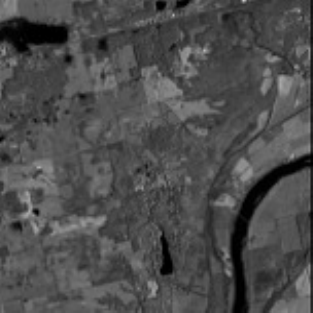}
        \end{minipage}
        \begin{minipage}{0.115\hsize}
            \includegraphics[keepaspectratio, scale = 0.37]{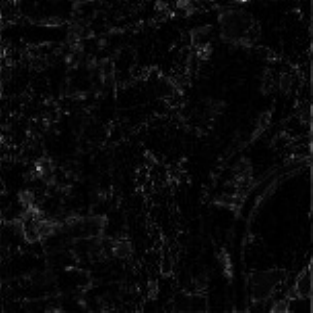}
        \end{minipage}
        \begin{minipage}{0.115\hsize}
            \includegraphics[keepaspectratio, scale = 0.37]{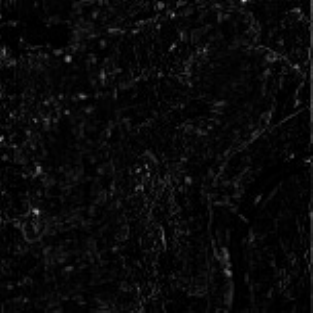}
        \end{minipage}
        \begin{minipage}{0.115\hsize}
            \includegraphics[keepaspectratio, scale = 0.37]{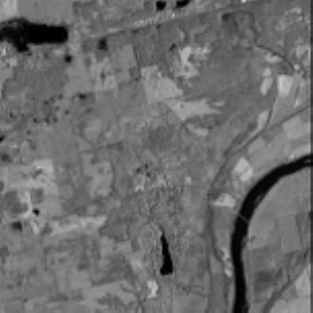}
        \end{minipage}
        \begin{minipage}{0.115\hsize}
            \includegraphics[keepaspectratio, scale = 0.37]{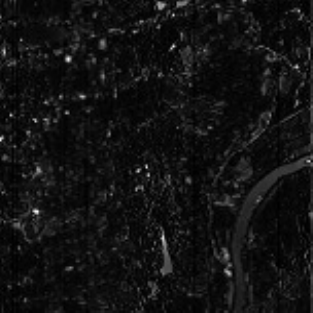}
        \end{minipage}
        \begin{minipage}{0.115\hsize}
            \includegraphics[keepaspectratio, scale = 0.37]{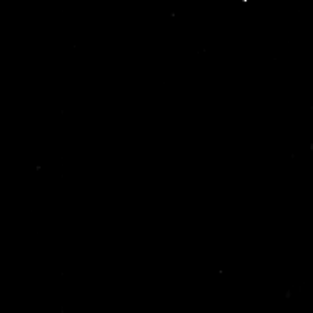}
        \end{minipage}
    
        \vspace{0.5mm}
    
        \begin{minipage}{0.115\hsize}
            \centerline{\footnotesize{LSDM-MoG}}
        \end{minipage}
        \begin{minipage}{0.115\hsize}
            \centerline{\footnotesize{PCA-TLRSR}}
        \end{minipage}
        \begin{minipage}{0.115\hsize}
            \centerline{\footnotesize{AHMID}}
        \end{minipage}
        \begin{minipage}{0.115\hsize}
            \centerline{\footnotesize{MTVLRR}}
        \end{minipage}
        \begin{minipage}{0.115\hsize}
            \centerline{\footnotesize{Ours (HTV)}}
        \end{minipage}
        \begin{minipage}{0.115\hsize}
            \centerline{\footnotesize{Ours (SSTV)}}
        \end{minipage}
        \begin{minipage}{0.115\hsize}
            \centerline{\footnotesize{Ours (HSSTV)}}
        \end{minipage}
        \begin{minipage}{0.115\hsize}
            \centerline{\footnotesize{Ours (Nuclear)}}
        \end{minipage}
    \end{minipage}

    \caption{Resulting detection maps for Hyperion in Case 1 generated by all the detectors.}
    \label{fig:DetectionMap_Case1_Hyperion}

\end{figure*}

\begin{figure*}[!t]
    \centering

    \begin{minipage}{0.03\hsize}
        \centerline{{\rotatebox{90}{\small{\shortstack{(a) Pavia Centre (Case 1)}}}}}
    \end{minipage}
    \begin{minipage}{0.95\hsize}
        \begin{minipage}{0.115\hsize}
            \includegraphics[keepaspectratio, scale = 0.37]{figs/Original/Beach4/pseudocolor-eps-converted-to.pdf}
        \end{minipage}
        \begin{minipage}{0.115\hsize}
            \includegraphics[keepaspectratio, scale = 0.37]{figs/Original/Beach4/GT-eps-converted-to.pdf}
        \end{minipage}
        \begin{minipage}{0.115\hsize}
            \includegraphics[keepaspectratio, scale = 0.37]{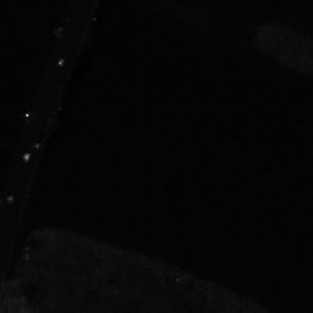}
        \end{minipage}
        \begin{minipage}{0.115\hsize}
            \includegraphics[keepaspectratio, scale = 0.37]{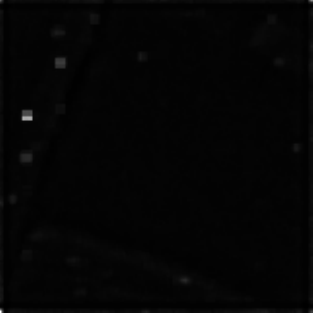}
        \end{minipage}
        \begin{minipage}{0.115\hsize}
            \includegraphics[keepaspectratio, scale = 0.37]{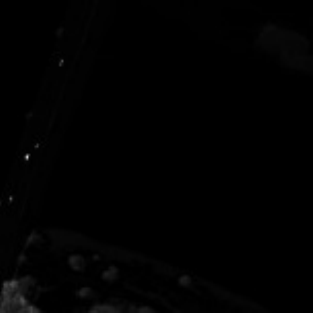}
        \end{minipage}
        \begin{minipage}{0.115\hsize}
            \includegraphics[keepaspectratio, scale = 0.37]{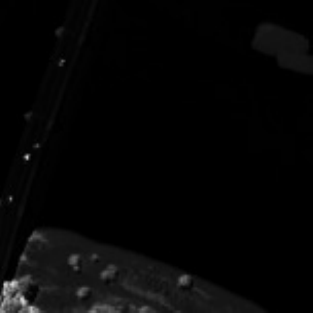}
        \end{minipage}
        \begin{minipage}{0.115\hsize}
            \includegraphics[keepaspectratio, scale = 0.37]{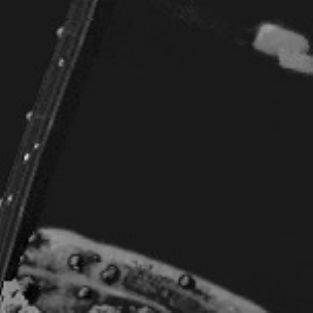}
        \end{minipage}
        \begin{minipage}{0.115\hsize}
            \includegraphics[keepaspectratio, scale = 0.37]{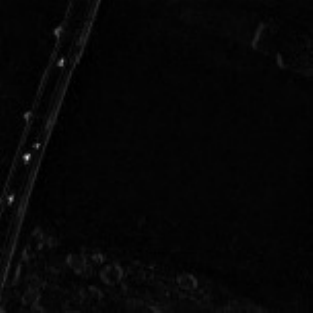}
        \end{minipage}
    
        \vspace{0.5mm}
    
        \begin{minipage}{0.115\hsize}
            \centerline{\footnotesize{Pseudocolor}}
        \end{minipage}
        \begin{minipage}{0.115\hsize}
            \centerline{\footnotesize{Ground Truth}}
        \end{minipage}
        \begin{minipage}{0.115\hsize}
            \centerline{\footnotesize{GRX}}
        \end{minipage}
        \begin{minipage}{0.115\hsize}
            \centerline{\footnotesize{2S-GLRT}}
        \end{minipage}
        \begin{minipage}{0.115\hsize}
            \centerline{\footnotesize{GAED}}
        \end{minipage}
        \begin{minipage}{0.115\hsize}
            \centerline{\footnotesize{RGAE}}
        \end{minipage}
        \begin{minipage}{0.115\hsize}
            \centerline{\footnotesize{ADLR}}
        \end{minipage}
        \begin{minipage}{0.115\hsize}
            \centerline{\footnotesize{GTVLRR}}
        \end{minipage}
    
        \vspace{1mm}
    
        \begin{minipage}{0.115\hsize}
            \includegraphics[keepaspectratio, scale = 0.37]{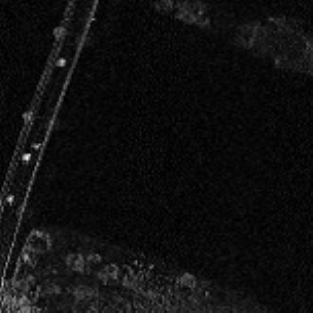}
        \end{minipage}
        \begin{minipage}{0.115\hsize}
            \includegraphics[keepaspectratio, scale = 0.37]{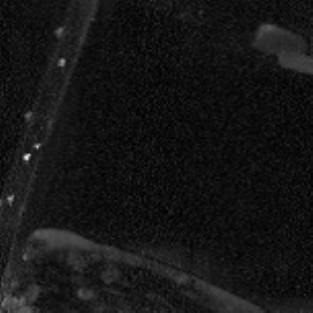}
        \end{minipage}
        \begin{minipage}{0.115\hsize}
            \includegraphics[keepaspectratio, scale = 0.37]{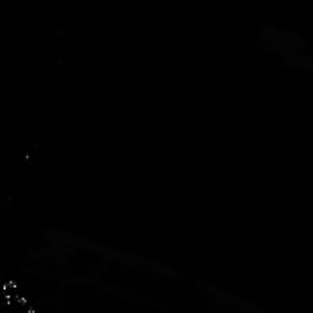}
        \end{minipage}
        \begin{minipage}{0.115\hsize}
            \includegraphics[keepaspectratio, scale = 0.37]{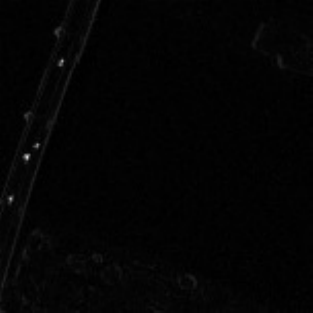}
        \end{minipage}
        \begin{minipage}{0.115\hsize}
            \includegraphics[keepaspectratio, scale = 0.37]{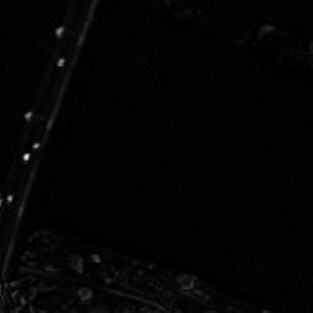}
        \end{minipage}
        \begin{minipage}{0.115\hsize}
            \includegraphics[keepaspectratio, scale = 0.37]{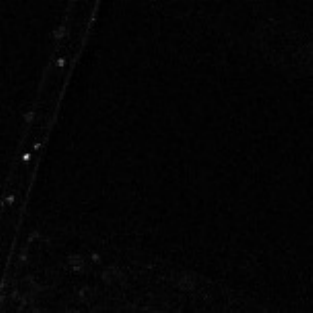}
        \end{minipage}
        \begin{minipage}{0.115\hsize}
            \includegraphics[keepaspectratio, scale = 0.37]{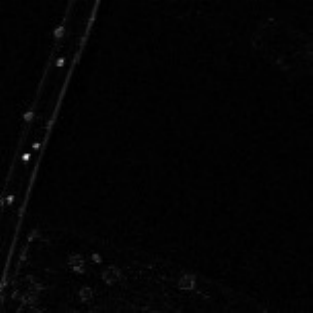}
        \end{minipage}
        \begin{minipage}{0.115\hsize}
            \includegraphics[keepaspectratio, scale = 0.37]{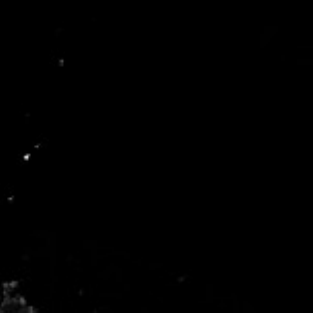}
        \end{minipage}
    
        \vspace{0.5mm}
    
        \begin{minipage}{0.115\hsize}
            \centerline{\footnotesize{LSDM-MoG}}
        \end{minipage}
        \begin{minipage}{0.115\hsize}
            \centerline{\footnotesize{PCA-TLRSR}}
        \end{minipage}
        \begin{minipage}{0.115\hsize}
            \centerline{\footnotesize{AHMID}}
        \end{minipage}
        \begin{minipage}{0.115\hsize}
            \centerline{\footnotesize{MTVLRR}}
        \end{minipage}
        \begin{minipage}{0.115\hsize}
            \centerline{\footnotesize{Ours (HTV)}}
        \end{minipage}
        \begin{minipage}{0.115\hsize}
            \centerline{\footnotesize{Ours (SSTV)}}
        \end{minipage}
        \begin{minipage}{0.115\hsize}
            \centerline{\footnotesize{Ours (HSSTV)}}
        \end{minipage}
        \begin{minipage}{0.115\hsize}
            \centerline{\footnotesize{Ours (Nuclear)}}
        \end{minipage}
    \end{minipage}

    \vspace{2mm}

    \begin{minipage}{0.03\hsize}
        \centerline{{\rotatebox{90}{\small{\shortstack{(b) Pavia Centre (Case 2)}}}}}
    \end{minipage}
    \begin{minipage}{0.95\hsize}
        \begin{minipage}{0.115\hsize}
            \includegraphics[keepaspectratio, scale = 0.37]{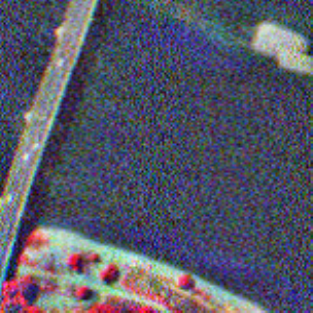}
        \end{minipage}
        \begin{minipage}{0.115\hsize}
            \includegraphics[keepaspectratio, scale = 0.37]{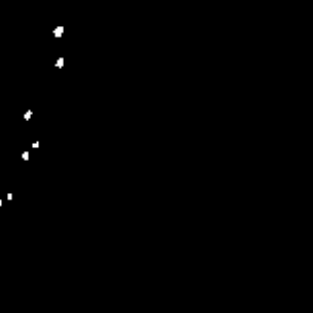}
        \end{minipage}
        \begin{minipage}{0.115\hsize}
            \includegraphics[keepaspectratio, scale = 0.37]{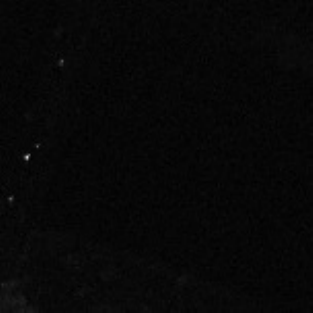}
        \end{minipage}
        \begin{minipage}{0.115\hsize}
            \includegraphics[keepaspectratio, scale = 0.37]{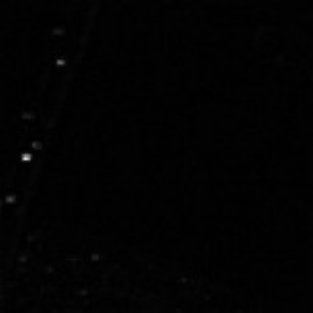}
        \end{minipage}
        \begin{minipage}{0.115\hsize}
            \includegraphics[keepaspectratio, scale = 0.37]{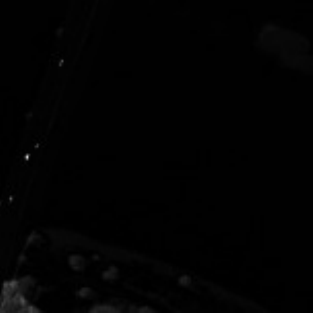}
        \end{minipage}
        \begin{minipage}{0.115\hsize}
            \includegraphics[keepaspectratio, scale = 0.37]{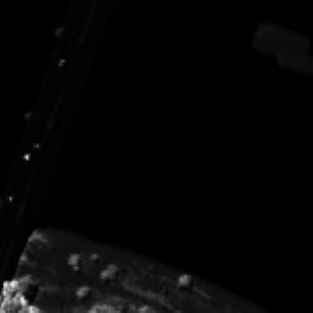}
        \end{minipage}
        \begin{minipage}{0.115\hsize}
            \includegraphics[keepaspectratio, scale = 0.37]{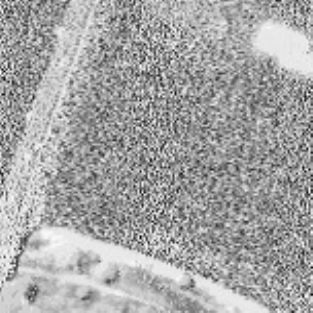}
        \end{minipage}
        \begin{minipage}{0.115\hsize}
            \includegraphics[keepaspectratio, scale = 0.37]{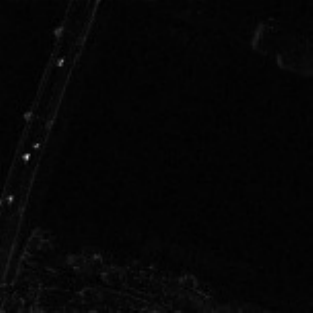}
        \end{minipage}
    
        \vspace{0.5mm}
    
        \begin{minipage}{0.115\hsize}
            \centerline{\footnotesize{Pseudocolor}}
        \end{minipage}
        \begin{minipage}{0.115\hsize}
            \centerline{\footnotesize{Ground Truth}}
        \end{minipage}
        \begin{minipage}{0.115\hsize}
            \centerline{\footnotesize{GRX}}
        \end{minipage}
        \begin{minipage}{0.115\hsize}
            \centerline{\footnotesize{2S-GLRT}}
        \end{minipage}
        \begin{minipage}{0.115\hsize}
            \centerline{\footnotesize{GAED}}
        \end{minipage}
        \begin{minipage}{0.115\hsize}
            \centerline{\footnotesize{RGAE}}
        \end{minipage}
        \begin{minipage}{0.115\hsize}
            \centerline{\footnotesize{ADLR}}
        \end{minipage}
        \begin{minipage}{0.115\hsize}
            \centerline{\footnotesize{GTVLRR}}
        \end{minipage}
    
        \vspace{1mm}
    
        \begin{minipage}{0.115\hsize}
            \includegraphics[keepaspectratio, scale = 0.37]{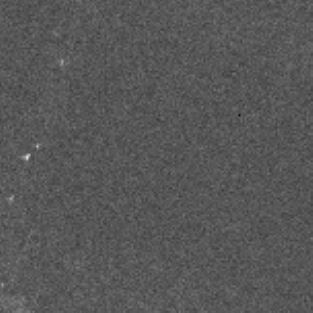}
        \end{minipage}
        \begin{minipage}{0.115\hsize}
            \includegraphics[keepaspectratio, scale = 0.37]{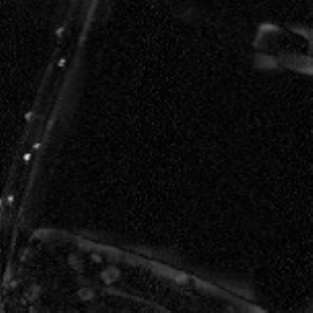}
        \end{minipage}
        \begin{minipage}{0.115\hsize}
            \includegraphics[keepaspectratio, scale = 0.37]{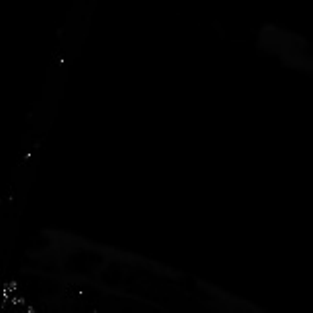}
        \end{minipage}
        \begin{minipage}{0.115\hsize}
            \includegraphics[keepaspectratio, scale = 0.37]{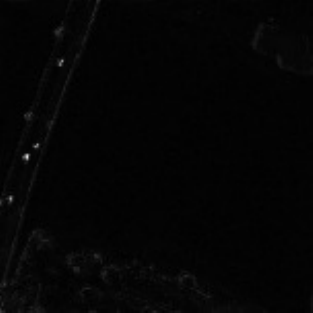}
        \end{minipage}
        \begin{minipage}{0.115\hsize}
            \includegraphics[keepaspectratio, scale = 0.37]{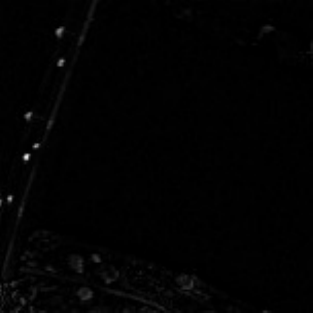}
        \end{minipage}
        \begin{minipage}{0.115\hsize}
            \includegraphics[keepaspectratio, scale = 0.37]{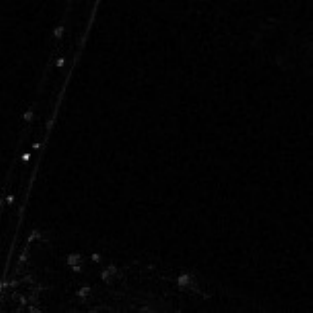}
        \end{minipage}
        \begin{minipage}{0.115\hsize}
            \includegraphics[keepaspectratio, scale = 0.37]{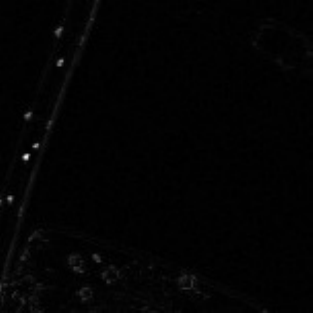}
        \end{minipage}
        \begin{minipage}{0.115\hsize}
            \includegraphics[keepaspectratio, scale = 0.37]{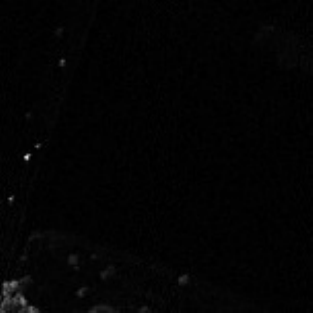}
        \end{minipage}
    
        \vspace{0.5mm}
    
        \begin{minipage}{0.115\hsize}
            \centerline{\footnotesize{LSDM-MoG}}
        \end{minipage}
        \begin{minipage}{0.115\hsize}
            \centerline{\footnotesize{PCA-TLRSR}}
        \end{minipage}
        \begin{minipage}{0.115\hsize}
            \centerline{\footnotesize{AHMID}}
        \end{minipage}
        \begin{minipage}{0.115\hsize}
            \centerline{\footnotesize{MTVLRR}}
        \end{minipage}
        \begin{minipage}{0.115\hsize}
            \centerline{\footnotesize{Ours (HTV)}}
        \end{minipage}
        \begin{minipage}{0.115\hsize}
            \centerline{\footnotesize{Ours (SSTV)}}
        \end{minipage}
        \begin{minipage}{0.115\hsize}
            \centerline{\footnotesize{Ours (HSSTV)}}
        \end{minipage}
        \begin{minipage}{0.115\hsize}
            \centerline{\footnotesize{Ours (Nuclear)}}
        \end{minipage}
    \end{minipage}

    \vspace{2mm}

    \begin{minipage}{0.03\hsize}
        \centerline{{\rotatebox{90}{\small{\shortstack{(c) Pavia Centre (Case 3)}}}}}
    \end{minipage}
    \begin{minipage}{0.95\hsize}
        \begin{minipage}{0.115\hsize}
            \includegraphics[keepaspectratio, scale = 0.37]{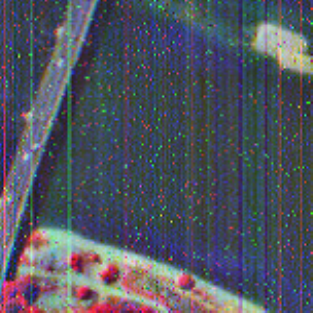}
        \end{minipage}
        \begin{minipage}{0.115\hsize}
            \includegraphics[keepaspectratio, scale = 0.37]{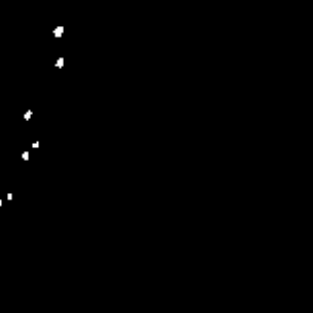}
        \end{minipage}
        \begin{minipage}{0.115\hsize}
            \includegraphics[keepaspectratio, scale = 0.37]{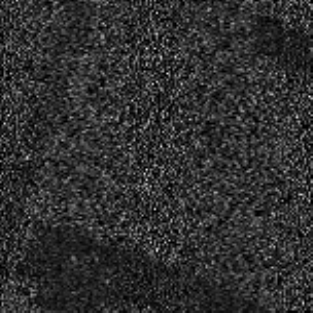}
        \end{minipage}
        \begin{minipage}{0.115\hsize}
            \includegraphics[keepaspectratio, scale = 0.37]{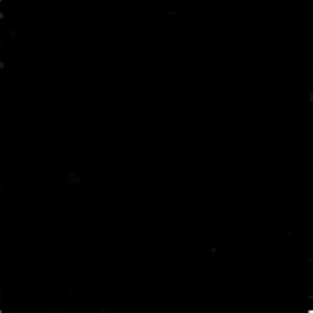}
        \end{minipage}
        \begin{minipage}{0.115\hsize}
            \includegraphics[keepaspectratio, scale = 0.37]{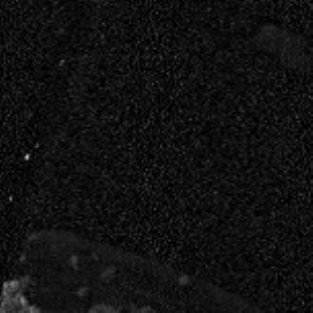}
        \end{minipage}
        \begin{minipage}{0.115\hsize}
            \includegraphics[keepaspectratio, scale = 0.37]{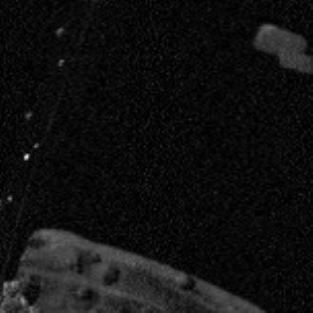}
        \end{minipage}
        \begin{minipage}{0.115\hsize}
            \includegraphics[keepaspectratio, scale = 0.37]{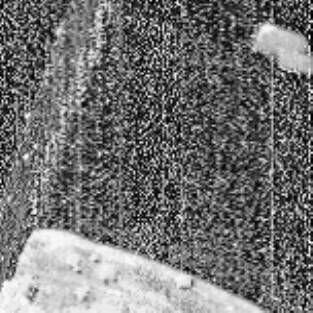}
        \end{minipage}
        \begin{minipage}{0.115\hsize}
            \includegraphics[keepaspectratio, scale = 0.37]{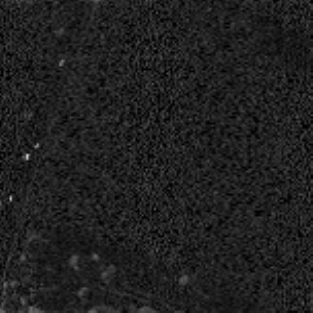}
        \end{minipage}
    
        \vspace{0.5mm}
    
        \begin{minipage}{0.115\hsize}
            \centerline{\footnotesize{Pseudocolor}}
        \end{minipage}
        \begin{minipage}{0.115\hsize}
            \centerline{\footnotesize{Ground Truth}}
        \end{minipage}
        \begin{minipage}{0.115\hsize}
            \centerline{\footnotesize{GRX}}
        \end{minipage}
        \begin{minipage}{0.115\hsize}
            \centerline{\footnotesize{2S-GLRT}}
        \end{minipage}
        \begin{minipage}{0.115\hsize}
            \centerline{\footnotesize{GAED}}
        \end{minipage}
        \begin{minipage}{0.115\hsize}
            \centerline{\footnotesize{RGAE}}
        \end{minipage}
        \begin{minipage}{0.115\hsize}
            \centerline{\footnotesize{ADLR}}
        \end{minipage}
        \begin{minipage}{0.115\hsize}
            \centerline{\footnotesize{GTVLRR}}
        \end{minipage}
    
        \vspace{1mm}
    
        \begin{minipage}{0.115\hsize}
            \includegraphics[keepaspectratio, scale = 0.37]{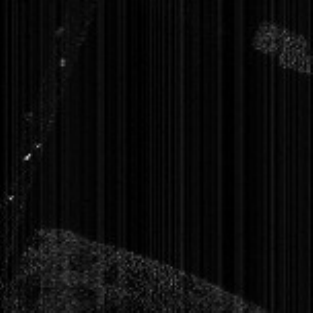}
        \end{minipage}
        \begin{minipage}{0.115\hsize}
            \includegraphics[keepaspectratio, scale = 0.37]{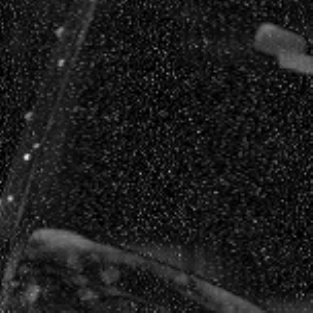}
        \end{minipage}
        \begin{minipage}{0.115\hsize}
            \includegraphics[keepaspectratio, scale = 0.37]{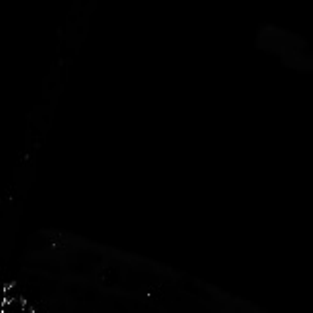}
        \end{minipage}
        \begin{minipage}{0.115\hsize}
            \includegraphics[keepaspectratio, scale = 0.37]{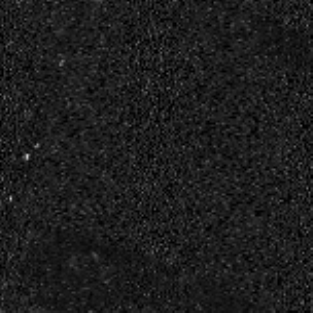}
        \end{minipage}
        \begin{minipage}{0.115\hsize}
            \includegraphics[keepaspectratio, scale = 0.37]{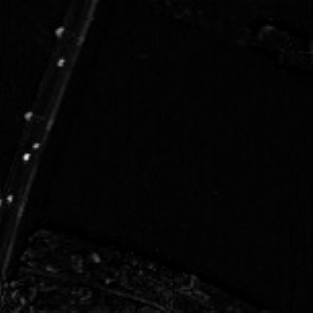}
        \end{minipage}
        \begin{minipage}{0.115\hsize}
            \includegraphics[keepaspectratio, scale = 0.37]{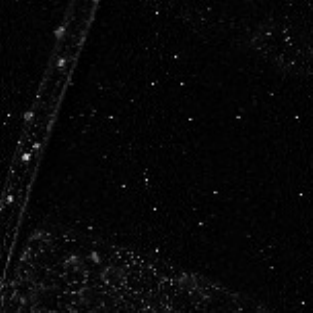}
        \end{minipage}
        \begin{minipage}{0.115\hsize}
            \includegraphics[keepaspectratio, scale = 0.37]{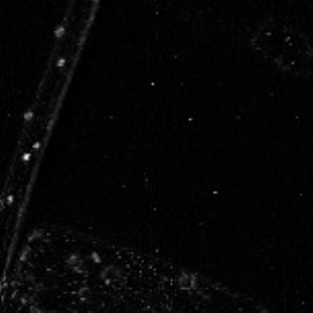}
        \end{minipage}
        \begin{minipage}{0.115\hsize}
            \includegraphics[keepaspectratio, scale = 0.37]{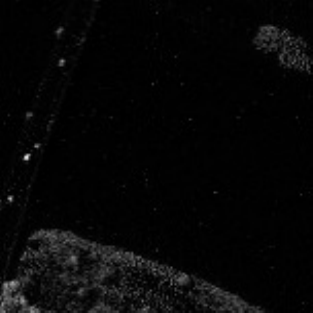}
        \end{minipage}
    
        \vspace{0.5mm}
    
        \begin{minipage}{0.115\hsize}
            \centerline{\footnotesize{LSDM-MoG}}
        \end{minipage}
        \begin{minipage}{0.115\hsize}
            \centerline{\footnotesize{PCA-TLRSR}}
        \end{minipage}
        \begin{minipage}{0.115\hsize}
            \centerline{\footnotesize{AHMID}}
        \end{minipage}
        \begin{minipage}{0.115\hsize}
            \centerline{\footnotesize{MTVLRR}}
        \end{minipage}
        \begin{minipage}{0.115\hsize}
            \centerline{\footnotesize{Ours (HTV)}}
        \end{minipage}
        \begin{minipage}{0.115\hsize}
            \centerline{\footnotesize{Ours (SSTV)}}
        \end{minipage}
        \begin{minipage}{0.115\hsize}
            \centerline{\footnotesize{Ours (HSSTV)}}
        \end{minipage}
        \begin{minipage}{0.115\hsize}
            \centerline{\footnotesize{Ours (Nuclear)}}
        \end{minipage}
    \end{minipage}

    \vspace{2mm}

    \begin{minipage}{0.03\hsize}
        \centerline{{\rotatebox{90}{\small{\shortstack{(d) Pavia Centre (Case 4)}}}}}
    \end{minipage}
    \begin{minipage}{0.95\hsize}
        \begin{minipage}{0.115\hsize}
            \includegraphics[keepaspectratio, scale = 0.37]{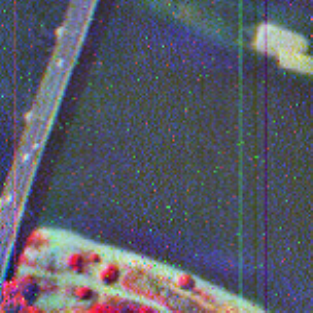}
        \end{minipage}
        \begin{minipage}{0.115\hsize}
            \includegraphics[keepaspectratio, scale = 0.37]{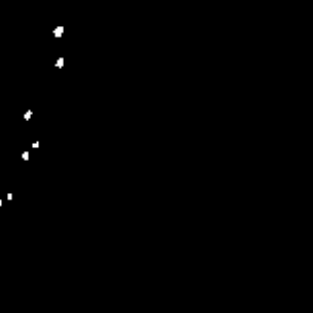}
        \end{minipage}
        \begin{minipage}{0.115\hsize}
            \includegraphics[keepaspectratio, scale = 0.37]{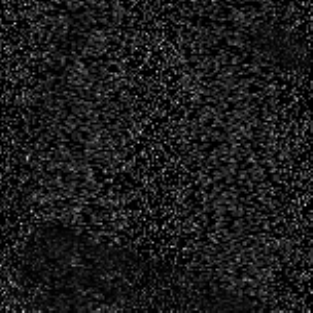}
        \end{minipage}
        \begin{minipage}{0.115\hsize}
            \includegraphics[keepaspectratio, scale = 0.37]{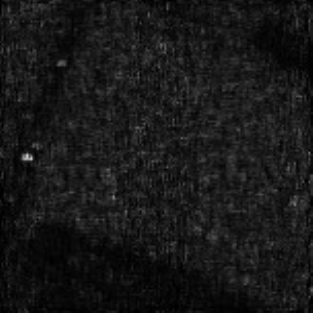}
        \end{minipage}
        \begin{minipage}{0.115\hsize}
            \includegraphics[keepaspectratio, scale = 0.37]{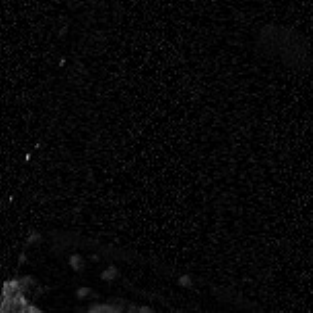}
        \end{minipage}
        \begin{minipage}{0.115\hsize}
            \includegraphics[keepaspectratio, scale = 0.37]{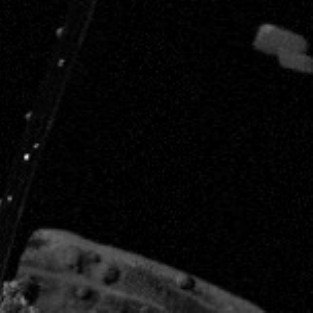}
        \end{minipage}
        \begin{minipage}{0.115\hsize}
            \includegraphics[keepaspectratio, scale = 0.37]{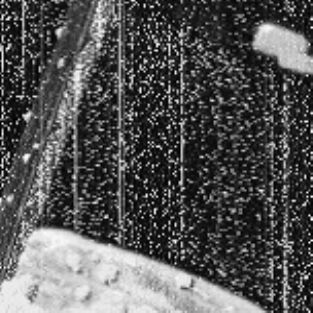}
        \end{minipage}
        \begin{minipage}{0.115\hsize}
            \includegraphics[keepaspectratio, scale = 0.37]{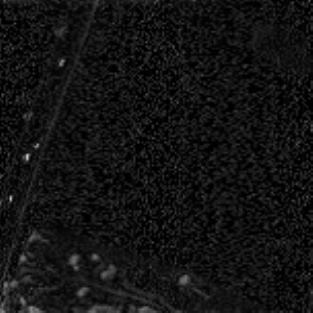}
        \end{minipage}
    
        \vspace{0.5mm}
    
        \begin{minipage}{0.115\hsize}
            \centerline{\footnotesize{Pseudocolor}}
        \end{minipage}
        \begin{minipage}{0.115\hsize}
            \centerline{\footnotesize{Ground Truth}}
        \end{minipage}
        \begin{minipage}{0.115\hsize}
            \centerline{\footnotesize{GRX}}
        \end{minipage}
        \begin{minipage}{0.115\hsize}
            \centerline{\footnotesize{2S-GLRT}}
        \end{minipage}
        \begin{minipage}{0.115\hsize}
            \centerline{\footnotesize{GAED}}
        \end{minipage}
        \begin{minipage}{0.115\hsize}
            \centerline{\footnotesize{RGAE}}
        \end{minipage}
        \begin{minipage}{0.115\hsize}
            \centerline{\footnotesize{ADLR}}
        \end{minipage}
        \begin{minipage}{0.115\hsize}
            \centerline{\footnotesize{GTVLRR}}
        \end{minipage}
    
        \vspace{1mm}
    
        \begin{minipage}{0.115\hsize}
            \includegraphics[keepaspectratio, scale = 0.37]{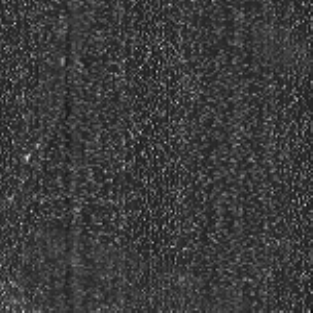}
        \end{minipage}
        \begin{minipage}{0.115\hsize}
            \includegraphics[keepaspectratio, scale = 0.37]{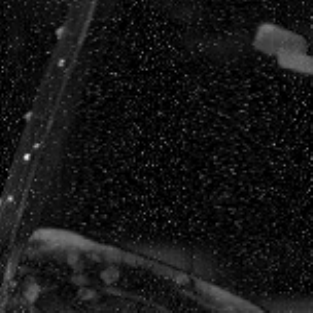}
        \end{minipage}
        \begin{minipage}{0.115\hsize}
            \includegraphics[keepaspectratio, scale = 0.37]{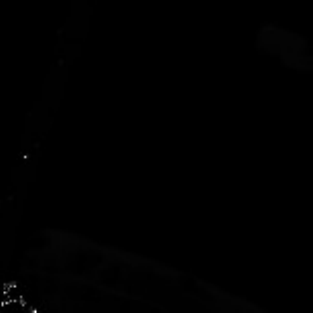}
        \end{minipage}
        \begin{minipage}{0.115\hsize}
            \includegraphics[keepaspectratio, scale = 0.37]{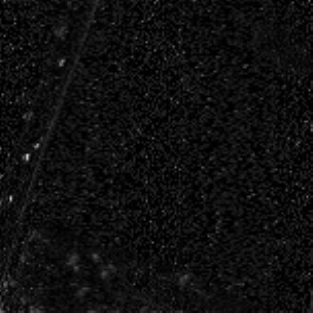}
        \end{minipage}
        \begin{minipage}{0.115\hsize}
            \includegraphics[keepaspectratio, scale = 0.37]{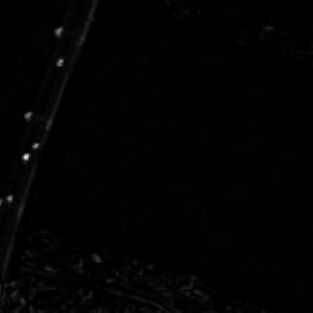}
        \end{minipage}
        \begin{minipage}{0.115\hsize}
            \includegraphics[keepaspectratio, scale = 0.37]{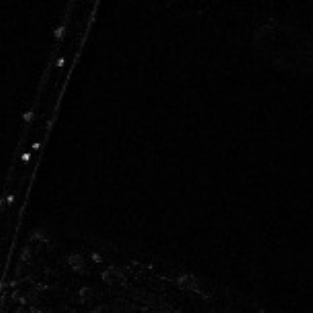}
        \end{minipage}
        \begin{minipage}{0.115\hsize}
            \includegraphics[keepaspectratio, scale = 0.37]{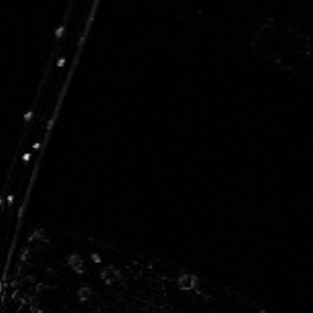}
        \end{minipage}
        \begin{minipage}{0.115\hsize}
            \includegraphics[keepaspectratio, scale = 0.37]{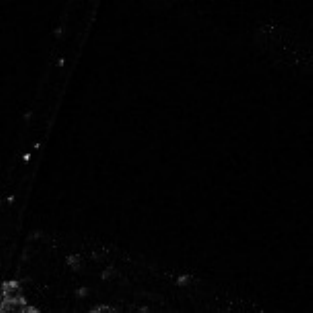}
        \end{minipage}
    
        \vspace{0.5mm}
    
        \begin{minipage}{0.115\hsize}
            \centerline{\footnotesize{LSDM-MoG}}
        \end{minipage}
        \begin{minipage}{0.115\hsize}
            \centerline{\footnotesize{PCA-TLRSR}}
        \end{minipage}
        \begin{minipage}{0.115\hsize}
            \centerline{\footnotesize{AHMID}}
        \end{minipage}
        \begin{minipage}{0.115\hsize}
            \centerline{\footnotesize{MTVLRR}}
        \end{minipage}
        \begin{minipage}{0.115\hsize}
            \centerline{\footnotesize{Ours (HTV)}}
        \end{minipage}
        \begin{minipage}{0.115\hsize}
            \centerline{\footnotesize{Ours (SSTV)}}
        \end{minipage}
        \begin{minipage}{0.115\hsize}
            \centerline{\footnotesize{Ours (HSSTV)}}
        \end{minipage}
        \begin{minipage}{0.115\hsize}
            \centerline{\footnotesize{Ours (Nuclear)}}
        \end{minipage}
    \end{minipage}

    \vspace{2mm}

    \begin{minipage}{0.03\hsize}
        \centerline{{\rotatebox{90}{\small{\shortstack{(e) Pavia Centre (Case 5)}}}}}
    \end{minipage}
    \begin{minipage}{0.95\hsize}
        \begin{minipage}{0.115\hsize}
            \includegraphics[keepaspectratio, scale = 0.37]{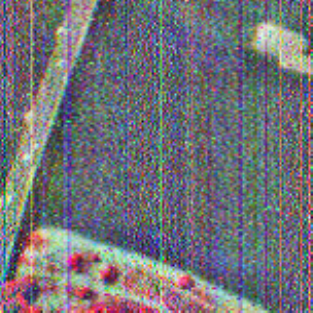}
        \end{minipage}
        \begin{minipage}{0.115\hsize}
            \includegraphics[keepaspectratio, scale = 0.37]{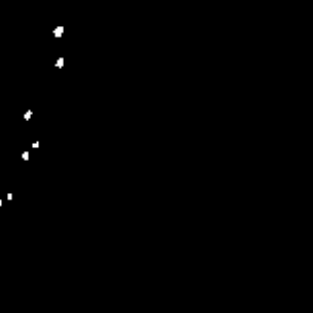}
        \end{minipage}
        \begin{minipage}{0.115\hsize}
            \includegraphics[keepaspectratio, scale = 0.37]{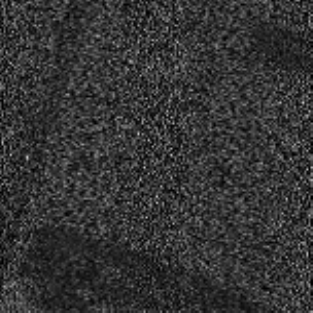}
        \end{minipage}
        \begin{minipage}{0.115\hsize}
            \includegraphics[keepaspectratio, scale = 0.37]{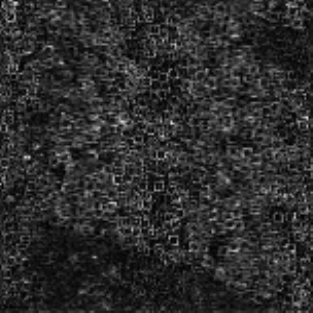}
        \end{minipage}
        \begin{minipage}{0.115\hsize}
            \includegraphics[keepaspectratio, scale = 0.37]{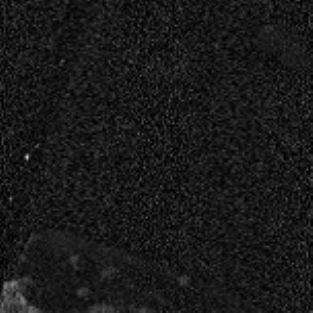}
        \end{minipage}
        \begin{minipage}{0.115\hsize}
            \includegraphics[keepaspectratio, scale = 0.37]{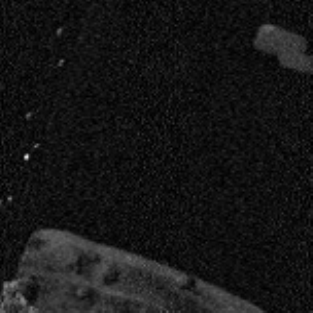}
        \end{minipage}
        \begin{minipage}{0.115\hsize}
            \includegraphics[keepaspectratio, scale = 0.37]{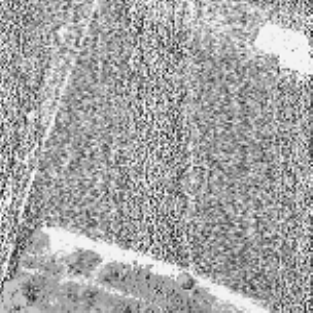}
        \end{minipage}
        \begin{minipage}{0.115\hsize}
            \includegraphics[keepaspectratio, scale = 0.37]{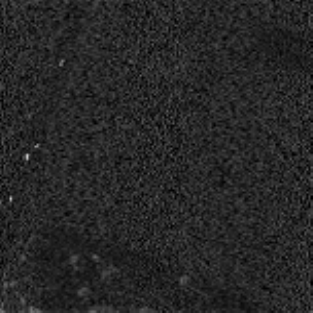}
        \end{minipage}
    
        \vspace{0.5mm}
    
        \begin{minipage}{0.115\hsize}
            \centerline{\footnotesize{Pseudocolor}}
        \end{minipage}
        \begin{minipage}{0.115\hsize}
            \centerline{\footnotesize{Ground Truth}}
        \end{minipage}
        \begin{minipage}{0.115\hsize}
            \centerline{\footnotesize{GRX}}
        \end{minipage}
        \begin{minipage}{0.115\hsize}
            \centerline{\footnotesize{2S-GLRT}}
        \end{minipage}
        \begin{minipage}{0.115\hsize}
            \centerline{\footnotesize{GAED}}
        \end{minipage}
        \begin{minipage}{0.115\hsize}
            \centerline{\footnotesize{RGAE}}
        \end{minipage}
        \begin{minipage}{0.115\hsize}
            \centerline{\footnotesize{ADLR}}
        \end{minipage}
        \begin{minipage}{0.115\hsize}
            \centerline{\footnotesize{GTVLRR}}
        \end{minipage}
    
        \vspace{1mm}
    
        \begin{minipage}{0.115\hsize}
            \includegraphics[keepaspectratio, scale = 0.37]{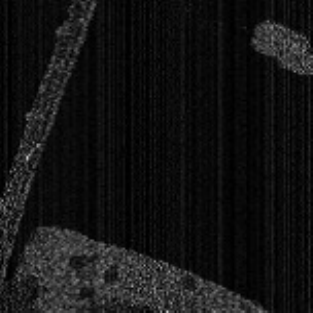}
        \end{minipage}
        \begin{minipage}{0.115\hsize}
            \includegraphics[keepaspectratio, scale = 0.37]{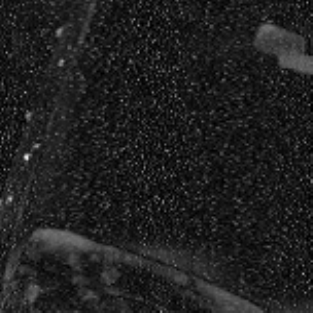}
        \end{minipage}
        \begin{minipage}{0.115\hsize}
            \includegraphics[keepaspectratio, scale = 0.37]{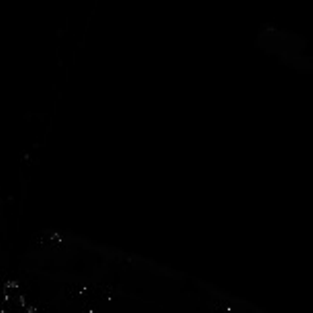}
        \end{minipage}
        \begin{minipage}{0.115\hsize}
            \includegraphics[keepaspectratio, scale = 0.37]{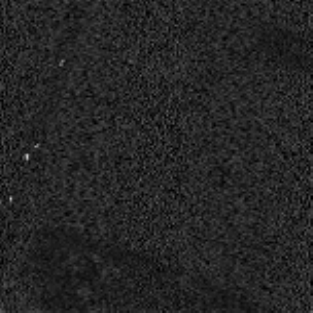}
        \end{minipage}
        \begin{minipage}{0.115\hsize}
            \includegraphics[keepaspectratio, scale = 0.37]{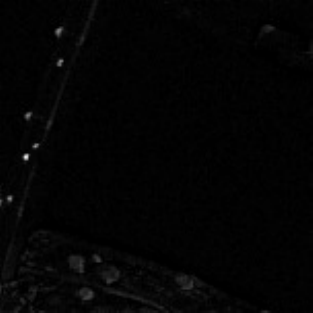}
        \end{minipage}
        \begin{minipage}{0.115\hsize}
            \includegraphics[keepaspectratio, scale = 0.37]{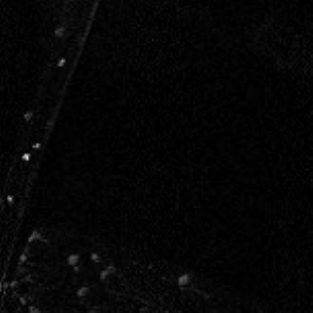}
        \end{minipage}
        \begin{minipage}{0.115\hsize}
            \includegraphics[keepaspectratio, scale = 0.37]{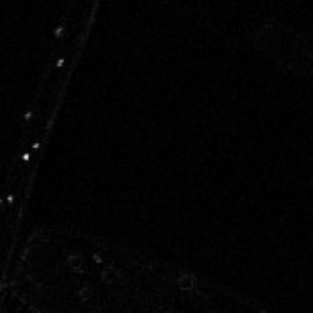}
        \end{minipage}
        \begin{minipage}{0.115\hsize}
            \includegraphics[keepaspectratio, scale = 0.37]{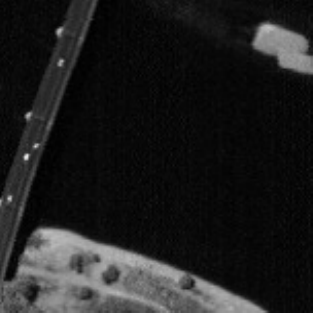}
        \end{minipage}
    
        \vspace{0.5mm}
    
        \begin{minipage}{0.115\hsize}
            \centerline{\footnotesize{LSDM-MoG}}
        \end{minipage}
        \begin{minipage}{0.115\hsize}
            \centerline{\footnotesize{PCA-TLRSR}}
        \end{minipage}
        \begin{minipage}{0.115\hsize}
            \centerline{\footnotesize{AHMID}}
        \end{minipage}
        \begin{minipage}{0.115\hsize}
            \centerline{\footnotesize{MTVLRR}}
        \end{minipage}
        \begin{minipage}{0.115\hsize}
            \centerline{\footnotesize{Ours (HTV)}}
        \end{minipage}
        \begin{minipage}{0.115\hsize}
            \centerline{\footnotesize{Ours (SSTV)}}
        \end{minipage}
        \begin{minipage}{0.115\hsize}
            \centerline{\footnotesize{Ours (HSSTV)}}
        \end{minipage}
        \begin{minipage}{0.115\hsize}
            \centerline{\footnotesize{Ours (Nuclear)}}
        \end{minipage}
    \end{minipage}

    \caption{Resulting detection maps for Pavia Centre in Case 1, 2, 3, 4, and 5 generated by all the detectors.}
    \label{fig:DetectionMap_pavia}

\end{figure*}

\begin{figure*}[!t]
    \centering
    \includegraphics[keepaspectratio, scale = 0.16]{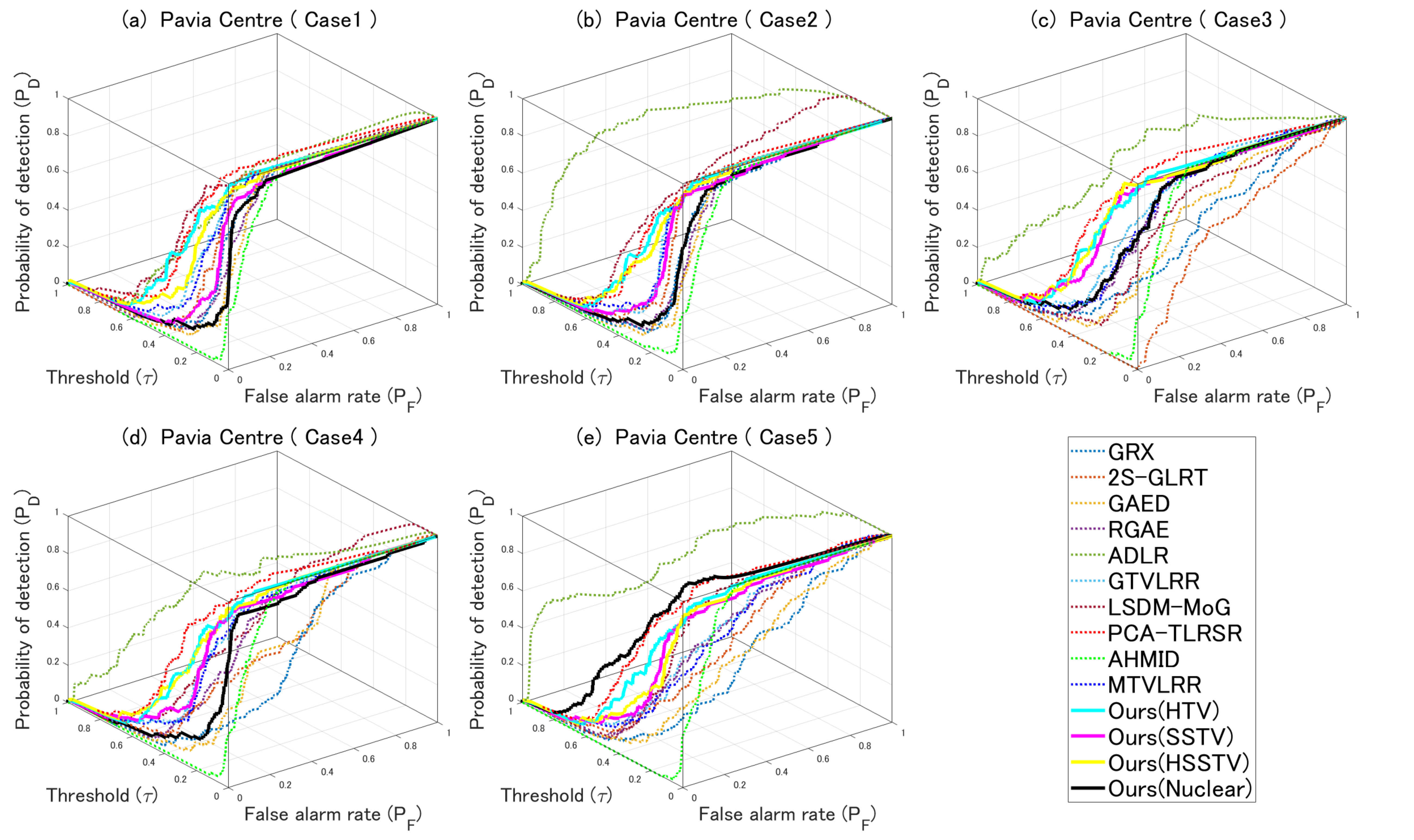}
    \caption{3D ROC curves of all the detectors for Pavia Centre in all cases.}
    \label{fig:semiROCs}
\end{figure*}

Table~\ref{tab:AUCs_Case1} summarizes the three types of AUC values for all detectors across each dataset in Case 1.
The best and second-best results are highlighted in bold and underlined, respectively. 
The detection performance of the proposed method is comparable to that of the existing state-of-the-art methods, even though it does not use a background dictionary and only uses HTV, SSTV, HSSTV, or the nuclear norm to characterize the background part.
In particular, the proposed method using HTV achieved the best $\mathrm{AUC}_{(P_D, P_F)}$ values in most of the datasets, along with sufficiently high $\mathrm{AUC}_{(P_D, \tau)}$ and low $\mathrm{AUC}_{(P_F, \tau)}$ values. 
This suggests that evaluating the spatial continuity of the background part only by the difference between neighboring pixels improves the detection performance. 
Regarding existing methods, 2S-GLRT achieved the best $\mathrm{AUC}_{(P_F, \tau)}$ values in most datasets, but its $\mathrm{AUC}_{(P_D, \tau)}$ values remained low, indicating a limited ability to detect anomalies.
In contrast, ADLR and LSDM-MoG achieved the best $\mathrm{AUC}_{(P_D, \tau)}$ values in most datasets, but exhibited high $\mathrm{AUC}_{(P_F, \tau)}$ values, reflecting insufficient background suppression.

Figs.~\ref{fig:DetectionMap_Case1}, ~\ref{fig:DetectionMap_Case1_Hyperion}, and~\ref{fig:DetectionMap_pavia}(a) show the detection maps of all detectors for each dataset in Case 1.
From these figures, 2S-GLRT failed to detect several anomalies, and those detected were not clearly highlighted.
GAED and RGAE also missed some anomalies, however, the anomalies they detected tended to be emphasized.
ADLR generated clear detection maps in Los Angels I, but failed to suppress background in the other datasets. 
GTVLRR, LSDM-MoG, PCA-TLRSR, AHMID, and MTVLRR detected almost all anomalies, but failed to distinguish between background and anomalies in certain regions.
In contrast, the proposed method detected almost all anomalies while suppressing the background in all datasets. 
In particular, the method using HTV succeeded in generating clear detection maps.

\subsubsection{Case 2 (Gaussian noise)}

\begin{table*}[!t]
    \centering
    \caption{$\mathrm{AUC}_{(P_D, P_F)}$, $\mathrm{AUC}_{(P_D, \tau)}$, And $\mathrm{AUC}_{(P_F, \tau)}$ Values of All The Detectors For Each Dataset In Case 2. \\ (The Best And Second-Best Values Are Highlighted in Bold And Underlined, Respectively.)}
    \label{tab:AUCs_Case2}
    \scalebox{0.75}{
    \begin{tabular}{cccccccccccccccc}
        \toprule
        Datasets & Metrics & \multicolumn{14}{c}{Methods} \\ 
        \cmidrule(lr){1-1} \cmidrule(lr){2-2} \cmidrule(lr){3-16}
        & & GRX & 2S-GLRT & GAED & RGAE & ADLR & GTVLRR & LSDM-MoG & PCA-TLRSR & AHMID & MTVLRR & Ours & Ours & Ours & Ours \\  
        & & \cite{GRX_1990} & \cite{2SGLRT_2022} & \cite{GAED_2022} & \cite{RGAE_2022} & \cite{ADLR_2018} &\cite{GTVLRR_2020} & \cite{LSDMMog_2021} & \cite{PCA-TLRSR_2023} & \cite{AHMID_2023} & \cite{MTVLRR_2024} & (HTV) & (SSTV) & (HSSTV) & (Nuclear) \\

        \midrule
        \multirow{3}{*}{\shortstack{Pavia \\ Centre}}
         & $\mathrm{AUC}_{(P_D, P_F)}$  & 0.9188 & \ValBest{0.9935} & 0.9430 & 0.9157 & 0.8579 & \ValFourth{0.9799} & 0.7996 & 0.9619 & 0.9171 & \ValFifth{0.9797} & \ValSecond{0.9888} & 0.9645 & \ValThird{0.9843} & 0.9367 \\ 
         & $\mathrm{AUC}_{(P_D, \tau)}$  & 0.1185 & 0.1676 & 0.0915 & 0.1352 & \ValBest{0.8607} & 0.2038 & \ValSecond{0.4407} & \ValThird{0.3159} & 0.0353 & 0.2156 & \ValFourth{0.3065} & 0.1963 & \ValFifth{0.2932} & 0.1266 \\ 
         & $\mathrm{AUC}_{(P_F, \tau)}$  & 0.0314 & \ValThird{0.0092} & \ValSecond{0.0080} & 0.0245 & 0.6534 & 0.0236 & 0.3458 & 0.0577 & \ValBest{0.0030} & 0.0192 & \ValFourth{0.0107} & 0.0134 & \ValFifth{0.0127} & 0.0151 \\ 
        \cmidrule(lr){1-16}
        
        \multirow{3}{*}{\shortstack{Texas \\ Coast}} 
         & $\mathrm{AUC}_{(P_D, P_F)}$  & 0.9016 & 0.9850 & 0.9811 & 0.9821 & 0.9799 & \ValFourth{0.9879} & 0.8252 & \ValFifth{0.9872} & 0.9791 & 0.9714 & \ValBest{0.9972} & 0.9836 & \ValSecond{0.9905} & \ValThird{0.9884} \\ 
         & $\mathrm{AUC}_{(P_D, \tau)}$  & 0.5553 & 0.1933 & 0.3680 & 0.3723 & \ValBest{0.9123} & \ValFourth{0.5936} & \ValSecond{0.8087} & \ValFifth{0.5571} & 0.3670 & 0.4845 & 0.4756 & \ValThird{0.6235} & 0.4470 & 0.5452 \\ 
         & $\mathrm{AUC}_{(P_F, \tau)}$  & 0.2689 & \ValBest{0.0140} & \ValThird{0.0186} & \ValSecond{0.0184} & 0.3941 & 0.0763 & 0.7222 & 0.1104 & 0.0974 & 0.0644 & \ValFourth{0.0191} & 0.2123 & \ValFifth{0.0306} & 0.0540 \\ 
        \cmidrule(lr){1-16}
        
        \multirow{3}{*}{\shortstack{Gainesville}}
         & $\mathrm{AUC}_{(P_D, P_F)}$  & 0.7806 & 0.9472 & \ValFifth{0.9616} & 0.8340 & 0.9551 & 0.9312 & 0.7458 & \ValSecond{0.9887} & \ValFourth{0.9690} & 0.8461 & \ValBest{0.9951} & 0.9446 & \ValThird{0.9768} & 0.9535 \\ 
         & $\mathrm{AUC}_{(P_D, \tau)}$  & \ValFifth{0.4254} & 0.2360 & 0.1277 & 0.0970 & \ValBest{0.9596} & 0.2677 & \ValSecond{0.7888} & \ValFourth{0.4282} & 0.2077 & 0.2261 & \ValThird{0.4290} & 0.3224 & 0.3770 & 0.1396 \\ 
         & $\mathrm{AUC}_{(P_F, \tau)}$  & 0.3227 & \ValFifth{0.0416} & \ValBest{0.0223} & \ValSecond{0.0368} & 0.5739 & 0.1259 & 0.7280 & 0.0930 & 0.0447 & 0.1048 & \ValBest{0.0223} & 0.0704 & 0.0506 & \ValFourth{0.0380} \\ 
        \cmidrule(lr){1-16}
        
        \multirow{3}{*}{\shortstack{Los \\ Angeles I}}
         & $\mathrm{AUC}_{(P_D, P_F)}$  & 0.7863 & 0.9712 & 0.9911 & 0.9922 & 0.9916 & \ValBest{0.9967} & 0.5614 & 0.9712 & \ValSecond{0.9965} & 0.9418 & \ValFifth{0.9932} & 0.9711 & \ValFourth{0.9935} & \ValThird{0.9962} \\ 
         & $\mathrm{AUC}_{(P_D, \tau)}$  & 0.0437 & 0.0950 & 0.0391 & 0.0412 & \ValBest{0.7337} & \ValThird{0.1992} & \ValSecond{0.5001} & 0.1212 & 0.0656 & 0.0633 & \ValFifth{0.1267} & 0.0549 & 0.1241 & \ValFourth{0.1758} \\ 
         & $\mathrm{AUC}_{(P_F, \tau)}$  & 0.0222 & 0.0112 & \ValBest{0.0015} & \ValSecond{0.0016} & 0.3149 & 0.0268 & 0.4905 & 0.0323 & \ValThird{0.0050} & 0.0091 & 0.0097 & \ValFourth{0.0086} & \ValFifth{0.0091} & 0.0328 \\ 
        \cmidrule(lr){1-16}
        
        \multirow{3}{*}{\shortstack{Los \\ Angeles I\hspace{-1.2pt}I}}
         & $\mathrm{AUC}_{(P_D, P_F)}$  & 0.7111 & 0.9406 & 0.9388 & 0.9545 & 0.9484 & 0.9502 & 0.5771 & \ValThird{0.9714} & \ValFifth{0.9644} & 0.8147 & \ValBest{0.9845} & \ValFourth{0.9694} & \ValSecond{0.9818} & 0.9622 \\ 
         & $\mathrm{AUC}_{(P_D, \tau)}$  & 0.3253 & 0.2064 & 0.2412 & 0.2507 & \ValBest{0.8932} & 0.2788 & \ValSecond{0.7279} & \ValThird{0.3954} & 0.0527 & 0.2070 & \ValFourth{0.3438} & 0.3098 & 0.3259 & \ValFifth{0.3378} \\ 
         & $\mathrm{AUC}_{(P_F, \tau)}$  & 0.2593 & 0.0306 & \ValThird{0.0209} & \ValSecond{0.0196} & 0.4454 & 0.0867 & 0.7118 & 0.1227 & \ValBest{0.0030} & 0.1338 & \ValFourth{0.0210} & 0.0297 & \ValFifth{0.0226} & 0.0459 \\ 
        \cmidrule(lr){1-16} 
        
        \multirow{3}{*}{\shortstack{San \\ Diego}}
         & $\mathrm{AUC}_{(P_D, P_F)}$  & 0.8461 & 0.9529 & \ValThird{0.9896} & \ValSecond{0.9902} & 0.9791 & \ValBest{0.9916} & 0.7374 & 0.9821 & 0.9335 & \ValFourth{0.9865} & \ValFifth{0.9844} & 0.9343 & 0.9484 & 0.9636 \\ 
         & $\mathrm{AUC}_{(P_D, \tau)}$  & 0.2227 & 0.1468 & 0.1920 & 0.1912 & \ValBest{0.9075} & \ValSecond{0.4051} & \ValThird{0.4049} & \ValFifth{0.3312} & 0.2141 & \ValFourth{0.3728} & 0.3243 & 0.1392 & 0.2424 & 0.2425 \\ 
         & $\mathrm{AUC}_{(P_F, \tau)}$  & 0.1299 & \ValThird{0.0152} & \ValBest{0.0119} & \ValSecond{0.0132} & 0.5123 & 0.0680 & 0.2965 & 0.0495 & 0.0283 & 0.0307 & \ValFifth{0.0234} & \ValFourth{0.0169} & 0.0302 & 0.0242 \\ 

        \cmidrule(lr){1-16} 

        \multirow{3}{*}{\shortstack{Hyperion}}
         & $\mathrm{AUC}_{(P_D, P_F)}$  & \ValFifth{0.9919} & 0.9887 & 0.9786 & 0.9408 & 0.8674 & \ValFourth{0.9940} & 0.8526 & \ValBest{0.9994} & 0.9843 & \ValThird{0.9976} & \ValSecond{0.9979} & 0.9817 & 0.9817 & 0.9832 \\ 
         & $\mathrm{AUC}_{(P_D, \tau)}$  & 0.2376 & 0.1474 & 0.2243 & 0.2395 & \ValBest{0.7053} & 0.4990 & \ValFifth{0.5368} & 0.3659 & 0.3569 & 0.4303 & 0.3183 & \ValSecond{0.6923} & \ValThird{0.6918} & \ValFourth{0.6114} \\ 
         & $\mathrm{AUC}_{(P_F, \tau)}$  & 0.0531 & \ValSecond{0.0143} & \ValThird{0.0187} & \ValFifth{0.0448} & 0.3762 & 0.0678 & 0.4002 & 0.0513 & 0.1593 & \ValFourth{0.0391} & \ValBest{0.0077} & 0.4317 & 0.4312 & 0.3384 \\ 

        \bottomrule
    \end{tabular}
    }
    \\
\end{table*}

Table~\ref{tab:AUCs_Case2} summarizes the three types of AUC values for all detectors across each dataset in Case 2.
These results indicate that the detection performances of GRX and LSDM-MoG degraded across all datasets relative to Case 1. 
While ADLR maintained stable detection performance, it failed in background suppression.
This is because the unmixing process, used as a preprocessing step for noise reduction, tends to assimilate subtle spectral signatures of anomalies into the background, leading to a loss of discriminability.
In contrast, the other existing methods (including, interestingly, those designed without explicit consideration of noise) exhibited minimal changes in their performance.
Most of these methods are designed to explicitly exploit spatial information, which likely mitigates the impact of Gaussian noise.
The proposed method achieved overall superior three types of AUC values because the second constraint in Prob.~\eqref{eq:proposed-optimization-problem} allows estimating the anomaly and background parts simultaneously while eliminating Gaussian noise. 
Among the variants, the proposed method using HTV achieved the best performance in almost all datasets.

Figs.~\ref{fig:DetectionMap_pavia}(b) and~\ref{fig:semiROCs}(b) show the detection maps and 3D-ROC curves for Pavia Centre in Case 2 generated by all the detectors, respectively.
The detection maps generated by ADLR and LSDM-MoG are noisy and the probability of detection for these methods is low.
In contrast, the detection maps generated by the other methods are similar to those in Case 1, and their 3D-ROC curves remain almost unchanged, indicating a certain degree of robustness to Gaussian noise.

\subsubsection{Case 3 (Non-Gaussian noise)}

\begin{table*}[!t]
    \centering
    \caption{$\mathrm{AUC}_{(P_D, P_F)}$, $\mathrm{AUC}_{(P_D, \tau)}$, And $\mathrm{AUC}_{(P_F, \tau)}$ Values of All The Detectors For Each Dataset In Case 3. \\ (The Best And Second-Best Values Are Highlighted in Bold And Underlined, Respectively.)}
    \label{tab:AUCs_Case3}
    \scalebox{0.75}{
    \begin{tabular}{cccccccccccccccc}
        \toprule
        Datasets & Metrics & \multicolumn{14}{c}{Methods} \\ 
        \cmidrule(lr){1-1} \cmidrule(lr){2-2} \cmidrule(lr){3-16}
        & & GRX & 2S-GLRT & GAED & RGAE & ADLR & GTVLRR & LSDM-MoG & PCA-TLRSR & AHMID & MTVLRR & Ours & Ours & Ours & Ours \\  
        & & \cite{GRX_1990} & \cite{2SGLRT_2022} & \cite{GAED_2022} & \cite{RGAE_2022} & \cite{ADLR_2018} &\cite{GTVLRR_2020} & \cite{LSDMMog_2021} & \cite{PCA-TLRSR_2023} & \cite{AHMID_2023} & \cite{MTVLRR_2024} & (HTV) & (SSTV) & (HSSTV) & (Nuclear) \\

        \midrule
        \multirow{3}{*}{\shortstack{Pavia \\ Centre}}
         & $\mathrm{AUC}_{(P_D, P_F)}$  & 0.5739 & 0.6411 & 0.7591 & 0.8756 & 0.7770 & 0.8543 & 0.8267 & \ValFourth{0.9545} & 0.9162 & 0.8442 & \ValBest{0.9910} & \ValThird{0.9675} & \ValSecond{0.9799} & \ValFifth{0.9282} \\ 
         & $\mathrm{AUC}_{(P_D, \tau)}$  & 0.2262 & 0.0020 & 0.1575 & 0.2317 & \ValBest{0.7075} & 0.2798 & 0.1586 & \ValSecond{0.4040} & 0.0155 & 0.2578 & \ValFourth{0.3402} & \ValFifth{0.3242} & \ValThird{0.3488} & 0.2255 \\ 
         & $\mathrm{AUC}_{(P_F, \tau)}$  & 0.1793 & \ValBest{0.0019} & 0.0684 & 0.0657 & 0.4419 & 0.1240 & 0.0447 & 0.1089 & \ValSecond{0.0028} & 0.1146 & \ValThird{0.0230} & 0.0304 & \ValFourth{0.0233} & \ValFifth{0.0283} \\ 
        \cmidrule(lr){1-16}
        
        \multirow{3}{*}{\shortstack{Texas \\ Coast}} 
         & $\mathrm{AUC}_{(P_D, P_F)}$  & 0.5662 & 0.8021 & 0.9221 & 0.9364 & \ValFifth{0.9841} & 0.9108 & 0.8269 & 0.9686 & 0.9711 & 0.8528 & \ValBest{0.9978} & \ValFourth{0.9852} & \ValSecond{0.9933} & \ValThird{0.9861} \\ 
         & $\mathrm{AUC}_{(P_D, \tau)}$  & 0.2984 & 0.0295 & 0.4450 & 0.4446 & \ValBest{0.8624} & \ValSecond{0.6491} & 0.4418 & 0.5593 & 0.2124 & 0.2743 & \ValFifth{0.5599} & 0.4026 & \ValFourth{0.5777} & \ValThird{0.5941} \\ 
         & $\mathrm{AUC}_{(P_F, \tau)}$  & 0.2757 & \ValBest{0.0102} & 0.1316 & 0.1182 & 0.2813 & 0.3505 & 0.2516 & 0.1435 & \ValFifth{0.0540} & 0.1698 & \ValThird{0.0504} & \ValFourth{0.0511} & \ValSecond{0.0452} & 0.1057 \\ 
        \cmidrule(lr){1-16}
        
        \multirow{3}{*}{\shortstack{Gainesville}}
         & $\mathrm{AUC}_{(P_D, P_F)}$  & 0.4659 & 0.7775 & 0.6533 & 0.6162 & 0.9364 & 0.5211 & 0.6450 & \ValThird{0.9510} & \ValFifth{0.9472} & 0.4883 & \ValBest{0.9937} & \ValFourth{0.9501} & \ValSecond{0.9781} & 0.9313 \\ 
         & $\mathrm{AUC}_{(P_D, \tau)}$  & 0.2496 & 0.0000 & 0.2400 & 0.1677 & \ValThird{0.4347} & \ValBest{0.4971} & 0.2979 & \ValFourth{0.4164} & 0.1941 & 0.2289 & \ValSecond{0.4877} & 0.2535 & \ValFifth{0.3540} & 0.2626 \\ 
         & $\mathrm{AUC}_{(P_F, \tau)}$  & 0.2610 & \ValBest{0.0003} & 0.1741 & 0.1281 & 0.2013 & 0.4773 & 0.2353 & 0.1272 & \ValFourth{0.0515} & 0.2299 & \ValFifth{0.0539} & \ValSecond{0.0474} & \ValSecond{0.0474} & 0.0563 \\ 
        \cmidrule(lr){1-16}
        
        \multirow{3}{*}{\shortstack{Los \\ Angeles I}}
         & $\mathrm{AUC}_{(P_D, P_F)}$  & 0.5224 & 0.6460 & 0.6633 & 0.6901 & 0.9522 & 0.8686 & 0.8138 & 0.7704 & \ValFourth{0.9937} & 0.6056 & \ValThird{0.9954} & \ValSecond{0.9960} & \ValBest{0.9961} & \ValFifth{0.9838} \\ 
         & $\mathrm{AUC}_{(P_D, \tau)}$  & \ValThird{0.2119} & 0.0188 & 0.0816 & 0.0818 & \ValBest{0.8480} & \ValSecond{0.2240} & 0.0637 & \ValFifth{0.1958} & 0.0479 & \ValFourth{0.1994} & 0.1691 & 0.1811 & 0.1804 & 0.1771 \\ 
         & $\mathrm{AUC}_{(P_F, \tau)}$  & 0.2008 & \ValSecond{0.0086} & 0.0493 & 0.0463 & 0.5296 & 0.1364 & 0.0292 & 0.1149 & \ValBest{0.0024} & 0.1671 & \ValThird{0.0165} & \ValFifth{0.0264} & \ValFourth{0.0261} & 0.0363 \\ 
        \cmidrule(lr){1-16}
        
        \multirow{3}{*}{\shortstack{Los \\ Angeles I\hspace{-1.2pt}I}}
         & $\mathrm{AUC}_{(P_D, P_F)}$  & 0.4752 & 0.6938 & 0.8244 & 0.8684 & 0.9103 & 0.6466 & 0.8162 & 0.8945 & \ValFourth{0.9612} & 0.5126 & \ValBest{0.9879} & \ValThird{0.9714} & \ValSecond{0.9838} & \ValFifth{0.9571} \\ 
         & $\mathrm{AUC}_{(P_D, \tau)}$  & 0.2629 & 0.0143 & 0.2997 & 0.3025 & \ValBest{0.8823} & \ValSecond{0.5347} & 0.3907 & \ValFourth{0.4511} & 0.3528 & 0.2083 & \ValFifth{0.4132} & 0.2907 & 0.4009 & \ValThird{0.4848} \\ 
         & $\mathrm{AUC}_{(P_F, \tau)}$  & 0.2799 & \ValBest{0.0053} & 0.1248 & 0.1033 & 0.6152 & 0.4582 & 0.2037 & 0.1685 & \ValFifth{0.0659} & 0.2058 & \ValThird{0.0353} & \ValFourth{0.0410} & \ValSecond{0.0317} & 0.1145 \\ 
        \cmidrule(lr){1-16}
        
        \multirow{3}{*}{\shortstack{San \\ Diego}}
         & $\mathrm{AUC}_{(P_D, P_F)}$  & 0.5544 & 0.6046 & 0.8132 & 0.8218 & \ValThird{0.9556} & 0.8844 & 0.6888 & \ValFifth{0.9418} & \ValFourth{0.9537} & 0.8909 & \ValBest{0.9881} & 0.9150 & \ValSecond{0.9716} & 0.9386 \\ 
         & $\mathrm{AUC}_{(P_D, \tau)}$  & 0.2402 & 0.0127 & 0.2802 & 0.2599 & \ValBest{0.7949} & \ValSecond{0.5096} & 0.4066 & \ValFifth{0.4178} & 0.2432 & 0.2690 & \ValThird{0.4465} & 0.1948 & \ValFourth{0.4293} & 0.2673 \\ 
         & $\mathrm{AUC}_{(P_F, \tau)}$  & 0.2265 & \ValBest{0.0109} & 0.1526 & 0.1312 & 0.3619 & 0.3022 & 0.3422 & 0.1411 & \ValSecond{0.0168} & 0.1308 & \ValFourth{0.0392} & 0.0547 & \ValFifth{0.0408} & \ValThird{0.0333} \\ 
        \cmidrule(lr){1-16} 

        \multirow{3}{*}{\shortstack{Hyperion}}
         & $\mathrm{AUC}_{(P_D, P_F)}$  & 0.6725 & 0.6739 & 0.8366 & 0.8510 & 0.9215 & 0.8983 & 0.7610 & \ValFifth{0.9698} & 0.7332 & 0.9213 & \ValBest{0.9973} & \ValThird{0.9929} & \ValSecond{0.9938} & \ValFourth{0.9847} \\ 
         & $\mathrm{AUC}_{(P_D, \tau)}$  & 0.3327 & 0.0004 & 0.3575 & 0.3591 & \ValFifth{0.5225} & \ValThird{0.5483} & 0.4843 & 0.4524 & 0.1841 & \ValFourth{0.5349} & 0.4104 & \ValBest{0.7738} & 0.4221 & \ValSecond{0.7603} \\ 
         & $\mathrm{AUC}_{(P_F, \tau)}$  & 0.2549 & \ValBest{0.0005} & 0.1884 & 0.1824 & 0.2960 & 0.2868 & 0.3827 & \ValFifth{0.1224} & \ValFourth{0.0615} & 0.2531 & \ValThird{0.0570} & 0.4977 & \ValSecond{0.0555} & 0.4978 \\ 

        \bottomrule
    \end{tabular}
    }
    \\
\end{table*}

Table~\ref{tab:AUCs_Case3} summarizes the three types of AUC values for all detectors across each dataset in Case 3.
As shown in this table, all the existing methods except for AHMID, which is designed to handle mixed noise, failed to detect anomalies.  
This is because these methods are designed without considering the effect of non-Gaussian noise, which makes it difficult to distinguish between background, anomalies, and such noise.
On the other hand, the proposed method achieved almost the same level of performance as in Case 1. 
This is due to the superior modeling of stripe noise by the third term and the first constraint in Prob.~\eqref{eq:proposed-optimization-problem}, and sparse noise by the third constraint. 
Among the different background regularizations, the proposed method using HTV demonstrated the best performance.

Figs.~\ref{fig:DetectionMap_pavia}(c) and~\ref{fig:semiROCs}(c) show the detection maps and 3D-ROC curves for Pavia Centre in Case 3 generated by all the detectors, respectively.
GRX, GAED, RGAE, ADLR, GTVLRR, LSDM-MoG, PCA-TLRSR, and MTVLRR are clearly affected by non-Gaussian noise.
In particular, stripe noise is noticeable in the detection maps of ADLR and LSDM-MoG.
Although the proposed method successfully detects all anomalies, sparse noise is not fully suppressed in the results of the methods using SSTV, HSSTV, or the nuclear norm.
This is because SSTV and HSSTV have limited ability to suppress sparse noise that lacks spectral continuity.
In addition, the nuclear norm does not explicitly enforce spatial smoothness, making it less effective in suppressing spatially localized sparse noise.

Besides the effect of background regularization, the 
separation between $\Acal$ and $\Scal$ can itself become 
ambiguous under certain conditions.
Compared with the result for Case~1 (see Fig.~\ref{fig:DetectionMap_pavia}(a)), 
slight residual artifacts remain in the detection map of the 
proposed method using HTV (see Fig.~\ref{fig:DetectionMap_pavia}(c)).
This is because both $\Acal$ and $\Scal$ are encouraged to be sparse, and their distinct sparsity patterns alone may not always be sufficient to distinguish them.
Introducing a regularization that more explicitly promotes 
the spectral continuity of the anomaly part may help 
mitigate this issue.

\subsubsection{Case 4 $\&$ 5 (Mixed noise)}

\begin{table*}[!t]
    \centering
    \caption{$\mathrm{AUC}_{(P_D, P_F)}$, $\mathrm{AUC}_{(P_D, \tau)}$, And $\mathrm{AUC}_{(P_F, \tau)}$ Values of All The Detectors For Each Dataset In Case 4. \\ (The Best And Second-Best Values Are Highlighted in Bold And Underlined, Respectively.)}
    \label{tab:AUC_Case4}
    \scalebox{0.75}{
    \begin{tabular}{cccccccccccccccc}
        \toprule
        Datasets & Metrics & \multicolumn{14}{c}{Methods} \\ 
        \cmidrule(lr){1-1} \cmidrule(lr){2-2} \cmidrule(lr){3-16}
        & & GRX & 2S-GLRT & GAED & RGAE & ADLR & GTVLRR & LSDM-MoG & PCA-TLRSR & AHMID & MTVLRR & Ours & Ours & Ours & Ours \\  
        & & \cite{GRX_1990} & \cite{2SGLRT_2022} & \cite{GAED_2022} & \cite{RGAE_2022} & \cite{ADLR_2018} &\cite{GTVLRR_2020} & \cite{LSDMMog_2021} & \cite{PCA-TLRSR_2023} & \cite{AHMID_2023} & \cite{MTVLRR_2024} & (HTV) & (SSTV) & (HSSTV) & (Nuclear) \\

        \midrule
        \multirow{3}{*}{\shortstack{Pavia \\ Centre}}
         & $\mathrm{AUC}_{(P_D, P_F)}$  & 0.6451 & 0.7300 & 0.7974 & 0.8937 & 0.8103 & 0.9041 & 0.7625 & \ValFourth{0.9553} & 0.9183 & 0.9176 & \ValBest{0.9904} & \ValThird{0.9583} & \ValSecond{0.9837} & \ValFifth{0.9328} \\ 
         & $\mathrm{AUC}_{(P_D, \tau)}$  & 0.1420 & 0.2122 & 0.1076 & 0.2169 & \ValBest{0.6832} & 0.2651 & \ValThird{0.3583} & \ValSecond{0.3924} & 0.0288 & 0.2457 & \ValFifth{0.3135} & 0.2584 & \ValFourth{0.3199} & 0.1339 \\ 
         & $\mathrm{AUC}_{(P_F, \tau)}$  & 0.0930 & 0.0968 & 0.0265 & 0.0446 & 0.3293 & 0.0713 & 0.2408 & 0.0826 & \ValBest{0.0028} & 0.0552 & \ValSecond{0.0098} & \ValFifth{0.0159} & \ValFourth{0.0133} & \ValThird{0.0113} \\ 
        \cmidrule(lr){1-16}
        
        \multirow{3}{*}{\shortstack{Texas \\ Coast}} 
         & $\mathrm{AUC}_{(P_D, P_F)}$  & 0.5277 & 0.9541 & 0.9404 & 0.9538 & 0.9713 & 0.9137 & 0.7063 & 0.9772 & \ValFifth{0.9800} & 0.8892 & \ValBest{0.9978} & \ValThird{0.9894} & \ValSecond{0.9954} & \ValFourth{0.9877} \\ 
         & $\mathrm{AUC}_{(P_D, \tau)}$  & 0.1690 & 0.2119 & 0.3843 & 0.3927 & \ValBest{0.8699} & \ValThird{0.6259} & 0.3536 & \ValFourth{0.5678} & 0.3183 & \ValFifth{0.5298} & 0.4994 & 0.4323 & 0.5034 & \ValSecond{0.6306} \\ 
         & $\mathrm{AUC}_{(P_F, \tau)}$  & 0.1737 & \ValFourth{0.0300} & 0.0561 & \ValFifth{0.0523} & 0.3395 & 0.2715 & 0.0958 & 0.1059 & 0.0888 & 0.2386 & \ValBest{0.0209} & \ValSecond{0.0283} & \ValThird{0.0298} & 0.0871 \\ 
        \cmidrule(lr){1-16}
        
        \multirow{3}{*}{\shortstack{Gainesville}}
         & $\mathrm{AUC}_{(P_D, P_F)}$  & 0.5857 & 0.8695 & 0.8009 & 0.7537 & 0.9539 & 0.9468 & 0.6865 & \ValThird{0.9723} & \ValFifth{0.9671} & 0.8360 & \ValBest{0.9952} & \ValFourth{0.9702} & \ValSecond{0.9817} & 0.9656 \\ 
         & $\mathrm{AUC}_{(P_D, \tau)}$  & 0.2044 & 0.2676 & 0.1590 & 0.1248 & \ValBest{0.6580} & 0.4126 & \ValFourth{0.4290} & \ValFifth{0.4279} & 0.2074 & \ValThird{0.4543} & \ValSecond{0.4556} & 0.4198 & 0.3311 & 0.1817 \\ 
         & $\mathrm{AUC}_{(P_F, \tau)}$  & 0.1770 & 0.1373 & 0.0721 & 0.0671 & 0.3428 & 0.1818 & 0.3536 & 0.0807 & \ValFourth{0.0486} & 0.2317 & \ValBest{0.0212} & \ValFifth{0.0539} & \ValThird{0.0303} & \ValSecond{0.0284} \\ 
        \cmidrule(lr){1-16}
        
        \multirow{3}{*}{\shortstack{Los \\ Angeles I}}
         & $\mathrm{AUC}_{(P_D, P_F)}$  & 0.5507 & 0.8150 & 0.7778 & 0.8146 & 0.8571 & 0.9719 & 0.6941 & 0.9177 & \ValBest{0.9966} & 0.7017 & \ValFourth{0.9959} & \ValFifth{0.9832} & \ValThird{0.9963} & \ValSecond{0.9965} \\ 
         & $\mathrm{AUC}_{(P_D, \tau)}$  & 0.1190 & 0.1463 & 0.0523 & 0.0548 & \ValBest{0.7798} & \ValThird{0.2138} & \ValFourth{0.1821} & 0.1487 & 0.0552 & 0.1114 & \ValFifth{0.1525} & 0.0980 & 0.1323 & \ValSecond{0.2373} \\ 
         & $\mathrm{AUC}_{(P_F, \tau)}$  & 0.1096 & 0.0486 & 0.0169 & \ValFifth{0.0162} & 0.5545 & 0.0780 & 0.1437 & 0.0463 & \ValBest{0.0037} & 0.0631 & \ValFourth{0.0125} & \ValSecond{0.0068} & \ValThird{0.0092} & 0.0539 \\ 
        \cmidrule(lr){1-16}
        
        \multirow{3}{*}{\shortstack{Los \\ Angeles I\hspace{-1.2pt}I}}
         & $\mathrm{AUC}_{(P_D, P_F)}$  & 0.5372 & 0.7573 & 0.8775 & 0.9122 & 0.9184 & 0.7272 & 0.6270 & 0.9371 & \ValFifth{0.9611} & 0.5829 & \ValBest{0.9881} & \ValThird{0.9788} & \ValSecond{0.9851} & \ValFourth{0.9631} \\ 
         & $\mathrm{AUC}_{(P_D, \tau)}$  & 0.1855 & 0.2069 & 0.2654 & 0.2804 & \ValBest{0.8146} & \ValThird{0.4416} & \ValFourth{0.4374} & \ValSecond{0.4898} & 0.3567 & 0.2992 & \ValFifth{0.4077} & 0.3105 & 0.3229 & 0.3525 \\ 
         & $\mathrm{AUC}_{(P_F, \tau)}$  & 0.1859 & 0.0942 & 0.0580 & \ValFifth{0.0503} & 0.3746 & 0.2864 & 0.3970 & 0.1602 & 0.0668 & 0.2576 & \ValSecond{0.0279} & \ValThird{0.0312} & \ValBest{0.0181} & \ValFourth{0.0390} \\ 
        \cmidrule(lr){1-16}
        
        \multirow{3}{*}{\shortstack{San \\ Diego}}
         & $\mathrm{AUC}_{(P_D, P_F)}$  & 0.5719 & 0.7756 & 0.8830 & 0.8890 & 0.8948 & 0.9315 & 0.6626 & \ValThird{0.9741} & 0.9105 & 0.9337 & \ValBest{0.9867} & \ValFifth{0.9443} & \ValFourth{0.9641} & \ValSecond{0.9766} \\ 
         & $\mathrm{AUC}_{(P_D, \tau)}$  & 0.1845 & 0.1366 & 0.2234 & 0.2212 & \ValBest{0.8953} & \ValFourth{0.4141} & \ValFifth{0.4022} & \ValThird{0.4190} & 0.0271 & \ValSecond{0.4246} & 0.3803 & 0.1186 & 0.3417 & 0.2660 \\ 
         & $\mathrm{AUC}_{(P_F, \tau)}$  & 0.1586 & 0.0493 & 0.0611 & 0.0587 & 0.5946 & 0.1205 & 0.3408 & 0.0930 & \ValBest{0.0029} & 0.1238 & \ValFifth{0.0259} & \ValFourth{0.0247} & \ValThird{0.0201} & \ValSecond{0.0171} \\ 
        \cmidrule(lr){1-16} 

        \multirow{3}{*}{\shortstack{Hyperion}}
         & $\mathrm{AUC}_{(P_D, P_F)}$  & 0.5824 & 0.8287 & 0.8578 & 0.8723 & 0.9843 & 0.9383 & 0.7680 & \ValThird{0.9909} & 0.8222 & 0.9541 & \ValBest{0.9977} & \ValFourth{0.9894} & \ValSecond{0.9939} & \ValFifth{0.9862} \\ 
         & $\mathrm{AUC}_{(P_D, \tau)}$  & 0.1952 & 0.4124 & 0.2602 & 0.2645 & \ValSecond{0.5610} & \ValThird{0.5562} & 0.2540 & \ValFifth{0.4586} & 0.0081 & \ValFourth{0.5307} & 0.3472 & \ValBest{0.7296} & 0.3323 & 0.2070 \\ 
         & $\mathrm{AUC}_{(P_F, \tau)}$  & 0.1436 & 0.2472 & 0.0941 & 0.0920 & 0.2339 & 0.1907 & \ValFifth{0.0621} & 0.0814 & \ValBest{0.0050} & 0.1579 & \ValSecond{0.0117} & 0.4523 & \ValFourth{0.0302} & \ValThird{0.0210} \\ 

        \bottomrule
    \end{tabular}
    }
    \\
\end{table*}
\begin{table*}[!t]
    \centering
    \caption{$\mathrm{AUC}_{(P_D, P_F)}$, $\mathrm{AUC}_{(P_D, \tau)}$, And $\mathrm{AUC}_{(P_F, \tau)}$ Values of All The Detectors For Each Dataset In Case 5. \\ (The Best And Second-Best Values Are Highlighted in Bold And Underlined, Respectively.)}
    \label{tab:AUCs_Case5}
    \scalebox{0.75}{
    \begin{tabular}{cccccccccccccccc}
        \toprule
        Datasets & Metrics & \multicolumn{14}{c}{Methods} \\ 
        \cmidrule(lr){1-1} \cmidrule(lr){2-2} \cmidrule(lr){3-16}
        & & GRX & 2S-GLRT & GAED & RGAE & ADLR & GTVLRR & LSDM-MoG & PCA-TLRSR & AHMID & MTVLRR & Ours & Ours & Ours & Ours \\  
        & & \cite{GRX_1990} & \cite{2SGLRT_2022} & \cite{GAED_2022} & \cite{RGAE_2022} & \cite{ADLR_2018} &\cite{GTVLRR_2020} & \cite{LSDMMog_2021} & \cite{PCA-TLRSR_2023} & \cite{AHMID_2023} & \cite{MTVLRR_2024} & (HTV) & (SSTV) & (HSSTV) & (Nuclear) \\

        \midrule
        \multirow{3}{*}{\shortstack{Pavia \\ Centre}}
         & $\mathrm{AUC}_{(P_D, P_F)}$  & 0.4904 & 0.6530 & 0.6366 & 0.8072 & 0.7553 & 0.7497 & 0.8975 & \ValFourth{0.9268} & \ValFifth{0.9131} & 0.7309 & \ValBest{0.9791} & \ValThird{0.9472} & \ValSecond{0.9703} & 0.9088 \\ 
         & $\mathrm{AUC}_{(P_D, \tau)}$  & 0.2112 & 0.2557 & 0.1677 & 0.2279 & \ValBest{0.8140} & \ValFifth{0.2725} & 0.2511 & \ValThird{0.3843} & 0.0110 & 0.2591 & \ValFourth{0.3008} & 0.2450 & 0.2436 & \ValSecond{0.4679} \\ 
         & $\mathrm{AUC}_{(P_F, \tau)}$  & 0.2044 & 0.1811 & 0.1017 & 0.0892 & 0.6572 & 0.1534 & \ValFifth{0.0785} & 0.1269 & \ValBest{0.0022} & 0.1483 & \ValThird{0.0164} & \ValFourth{0.0232} & \ValSecond{0.0148} & 0.1228 \\ 
        \cmidrule(lr){1-16}
        
        \multirow{3}{*}{\shortstack{Texas \\ Coast}} 
         & $\mathrm{AUC}_{(P_D, P_F)}$  & 0.5499 & 0.9494 & 0.8592 & 0.8819 & 0.9671 & 0.7910 & 0.8355 & 0.9514 & \ValFifth{0.9674} & 0.7819 & \ValBest{0.9951} & \ValThird{0.9850} & \ValSecond{0.9949} & \ValFourth{0.9833} \\ 
         & $\mathrm{AUC}_{(P_D, \tau)}$  & 0.4089 & 0.3572 & 0.4600 & 0.4556 & \ValBest{0.8195} & \ValSecond{0.6561} & \ValFourth{0.5524} & \ValThird{0.5850} & 0.2549 & 0.4967 & 0.4569 & 0.4376 & 0.4160 & \ValFifth{0.5294} \\ 
         & $\mathrm{AUC}_{(P_F, \tau)}$  & 0.3907 & \ValFifth{0.0914} & 0.1924 & 0.1699 & 0.3553 & 0.4618 & 0.4221 & 0.2158 & \ValFourth{0.0587} & 0.3467 & \ValBest{0.0371} & \ValThird{0.0518} & \ValSecond{0.0405} & 0.1241 \\ 
        \cmidrule(lr){1-16}
        
        \multirow{3}{*}{\shortstack{Gainesville}}
         & $\mathrm{AUC}_{(P_D, P_F)}$  & 0.5312 & 0.7539 & 0.6637 & 0.6299 & \ValThird{0.9581} & 0.5926 & 0.6126 & \ValFifth{0.9097} & 0.9035 & 0.5337 & \ValBest{0.9817} & 0.8593 & \ValSecond{0.9652} & \ValFourth{0.9129} \\ 
         & $\mathrm{AUC}_{(P_D, \tau)}$  & 0.3523 & 0.2705 & 0.3117 & 0.2200 & \ValBest{0.7071} & \ValSecond{0.5333} & 0.3501 & \ValFourth{0.4084} & 0.1662 & \ValThird{0.4683} & \ValFifth{0.3681} & 0.2177 & 0.2857 & 0.2953 \\ 
         & $\mathrm{AUC}_{(P_F, \tau)}$  & 0.3466 & 0.1551 & 0.2493 & 0.1822 & 0.3545 & 0.4949 & 0.3100 & 0.1714 & \ValThird{0.0557} & 0.4586 & \ValBest{0.0390} & \ValFourth{0.0745} & \ValSecond{0.0541} & \ValFifth{0.1364} \\ 
        \cmidrule(lr){1-16}
        
        \multirow{3}{*}{\shortstack{Los \\ Angeles I}}
         & $\mathrm{AUC}_{(P_D, P_F)}$  & 0.4823 & 0.7396 & 0.5940 & 0.6174 & \ValFifth{0.9698} & 0.8403 & 0.8846 & 0.6619 & \ValThird{0.9900} & 0.5790 & \ValSecond{0.9914} & 0.9219 & \ValFourth{0.9863} & \ValBest{0.9935} \\ 
         & $\mathrm{AUC}_{(P_D, \tau)}$  & \ValFifth{0.1968} & \ValBest{0.2487} & 0.0970 & 0.0953 & \ValThird{0.2264} & \ValSecond{0.2308} & 0.0666 & \ValFourth{0.2260} & 0.0396 & 0.1617 & 0.1655 & 0.0704 & 0.1112 & 0.1369 \\ 
         & $\mathrm{AUC}_{(P_F, \tau)}$  & 0.1969 & 0.1504 & 0.0735 & 0.0685 & 0.0837 & 0.1509 & \ValFifth{0.0278} & 0.1742 & \ValBest{0.0013} & 0.1371 & 0.0370 & \ValThird{0.0247} & \ValSecond{0.0221} & \ValFourth{0.0261} \\ 
        \cmidrule(lr){1-16}
        
        \multirow{3}{*}{\shortstack{Los \\ Angeles I\hspace{-1.2pt}I}}
         & $\mathrm{AUC}_{(P_D, P_F)}$  & 0.4808 & 0.8256 & 0.7779 & 0.8234 & \ValFifth{0.9342} & 0.6410 & 0.8522 & 0.8537 & 0.9178 & 0.4973 & \ValSecond{0.9789} & \ValFourth{0.9525} & \ValBest{0.9802} & \ValThird{0.9582} \\ 
         & $\mathrm{AUC}_{(P_D, \tau)}$  & 0.3101 & 0.2297 & 0.3944 & 0.3770 & \ValBest{0.9545} & \ValSecond{0.5115} & 0.3794 & \ValThird{0.5015} & 0.1572 & \ValFourth{0.4238} & 0.3437 & 0.3358 & 0.3514 & \ValFifth{0.4120} \\ 
         & $\mathrm{AUC}_{(P_F, \tau)}$  & 0.3187 & \ValFifth{0.0895} & 0.2158 & 0.1708 & 0.7486 & 0.4462 & 0.2135 & 0.2543 & \ValBest{0.0155} & 0.4256 & \ValSecond{0.0332} & \ValFourth{0.0685} & \ValThird{0.0383} & 0.0937 \\ 
        \cmidrule(lr){1-16}
        
        \multirow{3}{*}{\shortstack{San \\ Diego}}
         & $\mathrm{AUC}_{(P_D, P_F)}$  & 0.5859 & \ValFifth{0.8571} & 0.7590 & 0.7671 & 0.8243 & 0.8482 & 0.7372 & \ValThird{0.9282} & 0.8212 & 0.8474 & \ValBest{0.9814} & 0.8397 & \ValSecond{0.9507} & \ValFourth{0.9213} \\ 
         & $\mathrm{AUC}_{(P_D, \tau)}$  & 0.3128 & 0.3308 & 0.3724 & 0.3223 & \ValBest{0.8473} & \ValFourth{0.4154} & \ValSecond{0.4772} & \ValFifth{0.4034} & 0.1932 & \ValThird{0.4183} & 0.2966 & 0.1473 & 0.2359 & 0.2624 \\ 
         & $\mathrm{AUC}_{(P_F, \tau)}$  & 0.2868 & 0.1453 & 0.2461 & 0.1961 & 0.6011 & 0.2264 & 0.4221 & 0.1618 & \ValFifth{0.0821} & 0.2259 & \ValBest{0.0249} & \ValFourth{0.0618} & \ValSecond{0.0273} & \ValThird{0.0375} \\ 
        \cmidrule(lr){1-16} 

        \multirow{3}{*}{\shortstack{Hyperion}}
         & $\mathrm{AUC}_{(P_D, P_F)}$  & 0.6940 & 0.8445 & 0.8030 & 0.8174 & \ValFifth{0.9683} & 0.8652 & 0.7321 & 0.9628 & 0.6052 & 0.8899 & \ValBest{0.9968} & \ValThird{0.9911} & \ValSecond{0.9962} & \ValFourth{0.9869} \\ 
         & $\mathrm{AUC}_{(P_D, \tau)}$  & 0.4212 & 0.3098 & 0.4251 & 0.4314 & \ValBest{0.9204} & 0.5985 & \ValFourth{0.6791} & 0.5694 & 0.1145 & \ValFifth{0.6033} & 0.3522 & \ValSecond{0.7451} & 0.4010 & \ValThird{0.7423} \\ 
         & $\mathrm{AUC}_{(P_F, \tau)}$  & 0.3251 & \ValFourth{0.1202} & 0.2590 & 0.2542 & 0.6133 & 0.3418 & 0.6142 & \ValFifth{0.1810} & \ValThird{0.0670} & 0.3282 & \ValBest{0.0099} & 0.4496 & \ValSecond{0.0560} & 0.4120 \\ 

        \bottomrule
    \end{tabular}
    }
    \\
\end{table*}

The three types of AUC values for all detectors across each dataset in Cases 4 and 5 are summarized in Tables~\ref{tab:AUC_Case4} and~\ref{tab:AUCs_Case5}, respectively.
Figs.~\ref{fig:DetectionMap_pavia}(d)--(e) and \ref{fig:semiROCs}(d)--(e) show the detection maps and 3D-ROC curves for the Pavia Centre under these mixed-noise conditions.
These tables and figures illustrate that the detection performance of all the existing methods except for AHMID is significantly degraded compared to the results in Case 1.
This is because, as discussed in Case 3, these methods are designed without explicit consideration of noise or are based on the assumption of Gaussian noise, and thus they cannot accurately separate background, anomalies, and mixed noise.
In contrast, AHMID and the proposed method demonstrated robust detection performance.
However, the performance of AHMID tends to degrade as the noise intensity increases, as it relies on the quality of a pre-constructed dictionary. 
On the other hand, the proposed method maintains stable detection performance, comparable to that in Case 1.

\subsection{Computational Cost}
We measured the actual running times using MATLAB R2021a on a 64-bit Windows 11 PC with an Intel Core i9-10900K, 32GB of RAM, and an NVIDIA GeForce RTX 3090.
Table~\ref{tab:time} shows the average running times for each dataset across all cases.
The running time of the proposed method differed for each regularization employed, with the method using HTV being the fastest among them. 
This is due to the difference in computational complexity at each iteration, as described in Sec.~\ref{subsec:computational_complexity}, and the fact that HTV is designed to handle only first-order differences, leading to faster convergence to a solution compared to SSTV and HSSTV.

Compared to existing methods, the proposed method using HTV is slower than GRX and PCA-TLRSR.
This is because GRX is a computationally efficient statistical detector, while PCA-TLRSR significantly reduces the input dimensionality through PCA as a preprocessing step.
Nevertheless, the proposed method using HTV maintains a practical execution time and offers distinct advantages in terms of detection performance and robustness against various types of mixed noise.

\begin{table*}[!t]
    \centering
    \caption{Average Running Times [s] for Each Dataset Across All Cases. \\ (The Best And Second-Best Values Are Highlighted in Bold And Underlined, Respectively.)}
    \label{tab:time}
    \scalebox{0.75}{
    \begin{tabular}{ccccccccccccccc}
        \toprule
        Datasets & \multicolumn{14}{c}{Methods} \\ 
        \cmidrule(lr){1-1} \cmidrule(lr){2-15}
        & GRX & 2S-GLRT & GAED & RGAE & ADLR & GTVLRR & LSDM-MoG & PCA-TLRSR & AHMID & MTVLRR & Ours & Ours & Ours & Ours \\  
        & \cite{GRX_1990} & \cite{2SGLRT_2022} & \cite{GAED_2022} & \cite{RGAE_2022} & \cite{ADLR_2018} &\cite{GTVLRR_2020} & \cite{LSDMMog_2021} & \cite{PCA-TLRSR_2023} & \cite{AHMID_2023} & \cite{MTVLRR_2024} & (HTV) & (SSTV) & (HSSTV) & (Nuclear) \\

        \midrule
        
        Pavia Centre & \ValBest{0.0497} & 22.6441 & 76.0738 & 103.5264 & 25.3356 & 83.1759 & 141.4048 & \ValThird{5.4739} & 79.6451 & 142.0832 & \ValSecond{2.9628} & \ValFifth{14.2996} & \ValFourth{9.1461} & 25.7702 \\ 

        Texas Coast & \ValBest{0.0462} & 28.3905 & 30.9730 & 72.8239 & \ValFourth{5.9054} & 45.2060 & 64.4329 & \ValThird{2.7160} & 58.5828 & 91.8825 & \ValSecond{2.5071} & \ValFifth{6.8161} & 10.6848 & 20.2238 \\ 

        Gainesville & \ValBest{0.0442} & 57.4362 & 38.8748 & 68.5991 & \ValFourth{8.1800} & 40.8299 & 49.3530 & \ValSecond{1.7853} & 66.0467 & 88.0037 & \ValThird{2.6781} & \ValFifth{13.0949} & 13.8887 & 19.1869 \\ 

        Los Angeles I & \ValBest{0.0455} & 37.1338 & 31.2475 & 61.6298 & \ValFifth{5.7687} & 39.4142 & 125.6955 & \ValSecond{1.8181} & 65.3444 & 77.4855 & \ValThird{4.3601} & \ValFourth{5.6451} & 8.3482 & 23.5670 \\ 

        Los Angeles I\hspace{-1.2pt}I & \ValBest{0.0477} & 28.7896 & 31.1007 & 72.4988 & \ValFourth{7.9558} & 46.1639 & 180.9510 & \ValSecond{2.1876} & 53.1778 & 92.6639 & \ValThird{3.3243} & \ValFifth{13.4290} & 13.8006 & 22.5041 \\ 

        San Diego & \ValBest{0.0438} & 150.2518 & 28.7062 & 64.1044 & \ValFourth{7.0702} & 40.1116 & 70.6507 & \ValSecond{1.7870} & 56.1631 & 82.0864 & \ValThird{2.8180} & \ValFifth{13.7074} & 16.8641 & 16.1304 \\ 

        Hyperion & \ValBest{0.0786} & 50.3015 & 74.2780 & 96.3566 & \ValFifth{13.4488} & 79.4833 & 131.3323 & \ValFourth{4.8981} & 120.4911 & 140.8215 & \ValThird{4.1288} & \ValSecond{1.9766} & 17.2475 & 25.8543 \\ 
 
        \bottomrule
 
    \end{tabular}

    }
    \\
\end{table*}

\subsection{Analysis of Differences in Background Characterization Function}

\begin{figure}[!t]

    \begin{minipage}{0.06\hsize}
        ~
    \end{minipage}
    \begin{minipage}{0.44\hsize}
        \centerline{\footnotesize{Background Part}}
    \end{minipage}
    \begin{minipage}{0.01\hsize}
        ~
    \end{minipage}
    \begin{minipage}{0.44\hsize}
        \centerline{\footnotesize{Anomaly Part}}
    \end{minipage}

    \vspace{0.5mm}
    
    \begin{minipage}{0.06\hsize}
        \centerline{{\rotatebox{90}{\small{\shortstack{(a) Ours \\ (HTV)}}}}}
    \end{minipage}
    \begin{minipage}{0.22\hsize}
        \includegraphics[keepaspectratio, scale = 0.37]{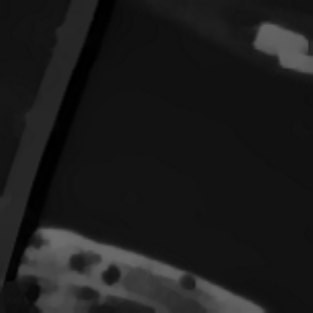}
    \end{minipage}
    \begin{minipage}{0.22\hsize}
        \includegraphics[keepaspectratio, scale = 0.37]{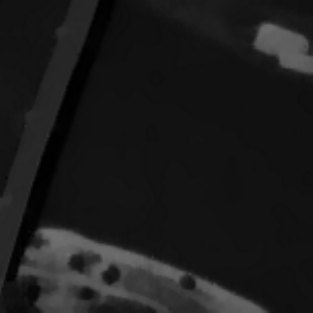}
    \end{minipage}
    \begin{minipage}{0.01\hsize}
        ~
    \end{minipage}
    \begin{minipage}{0.22\hsize}
        \includegraphics[keepaspectratio, scale = 0.37]{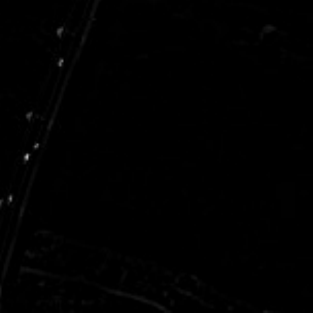}
    \end{minipage}
    \begin{minipage}{0.22\hsize}
        \includegraphics[keepaspectratio, scale = 0.37]{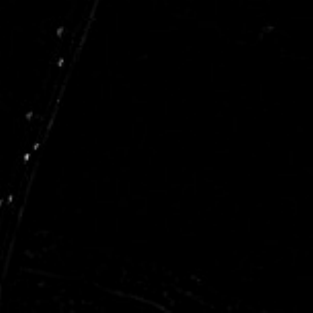}
    \end{minipage}

    \vspace{1mm}
    
    \begin{minipage}{0.06\hsize}
        \centerline{{\rotatebox{90}{\small{\shortstack{(b) Ours \\ (SSTV)}}}}}
    \end{minipage}
    \begin{minipage}{0.22\hsize}
        \includegraphics[keepaspectratio, scale = 0.37]{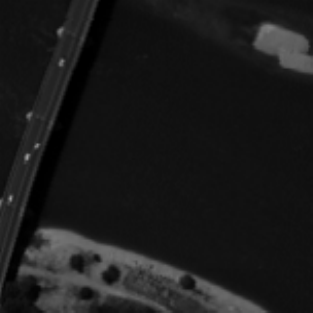}
    \end{minipage}
    \begin{minipage}{0.22\hsize}
        \includegraphics[keepaspectratio, scale = 0.37]{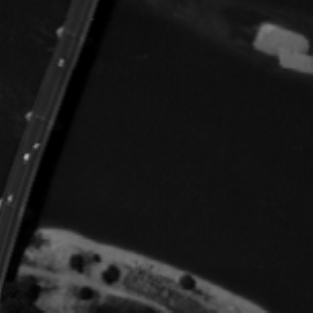}
    \end{minipage}
    \begin{minipage}{0.01\hsize}
        ~
    \end{minipage}
    \begin{minipage}{0.22\hsize}
        \includegraphics[keepaspectratio, scale = 0.37]{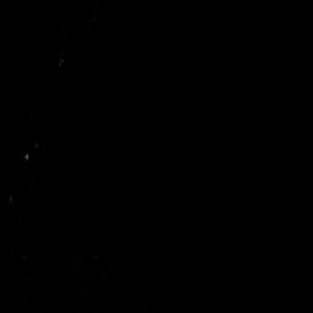}
    \end{minipage}
    \begin{minipage}{0.22\hsize}
        \includegraphics[keepaspectratio, scale = 0.37]{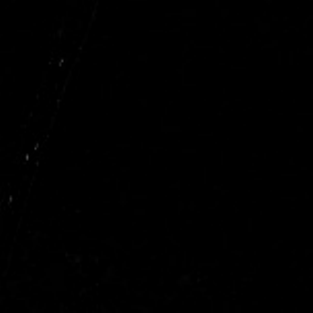}
    \end{minipage}

    \vspace{1mm}
    
    \begin{minipage}{0.06\hsize}
        \centerline{{\rotatebox{90}{\small{\shortstack{(c) Ours \\ (HSSTV)}}}}}
    \end{minipage}
    \begin{minipage}{0.22\hsize}
        \includegraphics[keepaspectratio, scale = 0.37]{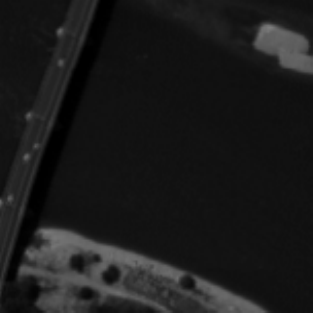}
    \end{minipage}
    \begin{minipage}{0.22\hsize}
        \includegraphics[keepaspectratio, scale = 0.37]{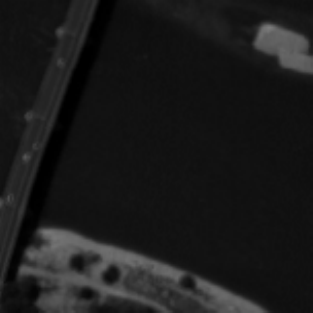}
    \end{minipage}
    \begin{minipage}{0.01\hsize}
        ~
    \end{minipage}
    \begin{minipage}{0.22\hsize}
        \includegraphics[keepaspectratio, scale = 0.37]{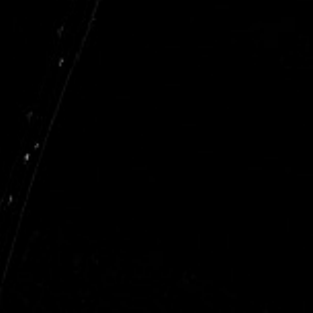}
    \end{minipage}
    \begin{minipage}{0.22\hsize}
        \includegraphics[keepaspectratio, scale = 0.37]{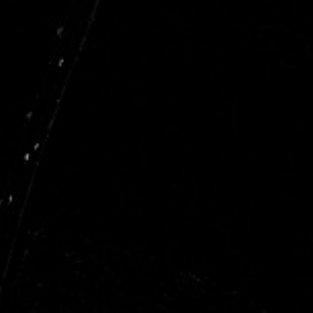}
    \end{minipage}

    \vspace{1mm}
    
    \begin{minipage}{0.06\hsize}
        \centerline{{\rotatebox{90}{\small{\shortstack{(d) Ours \\ (Nuclear)}}}}}
    \end{minipage}
    \begin{minipage}{0.22\hsize}
        \includegraphics[keepaspectratio, scale = 0.37]{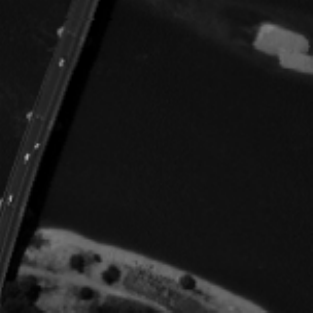}
    \end{minipage}
    \begin{minipage}{0.22\hsize}
        \includegraphics[keepaspectratio, scale = 0.37]{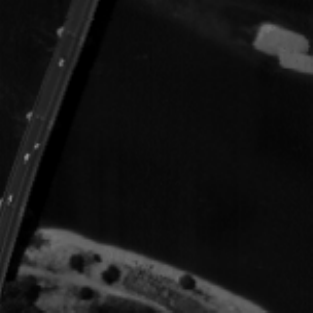}
    \end{minipage}
    \begin{minipage}{0.01\hsize}
        ~
    \end{minipage}
    \begin{minipage}{0.22\hsize}
        \includegraphics[keepaspectratio, scale = 0.37]{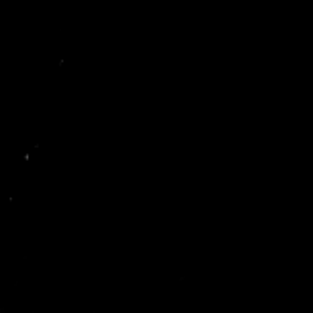}
    \end{minipage}
    \begin{minipage}{0.22\hsize}
        \includegraphics[keepaspectratio, scale = 0.37]{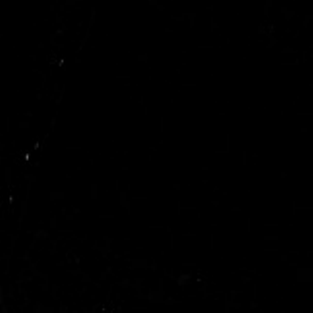}
    \end{minipage}
    
    \vspace{0.5mm}
    
    \begin{minipage}{0.06\hsize}
        ~
    \end{minipage}
    \begin{minipage}{0.22\hsize}
        \centerline{\footnotesize{Case 1}}
    \end{minipage}
    \begin{minipage}{0.22\hsize}
        \centerline{\footnotesize{Case 4}}
    \end{minipage}
    \begin{minipage}{0.01\hsize}
        ~
    \end{minipage}
    \begin{minipage}{0.22\hsize}
        \centerline{\footnotesize{Case 1}}
    \end{minipage}
    \begin{minipage}{0.22\hsize}
        \centerline{\footnotesize{Case 4}}
    \end{minipage}
    
    \caption{Results of the separation of background and anomaly parts with the 30th band.}
    \label{fig:separation_result}
\end{figure}

Fig.~\ref{fig:separation_result} shows the 30th band of the background and anomaly parts separated by the proposed method for the Pavia Centre in Cases 1 and 4.
In both cases, there are few differences, indicating that the proposed method achieves a noise-robust decomposition.
Among them, the method using HTV separates the background and anomaly parts most accurately.  
This is because HTV, which is designed to handle only vertical and horizontal neighborhood differences, best captures the feature that pixels in the background part are spectrally similar to the surrounding pixels.
From the results of the methods using SSTV and HSSTV, the use of spectral directional differences is likely to lead to false positives.

\begin{figure}[!t]
    \includegraphics[keepaspectratio, scale = 0.28]{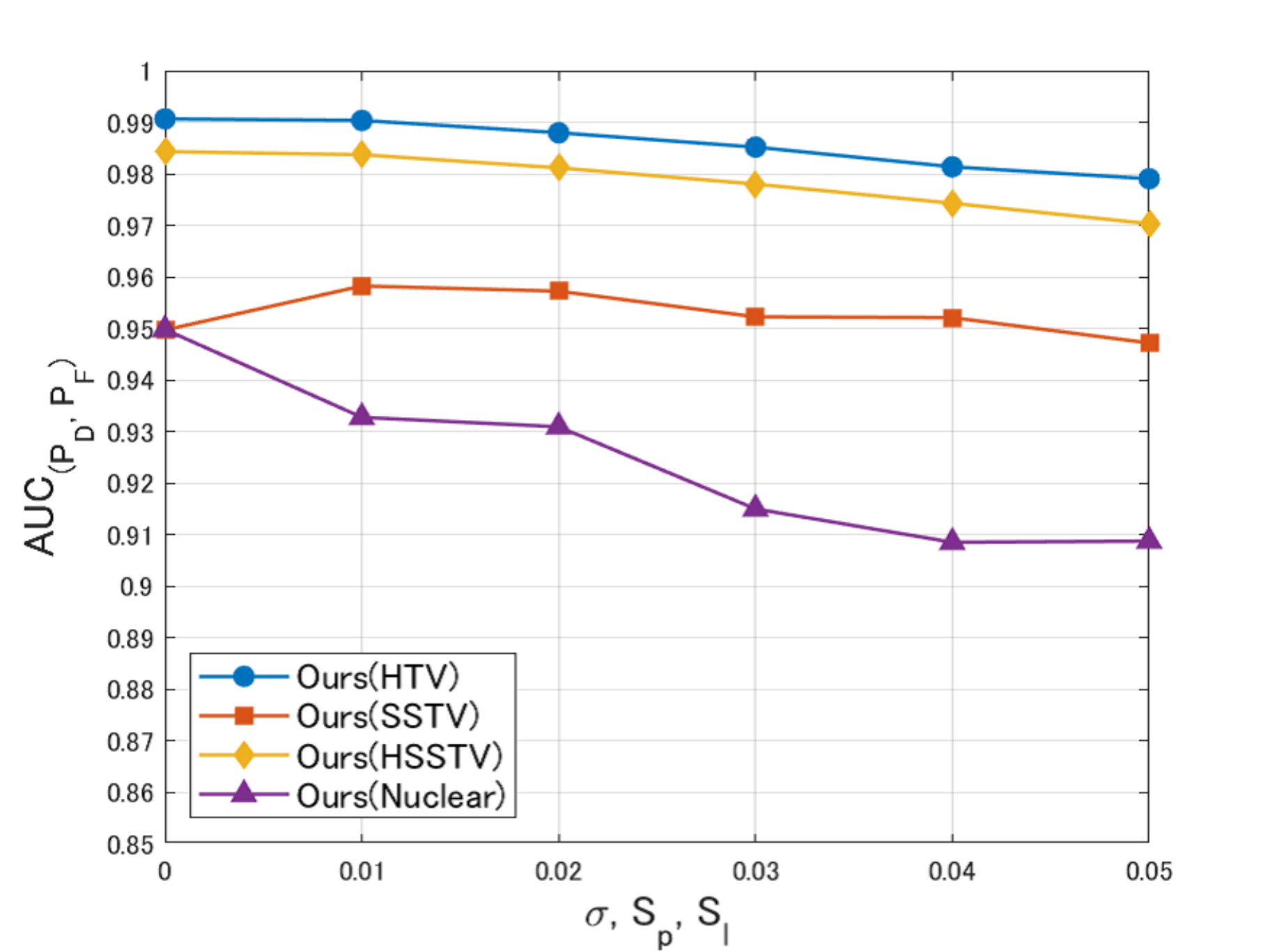}
    \caption{Detection performance variation of the proposed method with different noise levels.}
    \label{fig:noiselevel}
\end{figure}

Fig.~\ref{fig:noiselevel} shows how the detection performance of the proposed method changes when noise is added to Pavia Centre for $\sigma$, $S_p$, and $S_l$ values ranging from 0 to 0.05.
In each case, we set the hyperparameters of the proposed method to those with the maximum $\mathrm{AUC}_{(P_D, P_F)}$ value.
The method using HTV or HSSTV is robust to noise because there is little change in $\mathrm{AUC}_{(P_D, P_F)}$ value. 
In contrast, the method using SSTV showed a larger change in $\mathrm{AUC}_{(P_D, P_F)}$ value than the two methods. 
This is due to the limited ability of SSTV to reduce noise generated between consecutive bands. 
In addition, the method using the nuclear norm is not robust. 
This is because it cannot directly handle spatial smoothness, making the removal of locally concentrated noise difficult.

Based on these results, we provide practical guidelines for selecting the background characterization function.
The seven datasets cover diverse scenes, including airport, beach, urban, and agricultural environments.
Despite this diversity, the method using HTV achieves the best overall detection performance across most scenarios (see Tables~\ref{tab:AUCs_Case1}--\ref{tab:AUCs_Case5}).
These results suggest that the choice of background characterization function is more closely related to the structural properties of the background part than to the nominal scene category itself. 
In particular, piecewise spatial smoothness serves as an effective and broadly applicable prior for characterizing background structures. 
The method using HTV also maintains stable performance under various noise conditions (see Fig.~\ref{fig:noiselevel}), whereas the methods using SSTV and the nuclear norm tend to be more sensitive to noise contamination.
In addition, the method using HTV is the fastest or second-fastest among the four variants for most datasets (see Table~\ref{tab:time}).
Therefore, we recommend HTV as the default choice in practical applications.

However, when the background part contains prominent edges or complex structural patterns, the method using HTV may falsely detect such structures as anomalies.
In such cases, the nuclear norm, which does not explicitly enforce spatial smoothness, can serve as an alternative.
When strong spectral continuity is expected in the background part, the method using SSTV or HSSTV may also be considered. 
However, it tends to exhibit lower detection performance and reduced noise robustness compared to the method using HTV.

\subsection{Parameter Analysis}
In the proposed method, the four parameters, $\lambda_1$, $\lambda_2$, $\varepsilon$, and $\alpha$, affect the detection performance. 
Therefore, we describe the parameter analysis of the proposed method using HTV, which achieved the best detection performance among the specific designs of background characterization functions.

\begin{figure}[!t]
    \includegraphics[keepaspectratio, scale = 0.28]{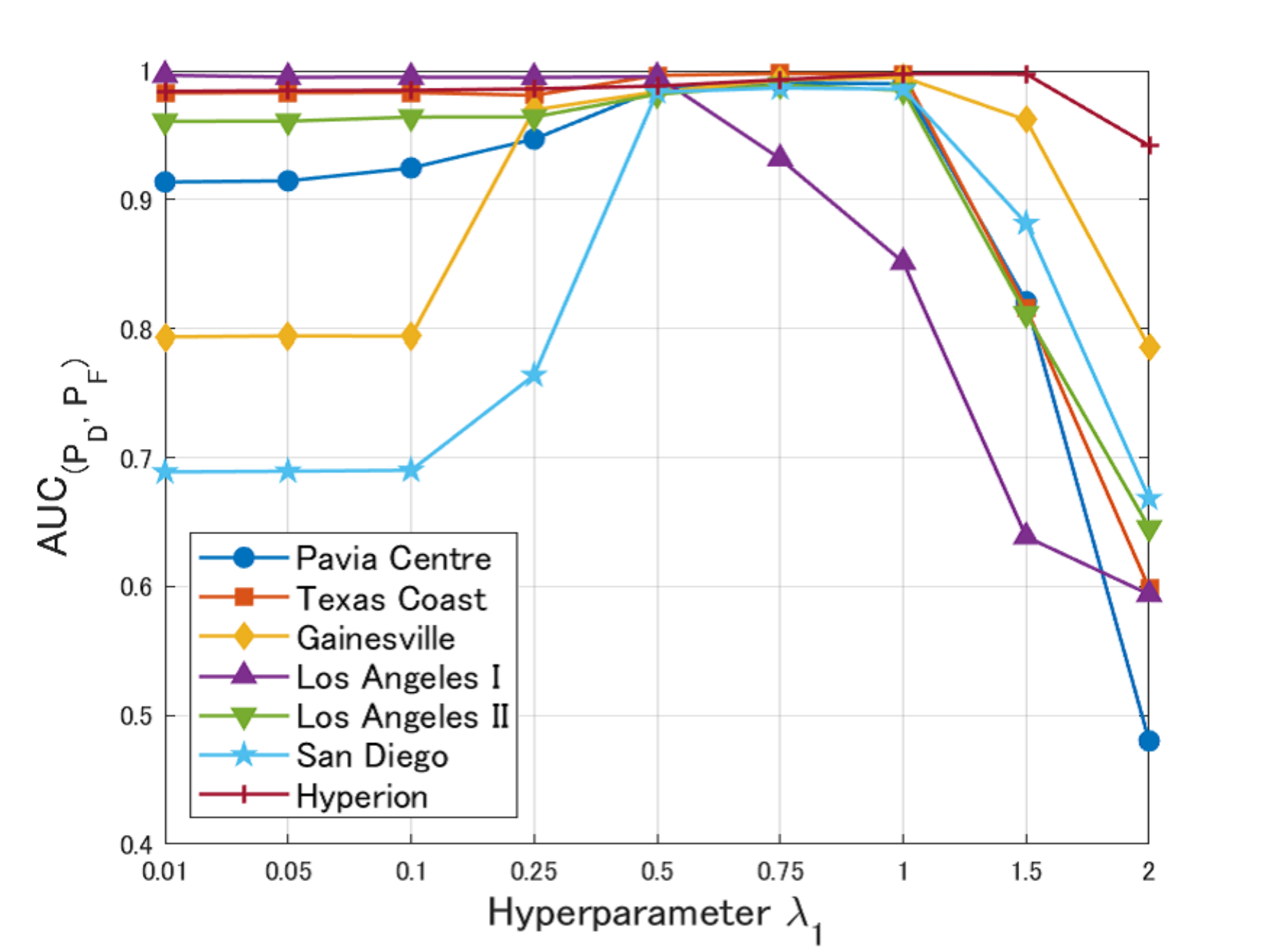}
    \caption{Parameter analysis of $\lambda_1$ for all datasets in Case 1.}
    \label{fig:parameter_analysis_case1}
\end{figure}

To investigate the influence of $\lambda_1$, we set $\varepsilon = \alpha = \lambda_2 =0$ and conducted experiments using data from Case 1.
Fig.~\ref{fig:parameter_analysis_case1} shows the $\mathrm{AUC}_{(P_D, P_F)}$ values with different $\lambda_1$ in each dataset.
For small values of $\lambda_1$, the sparsity constraint on the anomaly part is insufficiently enforced, causing background pixels to be falsely extracted as anomalies and increasing false alarms.
Conversely, for large values of $\lambda_1$, the sparsity is excessively enforced, suppressing true anomalies and causing missed detections.
These effects are well balanced when $\lambda_1$ is set between 0.5 and 1, yielding consistently high detection performance across most datasets with little variation.
Therefore, we recommend setting $\lambda_1$ to 0.5, 0.75, or 1.

\begin{figure}[!t]
    \includegraphics[keepaspectratio, scale = 0.28]{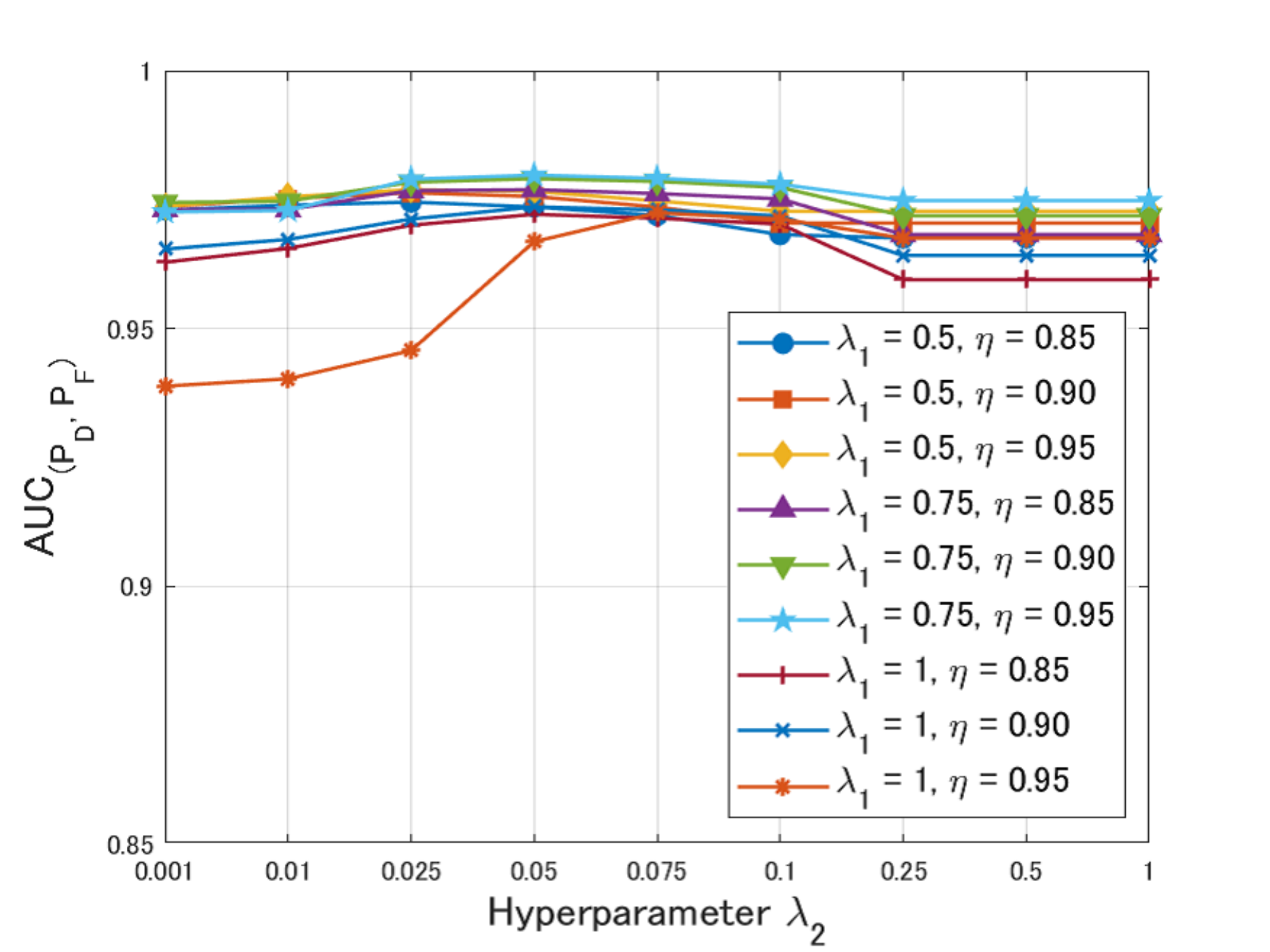}
    \caption{Parameter analysis of $\lambda_2$ and $\eta$ for Pavia Centre in Case 5.}
    \label{fig:parameter_analysis_case5}
\end{figure}

To investigate the influence of $\lambda_2$ and $\eta$, i.e., $\varepsilon$ and $\alpha$ (see Eq.~\eqref{eq:epsilon_alpha}), we fixed $\lambda_1$ to 0.5, 0.75, or 1 and conducted experiments using data from Case 5.
Fig.~\ref{fig:parameter_analysis_case5} shows the $\mathrm{AUC}_{(P_D, P_F)}$ values for different $\lambda_2$ and $\eta$ for Pavia Centre in Case 5.
For different values of $\lambda_2$, the detection performance varies little, indicating low sensitivity to this parameter.
Although $\lambda_2$ controls the sparsity of the stripe noise, the structural property of stripe noise is primarily governed by the flatness constraint, which enforces a constant structure along the vertical direction.
Consequently, moderate changes in $\lambda_2$ do not significantly affect the separation between the background and anomaly parts.
Slightly better performance is observed when $\lambda_2$ is set between 0.025 and 0.075, and similar trends are observed across other datasets.
Therefore, we recommend setting $\lambda_2$ within this range.

The parameter $\eta$ controls the tolerance of the constraint sets in Prob.~\eqref{eq:proposed-optimization-problem} by scaling the upper bounds $\varepsilon$ and $\alpha$.
For small values of $\eta$, the constraint sets become overly restrictive, causing noise to leak into the background and anomaly parts and degrading the separation accuracy. 
Conversely, for large values of $\eta$, the constraint sets become too permissive, allowing the background and anomaly parts to be absorbed into the noise parts, which also degrades performance. 
In our experiments, the best performance is consistently obtained around $\eta = 0.9$ with little variation. 
Therefore, we recommend setting $\eta = 0.9$.

\subsection{Analysis of the Flatness Constraint}
In the proposed method, a flatness constraint is imposed on stripe noise based on the assumption that stripe noise is superimposed with constant intensity along the vertical direction. 
However, in practical applications, stripe noise is not necessarily perfectly constant and may not be strictly aligned with the vertical direction.
Therefore, to investigate whether this strict assumption is appropriate, we compare the original formulation with a relaxed version, in which $\mathfrak{D}_v(\Lcal) = \Ocal$ is replaced by $\| \mathfrak{D}_v(\Lcal) \|_F \leq \delta \sqrt{HWB}$.
Here, $\delta$ is varied over $0$, $10^{-3}$, $10^{-2}$, and $10^{-1}$.
Note that the case of $\delta = 0$ corresponds to the original constraint $\mathfrak{D}_v(\Lcal) = \Ocal$.

\begin{figure}[!t]

    \begin{minipage}{0.06\hsize}
        \centerline{{\rotatebox{90}{\small{\shortstack{Pavia Centre \\ (Case 4)}}}}}
    \end{minipage}
    \begin{minipage}{0.22\hsize}
        \includegraphics[keepaspectratio, scale = 0.37]{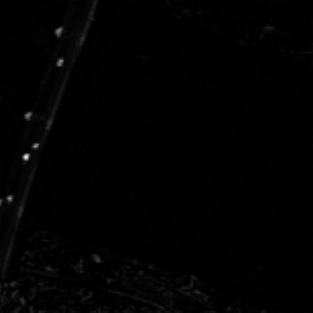}
    \end{minipage}
    \begin{minipage}{0.22\hsize}
        \includegraphics[keepaspectratio, scale = 0.37]{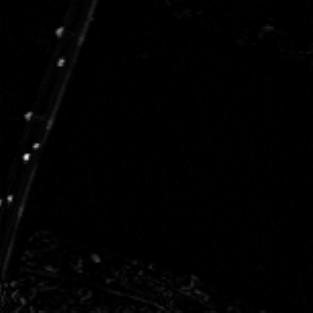}
    \end{minipage}
    \begin{minipage}{0.22\hsize}
        \includegraphics[keepaspectratio, scale = 0.37]{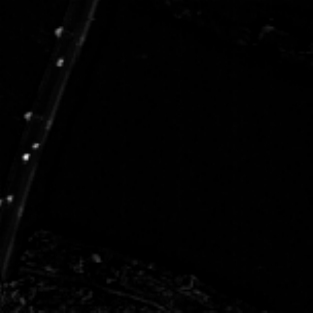}
    \end{minipage}
    \begin{minipage}{0.22\hsize}
        \includegraphics[keepaspectratio, scale = 0.37]{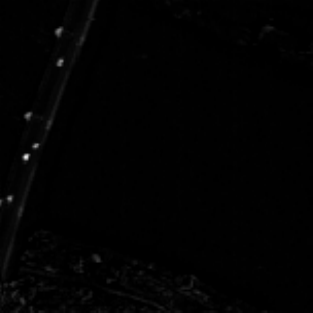}
    \end{minipage}

    \vspace{1mm}
    
    \begin{minipage}{0.06\hsize}
        \centerline{{\rotatebox{90}{\small{\shortstack{Pavia Centre \\ (Case 5)}}}}}
    \end{minipage}
    \begin{minipage}{0.22\hsize}
        \includegraphics[keepaspectratio, scale = 0.37]{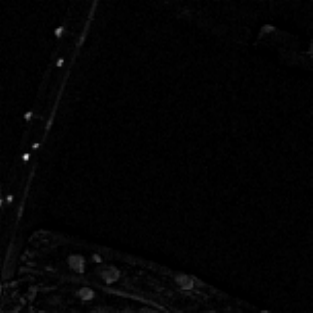}
    \end{minipage}
    \begin{minipage}{0.22\hsize}
        \includegraphics[keepaspectratio, scale = 0.37]{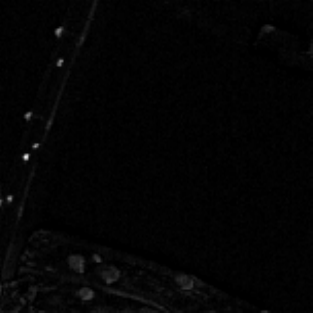}
    \end{minipage}
    \begin{minipage}{0.22\hsize}
        \includegraphics[keepaspectratio, scale = 0.37]{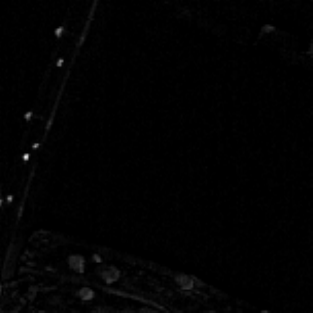}
    \end{minipage}
    \begin{minipage}{0.22\hsize}
        \includegraphics[keepaspectratio, scale = 0.37]{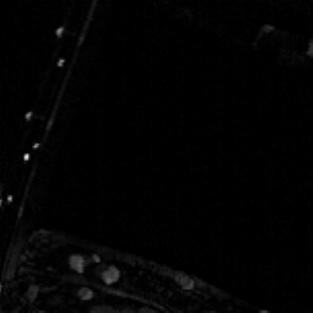}
    \end{minipage}

    \vspace{1mm}
    
    \begin{minipage}{0.06\hsize}
        ~
    \end{minipage}
    \begin{minipage}{0.22\hsize}
        \centerline{\footnotesize{\shortstack{Ours (HTV) \\ ($\delta = 0$)}}}
    \end{minipage}
    \begin{minipage}{0.22\hsize}
        \centerline{\footnotesize{\shortstack{Ours (HTV) \\ ($\delta = 10^{-3}$)}}}
    \end{minipage}
    \begin{minipage}{0.22\hsize}
        \centerline{\footnotesize{\shortstack{Ours (HTV) \\ ($\delta = 10^{-2}$)}}}
    \end{minipage}
    \begin{minipage}{0.22\hsize}
        \centerline{\footnotesize{\shortstack{Ours (HTV) \\ ($\delta = 10^{-1}$)}}}
    \end{minipage}

    \caption{Resulting detection maps for the Pavia Centre in Cases 4 and 5 generated by the proposed method using HTV under different relaxation levels of the flatness constraint.}
    \label{fig:FlatnessConstraint}
\end{figure}

Table~\ref{tab:AUCs_Dv_ablation} summarizes the three types of AUC values under different relaxation levels of the flatness constraint for the proposed method using HTV on Pavia Centre in Cases 4 and 5.
Fig.~\ref{fig:FlatnessConstraint} shows the corresponding detection maps.
These results show that the detection performance gradually degrades as $\delta$ increases.
This is because relaxing the constraint weakens the ability to exploit the directional structure of stripe noise, making the separation from other components less distinct.
These findings support the use of the strict flatness constraint when stripe noise is aligned with the vertical direction. 
However, when stripe noise is superimposed along oblique directions, the strict constraint may not hold, and the relaxed formulation can serve as a useful extension.

\begin{table}[!t]
    \centering
    \caption{$\mathrm{AUC}_{(P_D, P_F)}$, $\mathrm{AUC}_{(P_D, \tau)}$, And $\mathrm{AUC}_{(P_F, \tau)}$ Values of the Proposed Method Using HTV Under Different Relaxation Levels of the Flatness Constraint for Pavia Centre in Cases 4 and 5. Note that $\delta = 0$ corresponds to the original strict constraint $(\mathfrak{D}_v(\Lcal) = \Ocal)$. \\ (The Best And Second-Best Values Are Highlighted in Bold And Underlined, Respectively.)}
    \label{tab:AUCs_Dv_ablation}
    \scalebox{0.75}{
    \begin{tabular}{cccccc}
        \toprule
        Datasets & Metrics & \multicolumn{4}{c}{Methods} \\ 
        \cmidrule(lr){1-1} \cmidrule(lr){2-2} \cmidrule(lr){3-6}
        & & Ours (HTV) & Ours (HTV) & Ours (HTV) & Ours (HTV) \\  
        & & ($\delta = 0$) & ($\delta = 10^{-3}$) & ($\delta = 10^{-2}$) & ($\delta = 10^{-1}$) \\

        \midrule
        \multirow{3}{*}{\shortstack{Pavia \\ Centre \\ (Case 4)}}
         & $\mathrm{AUC}_{(P_D, P_F)}$  & \ValBest{0.9904} & \ValSecond{0.9903} & \ValThird{0.9898} & \ValFourth{0.9897} \\ 
         & $\mathrm{AUC}_{(P_D, \tau)}$  & \ValBest{0.3135} & \ValFourth{0.3111} & \ValThird{0.3125} & \ValSecond{0.3125} \\ 
         & $\mathrm{AUC}_{(P_F, \tau)}$  & \ValFourth{0.0098} & \ValBest{0.0090} & \ValSecond{0.0094} & \ValThird{0.0094} \\ 
        \cmidrule(lr){1-6}
        
        \multirow{3}{*}{\shortstack{Pavia \\ Centre \\ (Case 5)}}
        & $\mathrm{AUC}_{(P_D, P_F)}$  & \ValBest{0.9791} & \ValSecond{0.9783} & \ValThird{0.9782} & \ValFourth{0.9753} \\ 
        & $\mathrm{AUC}_{(P_D, \tau)}$  & \ValBest{0.3008} & \ValSecond{0.2998} & \ValFourth{0.2859} & \ValThird{0.2890} \\ 
        & $\mathrm{AUC}_{(P_F, \tau)}$  & \ValThird{0.0164} & \ValFourth{0.0172} & \ValBest{0.0104} & \ValSecond{0.0134} \\ 
        \bottomrule
    \end{tabular}
    }
    \\
\end{table}

\subsection{Noise Level Estimation for Practical Applications}
In practical applications, the standard deviation of Gaussian noise, 
$\sigma$, and the ratio of sparse noise pixels, $S_p$, in Eq.~\eqref{eq:epsilon_alpha} are generally unknown. 
To address this issue, we propose a strategy to estimate these noise characteristics from a given HS image and automatically determine the corresponding parameters.

We first estimate $S_p$.
Sparse noise is typically caused by sensor malfunction and manifests as pixels with extreme intensity values that appear abruptly and independently in HS images.
To roughly identify sparse noise from a given HS image $\Vcal$, we define the following two conditions.
First, we require large spectral differences from neighboring bands:
\begin{align}
    \label{eq:sparse_spectral}
    |[\Vcal]_{i,j,k} - [\Vcal]_{i,j,k-1}| \geq 0.1 \;\; \text{and} \;\; 
    |[\Vcal]_{i,j,k} - [\Vcal]_{i,j,k+1}| \geq 0.1.
\end{align}
While spatial differences can also capture abrupt variations, they are sensitive to image structures such as edges and textures, making them unreliable for identifying sparse noise.
In contrast, the background and anomaly parts exhibit continuity along the spectral dimension, whereas sparse noise does not, making spectral differences a more reliable criterion.
Second, sparse noise tends to take extreme intensity values near the saturation limits of the sensor:
\begin{align}
    \label{eq:sparse_extreme}
    [\Vcal]_{i,j,k} \geq 0.99 \;\; \text{or} \;\; 
    [\Vcal]_{i,j,k} \leq 0.01.
\end{align}
Here, the thresholds in Eq.~\eqref{eq:sparse_spectral} and Eq.\eqref{eq:sparse_extreme} were empirically determined under the normalization of pixel values to $[0, 1]$ adopted in this article (see Section~IV-A).
A pixel satisfying both conditions is regarded as sparse noise, and $S_p$ is estimated as the proportion of such pixels.

Next, the pixels identified as sparse noise are replaced by the average of the two spectrally adjacent bands.
Then, the noise estimation method in~\cite{Hysime_2008} is applied to extract noise residuals $\Ncal$ from the interpolated image.
Since the obtained residuals may contain outliers, the Median Absolute Deviation (MAD) is used as a robust estimator of the Gaussian noise standard deviation for each band.
Specifically, the standard deviation of the $k$-th band is estimated as
\begin{align}
    \label{eq:MAD_sigma}
    \sigma_k = \frac{\mathrm{median}\bigl(|[\Ncal]_{:,:,k} - \mathrm{median}([\Ncal]_{:,:,k})|\bigr)}{0.6745},
\end{align}
and the final estimate $\sigma$ is given as the average of $\sigma_k$ over all bands.

\begin{figure}[!t]
    \centering

    \begin{minipage}{0.45\linewidth}
        \centering\small Pavia Centre (Case 4)
    \end{minipage}
    \begin{minipage}{0.02\linewidth}
        ~
    \end{minipage}
    \begin{minipage}{0.45\linewidth}
        \centering\small Pavia Centre (Case 5)
    \end{minipage}

    \vspace{1mm}

    \begin{minipage}{0.22\linewidth}
        \centering
        \includegraphics[width=\linewidth]{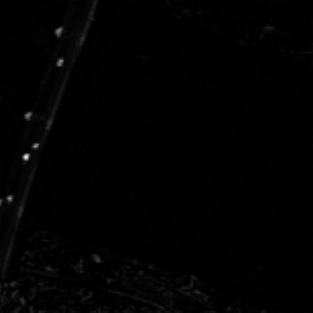}
    \end{minipage}
    \begin{minipage}{0.01\linewidth}
        ~
    \end{minipage}
    \begin{minipage}{0.22\linewidth}
        \centering
        \includegraphics[width=\linewidth]{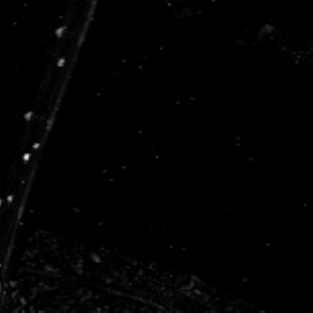}
    \end{minipage}
    \begin{minipage}{0.02\linewidth}
        ~
    \end{minipage}
    \begin{minipage}{0.22\linewidth}
        \centering
        \includegraphics[width=\linewidth]{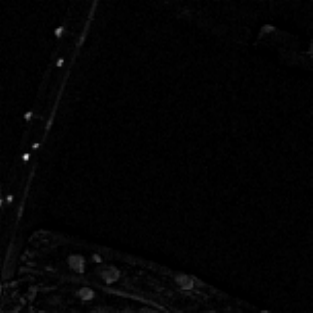}
    \end{minipage}
    \begin{minipage}{0.01\linewidth}
        ~
    \end{minipage}
    \begin{minipage}{0.22\linewidth}
        \centering
        \includegraphics[width=\linewidth]{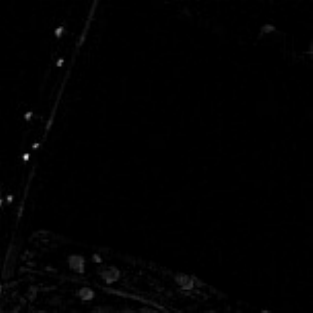}
    \end{minipage}

    \vspace{1mm}

    \begin{minipage}{0.22\linewidth}
        \centerline{\footnotesize{\shortstack{Ours (HTV) \\ ($\sigma = 0.01$, \\ $S_p = 0.01$)}}}
    \end{minipage}
    \begin{minipage}{0.01\linewidth}
        ~
    \end{minipage}
    \begin{minipage}{0.22\linewidth}
        \centerline{\footnotesize{\shortstack{Ours (HTV) \\ ($\sigma_{\mathrm{est}} = 0.01266$, \\ ${S_p}_{\mathrm{est}} = 0.008912$)}}}
    \end{minipage}
    \begin{minipage}{0.02\linewidth}
        ~
    \end{minipage}
    \begin{minipage}{0.22\linewidth}
        \centerline{\footnotesize{\shortstack{Ours (HTV) \\ ($\sigma = 0.05$, \\ $S_p = 0.05$)}}}
    \end{minipage}
    \begin{minipage}{0.01\linewidth}
        ~
    \end{minipage}
    \begin{minipage}{0.22\linewidth}
        \centerline{\footnotesize{\shortstack{Ours (HTV) \\ ($\sigma_{\mathrm{est}} = 0.05356$, \\ ${S_p}_{\mathrm{est}} = 0.05860$)}}}
    \end{minipage}

    \caption{Resulting detection maps for Pavia Centre in Cases 4 and 5 generated by the proposed method using HTV with oracle and estimated noise parameters.}
    \label{fig:est_sigma_Sp}
\end{figure}

Table~\ref{tab:AUCs_est_sigma_Sp} compares the oracle and estimated noise parameters and presents the three types of AUC values obtained by the proposed method with HTV using each set of parameters for Pavia Centre in Cases~4 and~5.
Fig.~\ref{fig:est_sigma_Sp} shows the corresponding detection maps.
The estimated values of $\sigma_{\mathrm{est}}$ and ${S_p}_{\mathrm{est}}$ are close to the oracle values, and the detection performance is comparable to the oracle case.
These results demonstrate that the proposed estimation strategy effectively eliminates the need for manual tuning of the noise parameters, enhancing the practical applicability of the method.

\begin{table}[!t]
    \centering
    \caption{$\mathrm{AUC}_{(P_D, P_F)}$, $\mathrm{AUC}_{(P_D, \tau)}$, And $\mathrm{AUC}_{(P_F, \tau)}$ Values of the Proposed Method Using HTV For Pavia Centre In Cases 4 and 5 with Oracle and Estimated Noise Parameters. \\ (The Best And Second-Best Values Are Highlighted in Bold And Underlined, Respectively.)}
    \label{tab:AUCs_est_sigma_Sp}
    \scalebox{0.75}{
    \begin{tabular}{ccccc}
        \toprule
        Datasets & Settings & $\mathrm{AUC}_{(P_D, P_F)}$ & $\mathrm{AUC}_{(P_D, \tau)}$ & $\mathrm{AUC}_{(P_F, \tau)}$ \\ 
        
        \midrule

        \multirow{4}{*}{\shortstack{Pavia \\ Centre \\ (Case 4)}}
        & \multirow{2}{*}{\shortstack{$\sigma = 0.01$ \\ $S_p = 0.01$}} & \multirow{2}{*}{\ValBest{0.9904}} & \multirow{2}{*}{\ValSecond{0.3135}} & \multirow{2}{*}{\ValBest{0.0098}} \\ 

        & & & & \\
        \cmidrule(lr){2-5}
        
        & \multirow{2}{*}{\shortstack{$\sigma_{\mathrm{est}} = 0.01266$ \\ ${S_p}_{\mathrm{est}} = 0.008912$}} & \multirow{2}{*}{\ValSecond{0.9877}} & \multirow{2}{*}{\ValBest{0.3170}} & \multirow{2}{*}{\ValSecond{0.0099}} \\ 

        & & & & \\

        \cmidrule(lr){1-5}

        \multirow{4}{*}{\shortstack{Pavia \\ Centre \\ (Case 5)}}
        & \multirow{2}{*}{\shortstack{$\sigma = 0.05$ \\ $S_p = 0.05$}} & \multirow{2}{*}{\ValBest{0.9791}} & \multirow{2}{*}{\ValBest{0.3008}} & \multirow{2}{*}{\ValSecond{0.0164}} \\ 

        & & & & \\
        \cmidrule(lr){2-5}
        
        & \multirow{2}{*}{\shortstack{$\sigma_{\mathrm{est}} = 0.05356$ \\ ${S_p}_{\mathrm{est}} = 0.05860$}} & \multirow{2}{*}{\ValSecond{0.9773}} & \multirow{2}{*}{\ValSecond{0.2796}} & \multirow{2}{*}{\ValBest{0.0067}} \\ 

        & & & & \\

        \bottomrule
    \end{tabular}
    }
    \\
\end{table}

\subsection{Summary}
We summarize the experimental discussion as follows:
\begin{itemize}
    \item The experimental results in Case 1 show that the proposed method achieves state-of-the-art detection performance.
          This is due to the proper modeling of the background and anomaly parts.

    \item The experimental results in Cases 2, 3, 4, and 5 show that the proposed method is robust to various types of noise.
          The reason for this is that the modeling of each noise is adequate, allowing for the estimation of the two parts simultaneously with noise removal.
          
    \item Among the background characterizations, the proposed method using HTV achieved the best detection performance.
          This is because it is most reasonable to characterize the spatial piecewise smoothness of the background part.
\end{itemize}

\section{Conclusion}
\label{sec:conclusion}
In this article, we have proposed a noise-robust HS anomaly detection method. 
To explicitly handle mixed noise, we have modeled three types of noise and formulated a constrained convex optimization problem that jointly estimates the background, anomaly, and noise components. 
We have then developed an efficient algorithm based on P-PDS. 
Experimental results on seven HS datasets demonstrate that the proposed method achieves detection performance comparable to state-of-the-art methods on the original datasets, and exhibits strong robustness under various types of noise.

However, several challenges remain.
First, the detection performance may degrade when the noise deviates from the assumed models, such as stripe noise superimposed along oblique directions. 
Relaxing the strict flatness constraint is a possible direction for handling such cases. 
Second, the separation between the anomaly part and the sparse noise can become ambiguous, since both are characterized by sparsity-inducing terms.
Incorporating a regularization that more explicitly 
promotes the spectral continuity of the anomaly part may 
help mitigate this issue.
Third, the background characterization may be enhanced by more expressive priors, such as non-convex regularization, to capture complex background structures. 
Addressing these challenges will be the subject of our future work.

\ifCLASSOPTIONcaptionsoff
  \newpage
\fi

\bibliographystyle{IEEEtran}
\bibliography{./bibtex/IEEEabrv,./bibtex/myrefs}

\begin{IEEEbiography}[{\includegraphics[width=1in,height=1.25in,clip,keepaspectratio]{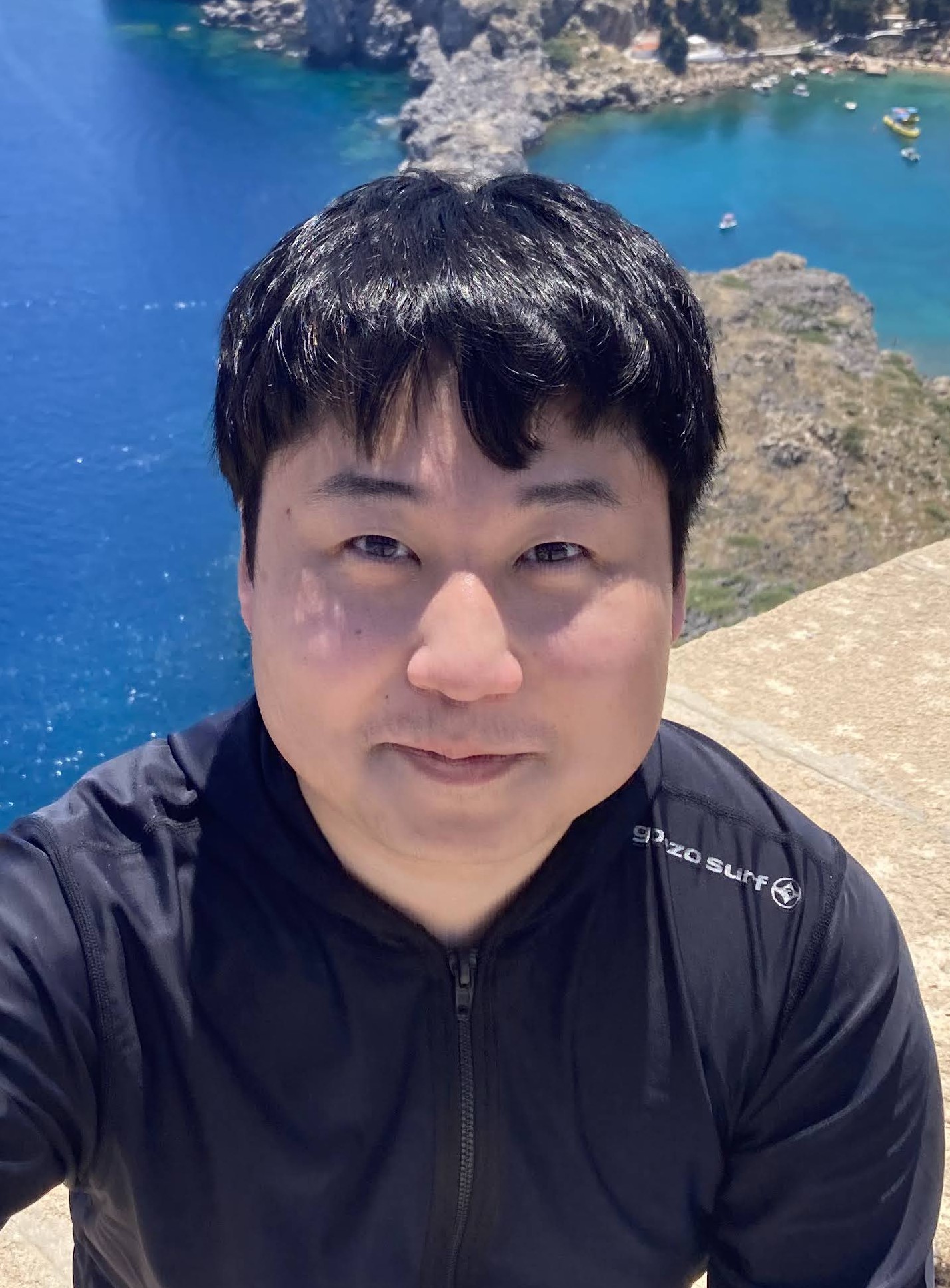}}] {Koyo Sato} (S’23) received B.E. and M.E. degrees in Information and Computer Science from the Tokyo Institute of Technology in 2022 and 2024, respectively. 
He is currently pursuing a Ph.D. degree with the Department of Computer Science at the Institute of Science Tokyo.
His current research interests include signal and image processing, mathematical optimization, and remote sensing.
Since April 2026, he has been a Research Fellow (DC2) with the Japan Society for the Promotion of Science (JSPS).  
\end{IEEEbiography}
  
\begin{IEEEbiography}[{\includegraphics[width=1in,height=1.25in,clip,keepaspectratio]{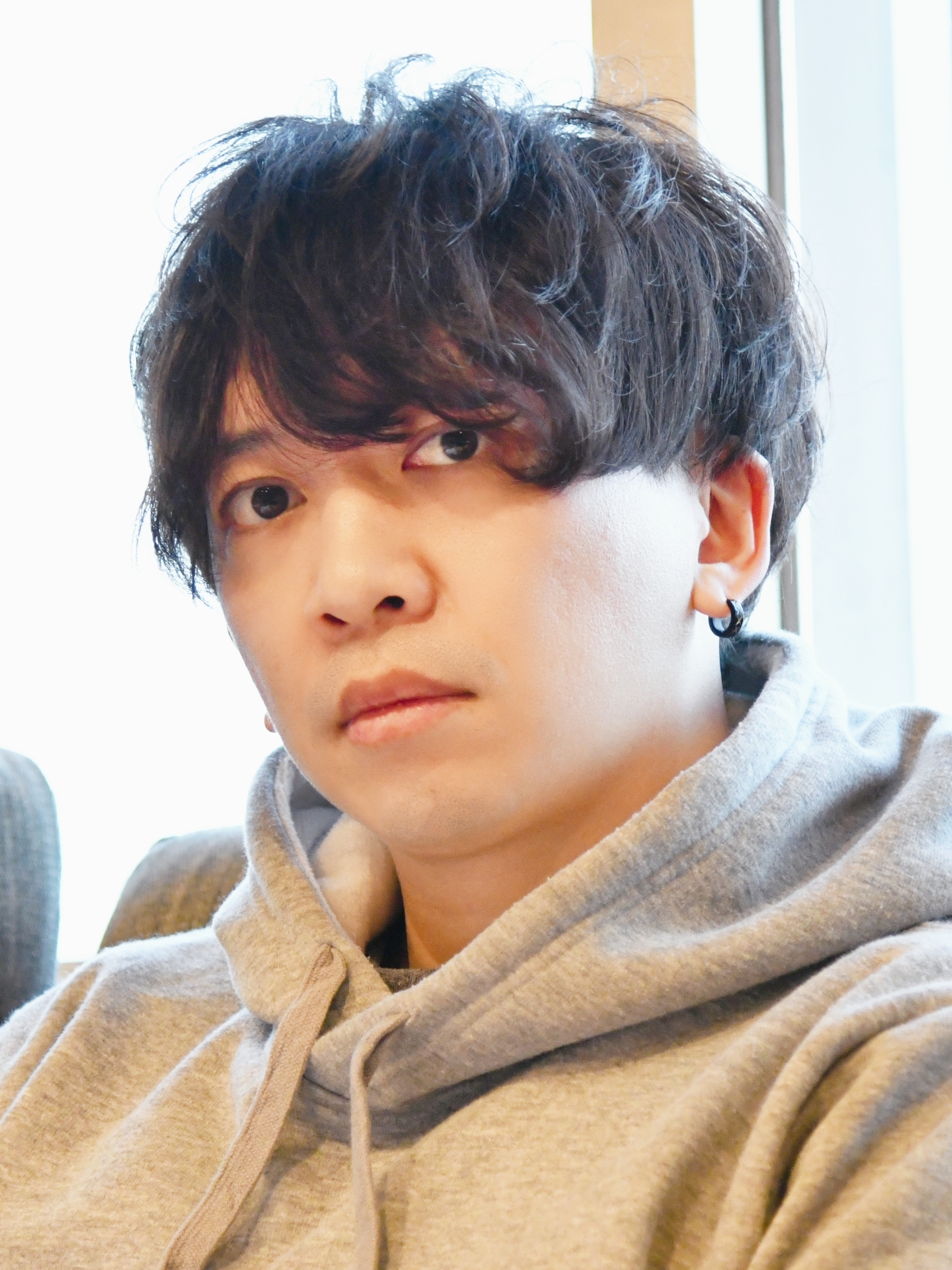}}]{Shunsuke Ono} (S’11--M’15--SM'23) received the B.E. degree in Computer Science in 2010 and the M.E. and Ph.D. degrees in Communications and Computer Engineering in 2012 and 2014, respectively, from the Tokyo Institute of Technology. From 2012 to 2014, he was a Research Fellow (DC1) of the Japan Society for the Promotion of Science (JSPS). He was an Assistant Professor and then an Associate Professor at the Tokyo Institute of Technology (TokyoTech), Tokyo, Japan, from 2014 to 2024. From 2016 to 2020, he was a Researcher with the Precursory Research for Embryonic Science and Technology (PRESTO), Japan Science and Technology Agency (JST), Tokyo, Japan. Following the institutional merger that established the Institute of Science Tokyo (Science Tokyo) in 2024, he continued as an Associate Professor at the new university until 2026, and has been a Professor in the same department since then. His research interests include signal processing, image analysis, optimization, remote sensing, and measurement informatics. He has served as an Associate Editor for IEEE TRANSACTIONS ON SIGNAL AND INFORMATION PROCESSING OVER NETWORKS (2019--2024). Dr. Ono received the Young Researchers' Award and the Excellent Paper Award from the IEICE in 2013 and 2014, respectively, the Outstanding Student Journal Paper Award and the Young Author Best Paper Award from the IEEE SPS Japan Chapter in 2014 and 2020, respectively, and the Best Paper Award at APSIPA ASC 2024. He also received the Funai Research Award in 2017, the Ando Incentive Prize in 2021, the MEXT Young Scientists' Award in 2022, the IEEE SPS Outstanding Editorial Board Member Award in 2023, the KDDI Foundation Award in 2025, and the Special Award for Science Tokyo Advanced Researchers (STAR) in 2026.
  \end{IEEEbiography}

\end{document}